\documentclass[a4paper,10pt,final]{amsart}

\usepackage[utf8]{inputenc}
\usepackage{amsthm, amssymb, mathtools, dsfont, bm}
\usepackage[margin=3.7cm]{geometry}
\usepackage[dvipsnames]{xcolor}
\usepackage[scr=rsfso]{mathalpha}
\usepackage[hide links]{hyperref}
\hypersetup{pdftitle={Haag Duality for 2D Quantum Spin Systems},
pdfauthor={Yoshiko Ogata, David P\'erez-Garc\'ia and Alberto Ruiz-de-Alarc\'on},
pdfkeywords={Quantum Spin Systems, Topological Order, Haag Duality, Tensor Networks, Quantum Phases of Matter, Operator Algebras}}

\usepackage[activate={true,nocompatibility},final,tracking=true,kerning=true,spacing=true,factor=1100,stretch=10,shrink=10]{microtype}
\SetTracking{encoding={*}, shape=sc}{10}
\microtypecontext{spacing=nonfrench}

\allowdisplaybreaks

\usepackage{xpatch}
\xpatchcmd{\proof}{\itshape}{\normalfont\bfseries}{}{}

\usepackage[noabbrev, capitalize, nameinlink]{cleveref}
\creflabelformat{equation}{#2\textup{#1}#3}
\crefname{enumi}{Item}{Items}
\newtheorem{theorem}{Theorem}[subsection]
\newtheorem{lemma}[theorem]{Lemma}
\newtheorem{corollary}[theorem]{Corollary}
\newtheorem{proposition}[theorem]{Proposition}
\newtheorem{claim}[theorem]{Claim}
\newtheorem{mainresult}{Main result}
\newtheorem*{theorem*}{Theorem}
\theoremstyle{definition}
\newtheorem{definition}[theorem]{Definition}
\newtheorem{definitions}[theorem]{Definitions}
\newtheorem{example}[theorem]{Example}
\newtheorem{remark}[theorem]{Remark}
\newtheorem*{remark*}{Remark}
\newtheorem*{notation}{Notation}
\newtheorem*{definition*}{Definition}
\newtheorem{setting}[theorem]{Setting}
\newtheorem{assumption}[theorem]{Assumption}

\newtheorem{caseHD}{Case}
\newtheorem{stepHD}{Step}
\newtheorem{stepPT}{Case}
\newtheorem{stepPTproof}{Item}
\newtheorem{stepSUFF}{Step}

\Crefname{theorem}{Theorem}{Theorems}
\Crefname{lemma}{Lemma}{Lemmas}
\Crefname{corollary}{Corollary}{Corollaries}
\Crefname{proposition}{Proposition}{Propositions}
\Crefname{claim}{Claim}{Claims}
\Crefname{definition}{Definition}{Definitions}
\Crefname{definitions}{Definition}{Definitions}
\Crefname{example}{Example}{Examples}
\Crefname{remark}{Remark}{Remarks}
\Crefname{setting}{Setting}{Settings}
\Crefname{assumption}{Assumption}{Assumptions}
\AddToHook{env/lemma/begin}{\crefalias{theorem}{lemma}}
\AddToHook{env/corollary/begin}{\crefalias{theorem}{corollary}}
\AddToHook{env/proposition/begin}{\crefalias{theorem}{proposition}}
\AddToHook{env/claim/begin}{\crefalias{theorem}{claim}}
\AddToHook{env/definition/begin}{\crefalias{theorem}{definition}}
\AddToHook{env/definitions/begin}{\crefalias{theorem}{definition}}
\AddToHook{env/example/begin}{\crefalias{theorem}{example}}
\AddToHook{env/remark/begin}{\crefalias{theorem}{remark}}
\AddToHook{env/setting/begin}{\crefalias{theorem}{setting}}
\AddToHook{env/assumption/begin}{\crefalias{theorem}{assumption}}

\newtheorem*{acknowledgements}{Acknowledgements}
\newtheorem*{conflictofinterest}{Conflict of interest statement}
\newtheorem*{dataavailability}{Data availability statement}

\newcommand{\eval}[2]{\langle \hspace{1pt} #1 \hspace{0.5pt} , \hspace{0.5pt} #2 \hspace{1pt} \rangle}
\newcommand{\mybra}[1]{\langle \, #1 \, |}
\newcommand{\myket}[1]{| \, #1 \, \rangle}
\newcommand{\dualPhi}{\bar{\Phi}}
\newcommand{\Stab}{Q}
\newcommand{\Interior}[1]{{#1}^{\circ}}
\newcommand{\unit}{\mathds{1}}
\newcommand{\lV}{\left\Vert}
\newcommand{\rV}{\right\Vert}

\newcommand{\lmk}{\left(}
\newcommand{\rmk}{\right)}

\newcommand{\caH}{{\mathscr{H}}}
\newcommand{\caA}{{\mathcal A}}
\newcommand{\caB}{{\mathcal B}}
\newcommand{\caM}{{\mathcal M}}
\newcommand{\caN}{{\mathcal N}}

\newcommand{\caL}{{\mathcal L}}

\newcommand{\ld}{\Lambda}
\newcommand{\braket}[2]{\left\langle#1,#2\right\rangle}
\newcommand{\bbC}{\mathbb{C}}
\newcommand{\bbR}{\mathbb{R}}
\newcommand{\bbN}{\mathbb{N}}
\newcommand{\bbZ}{\mathbb{Z}}
\newcommand{\bbQ}{\mathbb{Q}}

\newcommand{\Ad}{\mathop{\mathrm{Ad}}\nolimits}
\newcommand{\caU}{{\mathcal U}}

\title{Haag Duality for 2D Quantum Spin Systems}
\author[Y.~Ogata]{Yoshiko~Ogata$^1$}
\author[D.~P\'erez-Garc\'ia]{David~P\'erez-Garc\'ia$^{2,3}$}
\author[A.~Ruiz-de-Alarc\'on]{Alberto~Ruiz-de-Alarc\'on$^{2,4}$}
\date{}
\thanks{\\[-1em]%
$^1$\,Research Institute for Mathematical Sciences, Kyoto University, 606-8502 Kyoto, Japan.\\
$^2$\,Faculty of Mathematics, Complutense University of Madrid, 28040 Madrid, Spain.\\
$^3$\,Institute of Mathematical Sciences ICMAT, CSIC-UAM-UC3M-UCM, 28049 Madrid, Spain.\\
$^4$\,Department of Mathematics, CUNEF Universidad, 28040 Madrid, Spain.}

\begin{document}

\begin{abstract}
Haag duality is a fundamental locality property introduced in the pioneering formulation of algebraic quantum field theory by Haag and Kastler in the 1960s. Since then, it has played a central role, most notably in the classification of superselection sectors by Doplicher, Haag, and Roberts in the 1970s.
Over the past two decades, this concept has migrated from its relativistic origins to quantum spin systems, becoming a cornerstone of the operator-algebraic approach to the long-standing problem of classifying two-dimensional topological quantum phases of matter. In physics, it is widely conjectured that such phases are classified by their emergent anyons, a view supported by exactly solvable models exemplifying all known non-chiral phases: Kitaev’s quantum double models, Levin–Wen string-net models, and their slight generalizations. In these models, elementary excitations behave as quasi-particles, namely anyons, whose fusion and braiding properties form a tensor category expected to characterize the phase of matter.
A major open problem was to derive the emergence of anyons and the stability of their fusion and braiding beyond these solvable models. Recently, it has been shown that a weaker, phase-stable form of Haag duality resolves these questions. However, rigorous proofs of Haag duality in two dimensions were previously restricted to systems exhibiting abelian anyons.
In this work, we establish Haag duality for a broad class of tensor network models based on $C^*$-weak Hopf algebras, encompassing all Kitaev quantum double and Levin–Wen string-net models, and expected to include all non-chiral topological quantum phases of matter.
\end{abstract}

\maketitle

\section{Introduction}
\label{Section:Intro}

The notion of Haag duality emerged in the 1960s as a refinement of the locality principle within the algebraic formulation of quantum field theory developed by Haag and Kastler \cite{haag_algebraic_1964}.
In this approach, observables are assigned to regions of spacetime through operator algebras, and Haag duality requires that the algebra associated with a region is equal to the commutant of the algebra associated with its causal complement. This condition ensures that the local algebras are maximal, i.e.~no additional observables can be added without violating causal independence.

Furthermore, Haag duality is a crucial assumption in the Doplicher–Haag–Roberts framework \cite{doplicher_local_1971,doplicher_local_1974}, which offers a rigorous method for classifying superselection sectors in algebraic quantum field theory. This approach identifies physically distinct particle types or charge configurations by examining representations of the observable algebra that are locally equivalent to the vacuum.
Such representations encapsulate the notion of charges being spatially localized, as their differences from the vacuum are confined to specific regions of spacetime.

Haag duality was first rigorously established for free Bose scalar fields in double cones by Araki in 1964 \cite{araki1963lattice,araki1964neumann}.  This result was later simplified by Eckmann and Osterwalder using Tomita–Takesaki modular theory  \cite{eckmann1973application,leyland1978quality}.
If the local algebras are generated by Weightman fields, Bisognano and Wichmann \cite{bisognano1975duality} showed Haag duality with respect to wedge-shaped regions, sometimes referred to as \emph{essential} Haag duality.
In 1990 the work by Buchholz and Schulz-Mirbach proved Haag duality for intervals in conformal field theory \cite{buchholz1990haag}.

In the context of lattice quantum spin systems, Haag duality was rigorously established in one-dimensional infinite spin chains during the mid-2000s \cite{keyl2006entanglement, matsui2010spectral}.
For two-dimensional quantum spin systems, Haag duality was first established by Naaijkens for cone-like regions in the toric code model \cite{naaijkens_haag_2012}. Fiedler and Naaijkens later extended this result to Kitaev’s quantum double models for all finite abelian groups \cite{fiedler_haag_2015}.
These developments have led to Haag duality becoming a standard assumption in the analysis of two-dimensional gapped quantum phases using operator algebraic methods---a line that has been extremely successful for the mathematical classification of quantum phases of matter \cite{ogata_Z2index_2020,ogata_classification_2020,ogata_invariant_2022,bourne_classification_2021,ogata_H3valued_2021}. 

The underlying idea is the following. Topological phases are characterized by the categorical properties of their elementary excitations, known as \emph{anyons}, in a large class of exactly solvable  models, which are expected to include representatives of all non-chiral topological phases of matter. It is conjectured that, in these models, anyons are in one-to-one correspondence with the superselection sectors, and Haag duality seems pivotal to prove that. Indeed, {\it assuming} Haag duality, \cite{bols_category_2025} proved that anyons and superselection sectors are equivalent braided $C^*$-tensor categories for non abelian Kitaev's quantum double models, generalizing the abelian case previously proven in \cite{fiedler_haag_2015}. 

Recently, a weaker version of the Haag duality property, called \emph{approximate} Haag duality, was defined in \cite{ogata_derivation_2022} and was shown to be stable within each quantum phase of matter---that is, stable under quasi-local automorphisms.
Furthermore, it was shown that a spectral gap, together with the approximate Haag duality, implies that superselection sectors can be endowed with the structure of a braided $C^*$ tensor category---stable within each phase---that allows to interpret superselection sectors as anyons. This shows the power of the superselection sector approach, since it can be applied to recover anyons as the classifying objects for topological phases throughout the whole phase, and not just at particular exactly solvable models. Approximate Haag duality has also been used as a key assumption in the recent  proof that the Hall conductance is quantized in an infinite plane geometry \cite{bachmann_tensor_2024}. 

On top of all that, Haag duality has also appeared as a fundamental assumption in the recent program to extend quantum information, usually restricted to finite dimensional systems, to the von Neumann algebra setting \cite{van_luijk_schmidt_2024,luijk_pure_2024,luijk_embezzlement_2025,luijk_large-scale_2025,van_luijk_critical_2025}.

The key question which remained open so far was to prove (approximate) Haag duality in quantum spin systems beyond the very few known cases: 1D systems and 2D abelian models.
%
Since approximate Haag duality is stable within a phase, it is enough to check Haag duality for a representative of each phase.
Here, we will consider 2D $C^*$-weak Hopf algebra injective tensor networks as introduced in \cite{molnar_matrix_2022}, which are in addition renormalization fixed points \cite{ruiz_matrix_2024}. They can be seen as an algebraic formulation of the family of MPO-injective PEPS defined in \cite{sahinoglu_characterizing_2021, bultinck_anyons_2017}, which is  essentially equivalent to the construction in \cite{lootens_matrix_2021}. They include as particular cases both Levin-Wen string-net models \cite{levin_string-net_2005} and Kitaev quantum double models \cite{kitaev_fault_2003}.
Our main results are the following.

\begin{mainresult}[Corollary~\ref{Cor:MR1}]
$C^*$-weak Hopf algebra injective two-dimensional tensor network states which are renormalization fixed points fulfill the approximate Haag duality property.
\end{mainresult}

\begin{mainresult}[Corollary~\ref{Cor:MR2}]
They also satisfy Haag duality on the coarse-grained lattice obtained by grouping disjoint plaquettes of four sites of the rectangular lattice into single sites.
\end{mainresult}

Our proof requires a combination of operator algebraic and tensor network techniques. 
In first place, it takes the same strategy as that in the work of Naaijkens \cite{naaijkens_haag_2012}. Namely, we project local algebras onto
subspaces so that we obtain a cyclic vector, and apply the theorem by Rieffel and van Daele \cite{rieffel1975commutation}.
In \cite{naaijkens_haag_2012}, that subspace was given in terms of string operators of the Toric code model. In general, there are a priori no such operators. Instead of looking for alternative operators, we simply take zero-energy spaces inside and outside of the cones as the subspace we project onto. As we can see from the sufficient condition for the Haag duality provided here, the Haag duality holds if zero-energy subspaces on $\Lambda$ entangles with that of $\Lambda^c$ efficiently. This property will be checked later for topologically ordered renormalization fixed point tensor network states associated to biconnected $C^*$-weak Hopf algebras. For that we will strongly rely on the tensor network constructions and results made in \cite{molnar_matrix_2022, ruiz_matrix_2024}.

Let us introduce the concrete setting and definitions. We consider quantum spin systems defined on the two-dimensional lattice $\mathbb{Z}^2$, following the algebraic approach developed in the operator algebraic framework \cite{bratteli_operator_1987, bratteli_operator_1997}. This formalism is particularly suited for describing systems with infinitely many degrees of freedom, such as quantum lattice models in the thermodynamic limit.
Let $d \geq 2$ be a fixed natural number. The local degrees of freedom at each site are modeled by the $C^*$-algebra $\mathrm{M}_d$ of complex $d \times d$ matrices, which represents the observable algebra of a spin with $d$ internal states. For each site $\boldsymbol{x} \in \mathbb{Z}^2$, we consider a copy of the $C^*$-algebra $\mathrm{M}_d$, denoted by $\mathcal{A}_{\smash{\{\boldsymbol{x}\}}}$. Furthermore, given a finite subset $\Sigma \subset \mathbb{Z}^2$, the associated local observable algebra is given by the tensor product
\[
\mathcal{A}_\Sigma := \bigotimes_{\boldsymbol{x} \in \Sigma} \mathcal{A}_{\{\boldsymbol{x}\}},
\]
which acts on the finite-dimensional Hilbert space $\bigotimes_{\boldsymbol{x} \in \Sigma} \mathbb{C}^d$. This identification allows us to interpret $\mathcal{A}_\Sigma$ as the algebra of all bounded operators on the local Hilbert space associated with the region $\Sigma$.
These local algebras form a net, in the sense that for any inclusion of two finite subsets $\Sigma_1 \subset \Sigma_2$ there is a natural embedding $\mathcal{A}_{\Sigma_1} \hookrightarrow \mathcal{A}_{\Sigma_2}$ obtained by tensoring elements of $\mathcal{A}_{\Sigma_1}$ with identities on $\Sigma_2 \setminus \Sigma_1$. This net of algebras allows us to define the global observable algebra over arbitrary regions of the lattice.
For any, possibly infinite, region $\Gamma \subset \mathbb{Z}^2$, we define the associated quasilocal algebra as the $C^*$-inductive limit:
\[
\mathcal{A}_\Gamma := \overline{\bigcup_{\substack{\Sigma \subset \Gamma \\ |\Sigma| < \infty}} \mathcal{A}_\Sigma},
\]
where the closure is taken in the norm topology.
We also denote $\bigcup_{\Sigma \subset \Gamma , |\Sigma| < \infty} \mathcal{A}_\Sigma$
by $\caA_{\Gamma,\rm loc}$. In particular, we set $\caA_{\rm loc}=\caA_{\bbZ,\rm loc}$. The algebra $\mathcal{A}_\Gamma$ captures all observables localized in $\Gamma$, including limits of sequences of observables supported in finite subsets. When $\Gamma = \mathbb{Z}^2$, we simply write $\mathcal{A} := \mathcal{A}_{\mathbb{Z}^2}$.

In many physical situations, one is interested in spatial regions defined geometrically in $\mathbb{R}^2$, rather than by lattice subsets alone. For this purpose, we adopt the convention that for $\Gamma \subset \mathbb{R}^2$, the notation $\mathcal{A}_\Gamma$ refers to the algebra associated with the discrete intersection $\Gamma \cap \mathbb{Z}^2$. The complement $\Gamma^c := \mathbb{R}^2 \setminus \Gamma$ will also be used to describe exterior regions.
\\[5pt]
A central object of interest in the algebraic framework is the notion of a \emph{state}, which represent physical configurations or preparations of a quantum system. A linear functional $\omega: \mathcal{A} \to \mathbb{C}$ is said to be \emph{positive} if
\[
\omega(A^* A) \geq 0 \quad \text{for all } A \in \mathcal{A},
\]
and is called a \emph{state} if it is normalized, i.e., $\omega(\unit) = 1$, where $\unit$ stands for the unit element of $\mathcal{A}$. A state $\omega$ is said to be \emph{pure} if it cannot be expressed as a nontrivial convex combination of other states; physically, pure states correspond to extremal preparations, not decomposable into mixtures.
\\[5pt]
A \emph{representation} of $\mathcal{A}$ is a pair $(\mathscr{H}, \pi)$, where $\mathscr{H}$ is a Hilbert space and $\pi: \mathcal{A} \to \mathcal{B}(\mathscr{H})$ is a $*$-homomorphism into the algebra of bounded operators on $\mathscr{H}$. 
Given a state $\omega$ on the global algebra $\mathcal{A}$, one can associate a canonical representation via the Gel\-fand--Nai\-mark--Se\-gal (GNS) construction: there exists a triple $(\mathscr{H}, \pi, \Omega_\omega)$, where $\mathscr{H}$ is a Hilbert space, $\pi: \mathcal{A} \to \mathcal{B}(\mathscr{H})$ is a representation, and $\Omega_\omega \in \mathscr{H}$ is a unit vector,
such that
\begin{align} \label{gns}
\omega(A) = \braket{\Omega_\omega}{\pi(A)\Omega_\omega}, \quad \text{for all } A \in \mathcal{A}, \quad \text{and} \quad \mathscr{H} = \overline{\pi(\mathcal{A}) \Omega_\omega},
\end{align}
where the bar denotes closure with respect to the norm topology. The latter condition ensures that $\Omega_\omega$ is cyclic for the representation $\pi(\mathcal{A})$. The GNS triple is unique up to unitary equivalence, and throughout this paper, we follow the convention where the inner product is linear in the second variable.

We now turn to the von Neumann algebras generated by local algebras. Given a subset $\Gamma \subset \mathbb{R}^2$, we consider the von Neumann algebra $\pi(\mathcal{A}_\Gamma)''$ generated by the image of the local algebra $\mathcal{A}_\Gamma$ in the GNS representation. To formalize this, recall that for a set $\mathcal{M} \subset \mathcal{B}(\mathscr{H})$, its commutant is defined as
\[
\mathcal{M}' := \{ T \in \mathcal{B}(\mathscr{H}) : [T, M] = 0 \text{ for all } M \in \mathcal{M} \},
\]
and the bicommutant $\mathcal{M}''$ is called the von Neumann algebra generated by $\mathcal{M}$. A von Neumann algebra is thus a unital $*$-subalgebra of $\mathcal{B}(\mathscr{H})$ that is equal to its bicommutant.
The focus of this paper is the structure of such local von Neumann algebras, particularly when $\Gamma$ is a cone, i.e.~any subset of the form
\[
\Lambda_{{\boldsymbol{a}}, \theta, \varphi} := \left\{ {\boldsymbol{a}} + t(\cos\beta, \sin\beta) : t > 0,\ \beta \in (\theta - \varphi, \theta + \varphi) \right\},
\]
where ${\boldsymbol{a}} \in \mathbb{R}^2$ is the apex, $\theta \in \mathbb{R}$ is the central angle, and $\varphi \in (0, \pi)$ is the opening half-angle.
For $\Lambda_{{\boldsymbol{a}}, \theta, \varphi}$, we set $|\arg\Lambda|:=2\varphi$,
and $\bm e_{\ld}:=(\cos\theta,\sin\theta)$.
As aforementioned, these regions play a central role in the analysis of lattice systems, and a particularly significant property of such systems is Haag duality, which is defined as follows:

\begin{definition*}
Let $\omega$ be a pure state on $\mathcal{A}$ with GNS triple $(\mathscr{H}, \pi, \Omega)$. We say that $\omega$ satisfies \emph{Haag duality} if for every cone $\Lambda \subset \mathbb{R}^2$,
\[
    \pi( \mathcal{A}_{\smash{\Lambda^c}} )'
    =
    \pi( \mathcal{A}_{\smash{\Lambda}} )''.
\]
\end{definition*}

Note that the inclusion $\pi(\mathcal{A}_\Lambda)'' \subset \pi(\mathcal{A}_{\Lambda^c})'$ always holds by locality, since observables localized in disjoint regions commute. Haag duality requires this inclusion to be an equality, thus imposing a maximality condition on the algebras of observables.

For each cone $\ld=\Lambda_{\bm a, \theta,\varphi}$
and $\varepsilon>0$, we set the fattened cone
\begin{align}
\ld_\varepsilon
:=\Lambda_{\bm a, \theta,\varphi+\varepsilon}.
\end{align}

The definition of approximate Haag duality is the following.

\begin{definition*}
Let $\omega$ be a pure state on $\mathcal{A}$ with GNS triple $(\mathscr{H}, \pi, \Omega)$. We say that $\omega$ satisfies 
the \emph{approximate Haag duality} if
the following condition holds:
For any $\varphi\in (0,2\pi)$ and 
 $\varepsilon>0$ with
$\varphi+4\varepsilon<2\pi$,
there is some $R_{\varphi,\varepsilon}>0$ and decreasing
functions $f_{\varphi,\varepsilon,\delta}(t)$, $\delta>0$
on $\bbR_{\ge 0}$
with $\lim_{t\to\infty}f_{\varphi,\varepsilon,\delta}(t)=0$
such that
\begin{enumerate}
\item[(i)]
for any cone $\Lambda$ with $|\operatorname{arg}(\Lambda)|=\varphi$, there is a unitary 
$U_{\Lambda,\varepsilon}\in \caU(\caH)$
satisfying
\begin{align}\label{lem7p}
\pi\lmk\caA_{\Lambda^c}\rmk'\subset 
\Ad\lmk U_{\Lambda,\varepsilon}\rmk\lmk 
\pi\lmk \caA_{\lmk \Lambda-R_{\varphi,\varepsilon}\bm e_\Lambda\rmk_\varepsilon}\rmk''
\rmk,
\end{align}
\item[(ii)]
 for any $\delta>0$ and $t\ge 0$, there is a unitary 
 $\tilde U_{\Lambda,\varepsilon,\delta,t}\in \pi ( \caA_{\Lambda_{\varepsilon+\delta}-t\bm e_{\Lambda}} )''$
 satisfying
\begin{align}\label{uappro}
\|
U_{\Lambda,\varepsilon}-\tilde U_{\Lambda,\varepsilon,\delta,t}
\| \le f_{\varphi,\varepsilon,\delta}(t).
\end{align}
\end{enumerate}
\end{definition*}

This paper is organized as follows. In \cref{Section:HaagReduction} we provide a sufficient condition for the Haag duality property via finite systems. In \cref{Section:Construction} we revisit and reformulate the construction of topologically ordered two-dimensional tensor network states. More concretely, we review the algebraic framework on $C^*$-weak Hopf algebras and their representation theory, the structure of algebras of matrix product operators based on these algebraic structures, and revisit and reformulate the construction and properties of renormalization fixed points projected entangled pair states. Moreover, we provide explicit constructions of their parent Hamiltonians, which we prove in addition to be commuting, and prove that they satisfy local topological quantum order. Finally, in \cref{Section:HaagForPEPS} we use the characterization provided in \cref{Section:HaagReduction} to prove the Haag duality property in those systems.

\section{A sufficient condition for Haag duality}
\label{Section:HaagReduction}

\subsection{A sufficient condition for Haag duality}

For a $*$-algebra $\caB$, $\caB_\mathrm{h}$ means the set of all  self-adjoint elements in $\caB$. The following is a well-known result, see IV Lemma~5.7 of \cite{takesaki_theory_2001}.
\begin{theorem}\label{ar}
Let $\caH$ be a Hilbert space, let $\caM$ and $\caN$ be von Neumann algebras acting on $\caH$ and assume that $\caM\subset \caN'$
and there exists a vector $\Omega\in \caH$ cyclic for $\caM$.
Then the following statements are equivalent:
\begin{enumerate}
\item $\caM=\caN'$;
\item
\(
\overline{\lmk \caM_\mathrm{h}+i\caN_\mathrm{h}\rmk\Omega}=\caH
\).
\end{enumerate}
\end{theorem}
\begin{setting}\label{seta}
Let $\omega_0$ be a pure state
on $\caA$
with a GNS triple $(\caH,\pi,\Omega)$.
Suppose that there is a uniformly bounded finite range interaction $\Phi$ on 
$\caA$
such that
\begin{equation}
\Phi(X)\ge 0\quad \text{for all finite subsets } X \text{ of } \bbZ^2,
\end{equation}
and
\begin{equation}
\omega_0(\Phi(X))=0\quad \text{for all finite subsets } X \text{ of } \bbZ^2.
\end{equation}
For any finite subset $\Lambda\subset\bbZ^2$, let
\begin{equation}
    H_{\Phi}(\Lambda) \coloneq \sum_{X\subset\Lambda}\Phi(X),
\end{equation}
and denote by $Q_\Lambda$
the spectral projection of $H_{\Phi}(\Lambda)$ corresponding to $0$.
\end{setting}
We first note the following basic fact, which is obvious from the definition.
\begin{lemma}
Consider the Setting \ref{seta}.
For each $\Gamma\subset \bbZ^2$, we set
\begin{equation}
\mathscr{K}_{\Gamma} \coloneq \bigcap_{X\subset\Gamma} \ker\pi\lmk \Phi(X)\rmk,
\end{equation}
and let $P_\Gamma$ be the orthogonal projection onto $\mathscr{K}_{\Gamma}$.
Then,
\begin{equation}
P_\Gamma\in \pi(\caA_{\Gamma})'',
\end{equation}
and
\begin{equation}
P_{\Gamma}\Omega=\Omega.
\end{equation}
\end{lemma}
Using this fact, we obtain the following.
\begin{lemma}\label{infinite}
Consider the Setting \ref{seta}.
Let $\Gamma\subset\bbZ^2$.
Suppose that
\begin{align}\label{asuit}
\begin{split}
\overline{
\lmk P_\Gamma P_{\Gamma^c} \pi(\caA_{\Gamma})''_\mathrm{h} P_{\Gamma^c} P_{\Gamma}+
iP_{\Gamma}P_{\Gamma^c}
\pi\lmk \caA_{\Gamma^c}\rmk''_\mathrm{h} P_\Gamma P_{\Gamma^c}\rmk\Omega
}
=P_{\Gamma}P_{\Gamma^c}\caH
\end{split}
\end{align}
and
\begin{align}\label{nias}
\begin{split}
P_{\smash{\Gamma^c}}\caH=\overline{\pi(\caA_{\Gamma})\Omega}.
\end{split}
\end{align}
Then we have
\begin{align}
\begin{split}
\pi\lmk\caA_{\Gamma^c}\rmk'=\pi\lmk\caA_{\Gamma}\rmk''.
\end{split}
\end{align}
\end{lemma}
\begin{proof}
\begin{stepSUFF}\label{Step1Suff}
First we claim
\begin{align}\label{risu1}
\begin{split}
\overline{\lmk \pi\lmk\caA_{\Gamma}\rmk''_\mathrm{h}+iP_{\Gamma^c}\pi\lmk \caA_{\Gamma^c}\rmk''_\mathrm{h} P_{\Gamma^c}\rmk \Omega}
=P_{\Gamma^c}\caH.
\end{split}
\end{align}
The left-hand side is clearly included in the right-hand side
because $P_{\Gamma^c}\Omega=\Omega$ and
$P_{\Gamma^c}\in \pi(\caA_\Gamma)'$.
To see the opposite inclusion, set
\begin{align}
\caL \coloneq \lmk \pi\lmk\caA_{\Gamma}\rmk''_\mathrm{h}+iP_{\Gamma^c}\pi\lmk \caA_{\Gamma^c}\rmk''_\mathrm{h} P_{\Gamma^c}\rmk \Omega.
\end{align}
Because of the assumption \cref{nias}, it suffices to show
\begin{align}
\pi\lmk \caA_{\Gamma}\rmk\Omega\subset \overline{\caL}.
\end{align}
Then, by the definition of $\caL$, it suffices to show
\begin{align}\label{kuma}
i\pi\lmk\caA_{\Gamma}\rmk_\mathrm{h}\Omega
\subset \overline{\caL}.
\end{align}
To see this, take an arbitrary $b\in \pi(\caA_{\Gamma})_\mathrm{h}$.
Then we have
\begin{align}
\begin{split}
ib\Omega
=i\lmk bP_\Gamma-P_\Gamma b\rmk\Omega+iP_\Gamma b\Omega.
\end{split}
\end{align}
Because 
\begin{align}
i\lmk bP_\Gamma-P_\Gamma b\rmk\in \pi\lmk\caA_{\Gamma}\rmk''_\mathrm{h},
\end{align}
the first part of the right-hand side satisfies
\begin{align}\label{neko}
\begin{split}
i\lmk bP_\Gamma-P_\Gamma b\rmk\Omega
\in \pi\lmk\caA_{\Gamma}\rmk''_\mathrm{h}\Omega\subset \caL.
\end{split}
\end{align}
For the second term, because $b\in \pi(\caA_{\Gamma})$ and 
$P_{\Gamma^c}\in \pi\lmk\caA_{\Gamma^c}\rmk''$ commute, we have
\begin{align}
\notag{}iP_\Gamma b\Omega&=iP_\Gamma P_{\Gamma^c} b\Omega \\
&\in P_{\Gamma}P_{\Gamma^c}\caH
=\overline{
\lmk P_\Gamma P_{\Gamma^c} \pi(\caA_{\Gamma})''_\mathrm{h} P_{\Gamma^c} P_{\Gamma}+
iP_{\Gamma}P_{\Gamma^c}
\pi\lmk \caA_{\Gamma^c}\rmk''_\mathrm{h} P_\Gamma P_{\Gamma^c}\rmk\Omega
}\\
\notag{}&=\overline{
\lmk P_\Gamma \pi(\caA_{\Gamma})''_\mathrm{h} P_{\Gamma}+
iP_{\Gamma^c}
\pi\lmk \caA_{\Gamma^c}\rmk''_\mathrm{h} P_{\Gamma^c}\rmk\Omega
}
\subset \overline{\caL},
\end{align}
from the assumption \cref{asuit}.
For the last inclusion we used that $P_\Gamma\in \pi(\caA_{\Gamma})''$,
hence
\begin{align}
P_\Gamma \pi(\caA_{\Gamma})''_\mathrm{h} P_\Gamma 
\subset  \pi(\caA_{\Gamma})''_\mathrm{h}.
\end{align}
This proves $ib\Omega\in\overline{\caL}$ for an arbitrary $b\in \pi(\caA_{\Gamma})_\mathrm{h}$
and \cref{kuma} holds, completing the proof of \cref{risu1}.
\end{stepSUFF}
\begin{stepSUFF}
We apply Theorem \ref{ar} 
with
$\caH$, $\caM$, $\caN$, $\Omega$
replaced by
$P_{\Gamma^c}\caH$, 
$\pi\lmk\caA_{\Gamma}\rmk'' P_{\Gamma^c}$,
$P_{\Gamma^c}\pi\lmk\caA_{\Gamma^c}\rmk'' P_{\Gamma^c}$,
$\Omega$.
Because 
\begin{align}\label{inu}
P_{\Gamma^c}\in \pi\lmk\caA_{\Gamma^c}\rmk'' \subset
\pi\lmk\caA_{\Gamma}\rmk',
\end{align}
$\pi\lmk\caA_{\Gamma}\rmk'' P_{\Gamma^c}$ and 
$P_{\Gamma^c}\pi\lmk\caA_{\Gamma^c}\rmk'' P_{\Gamma^c}$
are von Neumann algebras acting on $P_{\Gamma^c}\caH$.
\cref{inu} also implies
\begin{align}
\pi\lmk\caA_{\Gamma}\rmk'' P_{\Gamma^c}
\subset 
\lmk P_{\Gamma^c}\pi\lmk\caA_{\Gamma^c}\rmk'' P_{\Gamma^c}\rmk'.
\end{align}
We also have $\Omega=P_{\Gamma^c}\Omega\in 
P_{\Gamma^c}\caH$
is cyclic for $\pi\lmk\caA_{\Gamma}\rmk'' P_{\Gamma^c}$
in $P_{\Gamma^c}\caH$
because of the assumption \cref{nias}.
From \cref{Step1Suff}, we know \cref{risu1}
and
\begin{align}
\begin{split}
&\overline{\lmk \lmk \pi\lmk\caA_{\Gamma}\rmk'' P_{\Gamma^c}\rmk_\mathrm{h}
+i\lmk
P_{\Gamma^c}\pi\lmk \caA_{\Gamma^c}\rmk'' P_{\Gamma^c}\rmk_\mathrm{h}\rmk \Omega}
\\  &\qquad =\overline{\lmk \pi\lmk\caA_{\Gamma}\rmk''_\mathrm{h}+iP_{\Gamma^c}\pi\lmk \caA_{\Gamma^c}\rmk''_\mathrm{h} P_{\Gamma^c}\rmk \Omega}
=P_{\Gamma^c}\caH.
\end{split}
\end{align}
Therefore, Item~2 of \cref{ar} holds.
Applying \cref{ar}, we obtain
\begin{align}
\pi\lmk\caA_{\Gamma}\rmk'' P_{\Gamma^c}
=\lmk P_{\Gamma^c}\pi\lmk\caA_{\Gamma^c}\rmk'' P_{\Gamma^c}\rmk'
=\pi\lmk\caA_{\Gamma^c}\rmk'P_{\Gamma^c}.
\end{align}
Note that $\pi(\caA_{\Gamma^c})'$ is a factor because $\omega$ is pure:
\begin{align}
\begin{split}
\pi(\caA_{\Gamma^c})'\cap \pi(\caA_{\Gamma^c})''
\subset 
\pi(\caA_{\Gamma^c})'\cap \pi(\caA_{\Gamma})'
=\bbC\mathds{1}.
\end{split}
\end{align}
Therefore, the $\sigma w$-continuous
$*$-homomorphism
\begin{align}
\pi(\caA_{\Gamma^c})'\ni a
\mapsto a P_{\Gamma^c} \in \pi(\caA_{\Gamma^c})'P_{\Gamma^c}
\end{align}
is an injection.
From $\pi\lmk\caA_{\Gamma}\rmk''\subset \pi(\caA_{\Gamma^c})'$ we have
\begin{align}
\pi\lmk\caA_{\Gamma^c}\rmk'=
\pi\lmk\caA_{\Gamma}\rmk''.
\end{align}
\end{stepSUFF}
\end{proof}

\subsection{Reduction of the condition to finite systems}
Now we introduce a sufficient condition for
\cref{asuit,nias} to hold in terms of objects in finite systems.

\begin{assumption}\label{finite}
Consider the \cref{seta}.
Let $\Gamma\subset \bbZ^2$ be an infinite connected subset of $\bbZ^2$ which will be taken as a cone-like region as defined in \cref{Def:comblikelkmk}.
For each $N\in\bbN$, let $\Lambda_N:=[-N,N]^2\cap\mathbb{Z}^2$.
For each $N\in\bbN$, we set $\Gamma_N^{(R)} \coloneq \Gamma\cap \Lambda_N$
and $\Gamma_N^{(L)} \coloneq \Gamma^c\cap \Lambda_N$.
We also set
\[
\caH_N \coloneq \bigotimes_{{\boldsymbol{x}}\in \Lambda_N}\bbC^d,\quad
\caH_N^{(L)} \coloneq \bigotimes_{{\boldsymbol{x}}\in \Gamma_N^{(L)}}\bbC^d,\quad
\caH_N^{(R)} \coloneq \bigotimes_{{\boldsymbol{x}}\in \Gamma_N^{(R)}}\bbC^d.
\]
We assume the following conditions:
\begin{enumerate}
\setlength{\itemsep}{3pt}
\item For each $N\in\bbN$, there is a unit vector $\Omega_N\in \caH_N$
such that
\[ {\Omega_N}\in Q_{\Lambda_N}\caH_{N}
\subset {Q_{\smash{\Gamma_{N}^{(L)}}}\caH_N^{(L)}}
\otimes {Q_{\smash{\Gamma_{N}^{(R)}}}\caH_N^{(R)}} . \]
\item $\omega_0$ is the unique frustration-free state on $\caA$, with respect to the interaction $\Phi$.
\item For each $\ell\in\bbN$, there exist $\bbR$-linear maps 
\[ S_\ell^{(1)}: ( \caA_{\smash{\Gamma_\ell^{(R)}}} )_\mathrm{h}\to ( \caA_{\smash{\Gamma_{2\ell}^{(R)}}} )_\mathrm{h} , \quad S_\ell^{(2)}: ( \caA_{\smash{\Gamma_\ell^{(R)}}} )_\mathrm{h}\to ( \caA_{\smash{\Gamma_{2\ell}^{(L)}}} )_\mathrm{h}. \]
\item For each $\ell\in\bbN$, there exists a linear map
$\Theta_\ell: \caA_{\Lambda_\ell}\to \caA_{\smash{\Gamma_{2\ell}^{(R)}}}$. 
\item There exists a bounded function $f: \bbN^{\times 2}\to [0,\infty)$
such that
\[ \lim_{s\to\infty}\limsup_{t\to\infty} f(t,s)= 0\]
satisfying
\begin{align}\label{Eq:QQQ}
\lV  \lmk iQ_{\smash{\Gamma_N^{(R)}}}A- \lmk{Q_{\smash{\Gamma_{N}^{(R)}}}} S_{\ell}^{(1)} (A)+i {Q_{\smash{\Gamma_{N}^{(L)}}}}S_{\ell}^{(2)}(A)\rmk\rmk\Omega_N\rV
\le \lV A\rV f(N,\ell)
\end{align}
for all $A\in  ( \caA_{\smash{\Gamma_\ell^{(R)}}} )_{\mathrm{h}}$,
and \begin{equation}\label{Eq:PropThetaL}
\lV
{Q_{\smash{\Gamma_{N}^{(L)}}}} B\Omega_N-\Theta_\ell(B)\Omega_N
\rV\le \lV B\rV f(N,\ell)
\end{equation}
for all $B\in  \caA_{\Lambda_\ell}$,
and
\begin{align}\label{QQ}
\begin{split}
&\lV {Q_{\smash{\Gamma_{N}^{(R)}}}}- {Q_{\smash{\Gamma_{3\ell}^{(R)}}}}
Q_{ \smash{( \Gamma_{2\ell}^{(R)} )^c\cap \Gamma_{N}^{(R)}}}
\rV \, \big( 1+\| S_\ell^{(1)}\|\big)\leq f(N,\ell),\\
& \lV {Q_{\smash{\Gamma_{N}^{(L)}}}}- {Q_{\smash{\Gamma_{3\ell}^{(L)}}}}
Q_{ \smash{( \Gamma_{2\ell}^{(L)}) ^c\cap \Gamma_{N}^{(L)} }}
\rV  \, \big( 1+\| S_\ell^{(2)} \| \big)
\leq  f(N,\ell).
\end{split}
\end{align}
\end{enumerate}
\end{assumption}

\begin{remark}\label{arem}
Pureness of $\omega_0$ follows from condition (2).
It also implies \[ \omega_0(A)=\lim_{N\to\infty} \lmk {\Omega_N}, 
{ A\Omega_N}\rmk \] for each $A\in \caA_{\rm loc}$.
In (3), (4), $2\ell$ can be replaced by some other function $h(\ell)$ of $\ell$ such that $h(\ell)-\ell\to\infty$ as $\ell\to\infty$.
For the model considered in this paper we may take $f(N,\ell)=0$, for $N-l$ large enough.
\end{remark}

\begin{lemma}\label{Lemma:HaagReduction}
Let $\Gamma\subset \bbZ^2$ be an infinite connected subset of $\bbZ^2$.
Consider the \cref{seta}.
If \cref{finite} holds for $\Gamma$, then
we have 
\begin{align}
\begin{split}
\pi\lmk\caA_{\Gamma^c}\rmk'=\pi\lmk\caA_{\Gamma}\rmk''.
\end{split}
\end{align}
\end{lemma}
\begin{proof}
We set $g(s):=\limsup_{t\to\infty} f(t,s)$.
From Lemma \ref{infinite}, our goal is to show \cref{asuit,nias}.
Now we show \cref{nias}, for which it suffices to prove that
\begin{align}\label{shika}
P_{\Gamma^c}\pi\lmk \caA_{\rm loc}\rmk\Omega\subset
\overline{\pi(\caA_\Gamma)\Omega},
\end{align}
because $\overline{\pi(\caA_{\rm loc})\Omega}=\caH$.
To prove this, take
an arbitary $B\in \caA_{\rm loc}$.
For any $\ell\in \bbN$ large enough so that 
\begin{align}
B\in \caA_{\Lambda_\ell},
\end{align}
we have
\begin{align*}
\|
P_{\Gamma^c}
\pi(B)
\Omega
&-\pi\lmk\Theta_\ell(B)\rmk\Omega
\|
=\lim_{N\to\infty}
\lV
\pi\lmk Q_{\smash{  \Gamma^{(L)}_N  }}\rmk
\pi(B)
\Omega
-\pi\lmk\Theta_\ell(B)\rmk\Omega
\rV\\
&\le
\limsup_{N\to\infty}
\lV
{Q_{ \smash{ \Gamma_{N}^{(L)}} }  }- 
{Q_{\smash{ \Gamma_{3\ell}^{(L)}} } }Q_{ \smash{ ( \Gamma_{2\ell}^{(L)} ) ^c\cap \Gamma_{N}^{(L)} }  }
\rV\lV B\rV
\\
& \phantom{\leq \limsup_{N\to\infty}\quad} +
\lV
\pi\lmk {Q_{\Gamma_{3\ell}^{(L)}}}Q_{\lmk \Gamma_{2\ell}^{(L)}\rmk^c\cap \Gamma_{N}^{(L)}}\rmk
\pi(B)
\Omega
-\pi\lmk\Theta_\ell(B)\rmk\Omega
\rV\\
&=
\limsup_{N\to\infty}
\lV
{Q_{\Gamma_{N}^{(L)}}}- 
{Q_{\Gamma_{3\ell}^{(L)}}}Q_{\lmk \Gamma_{2\ell}^{(L)}\rmk^c\cap \Gamma_{N}^{(L)}}
\rV\lV B\rV
\\
&\phantom{\leq \limsup_{N\to\infty}\quad}
+
\lV
\pi\lmk {Q_{\Gamma_{3\ell}^{(L)}}}\rmk
\pi(B)
\Omega
-\pi\lmk\Theta_\ell(B)\rmk\Omega
\rV\\
& \le \limsup_{N\to\infty}\lV B\rV\lmk f(N,\ell)\rmk
+
\lV
\pi\lmk {Q_{\Gamma_{3\ell}^{(L)}}}\rmk
\pi(B)
\Omega
-\pi\lmk\Theta_\ell(B)\rmk\Omega
\rV\\
&\le\lV B\rV g(\ell)
+\omega_0\lmk
\lmk {Q_{\Gamma_{3\ell}^{(L)}}}
B
-\Theta_\ell(B)\rmk^*
\lmk {Q_{\Gamma_{3\ell}^{(L)}}}
B
-\Theta_\ell(B)\rmk
\rmk^{\frac 12}
\\
&=\lV B\rV g(\ell)
+
\lim_{N\to\infty}
\lV
 {Q_{\Gamma_{3\ell}^{(L)}}}
B
\Omega_N
-\Theta_\ell(B)\Omega_N
\rV\\
&
=
\lV B\rV g(\ell)
+
\lim_{N\to\infty}
\lV
 {Q_{\Gamma_{3\ell}^{(L)}}}Q_{\lmk \Gamma_{2\ell}^{(L)}\rmk^c\cap \Gamma_{N}^{(L)}}
B
\Omega_N
-\Theta_\ell(B)\Omega_N
\rV\\
&\le
\lV B\rV g(\ell)+\limsup_{N\to\infty}
\lV
 {Q_{\Gamma_{N}^{(L)}}}
B
\Omega_N
-\Theta_\ell(B)\Omega_N
\rV
\\ &\phantom{\leq \limsup_{N\to\infty}\quad} +
\lV {Q_{\Gamma_{N}^{(L)}}}- {Q_{\Gamma_{3\ell}^{(L)}}}
Q_{\lmk \Gamma_{2\ell}^{(L)}\rmk^c\cap \Gamma_{N}^{(L)}}
\rV\lV B\rV\\
&\le  \lV B\rV g(\ell)+2\limsup_{N\to\infty}\lV B\rV\lmk f(N,\ell)\rmk
\\
&\le
 3\lV B\rV g(\ell).
\end{align*}
Hence for any $B\in \caA_{\rm loc}$ we have
\begin{align}
\begin{split}
P_{\Gamma^c}
\pi(B)
\Omega
=\lim_{\ell\to\infty}\pi\lmk\Theta_\ell(B)\rmk\Omega
\in \overline{\pi(\caA_\Gamma)\Omega},
\end{split}
\end{align}
showing \cref{shika} and hence \cref{nias}.

Next we show \cref{asuit}.
The inclusion
\begin{align}
\begin{split}
\overline{
\lmk P_\Gamma P_{\Gamma^c} \pi(\caA_{\Gamma})''_\mathrm{h} P_{\Gamma^c} P_{\Gamma}+
iP_{\Gamma}P_{\Gamma^c}
\pi\lmk \caA_{\Gamma^c}\rmk''_\mathrm{h} P_\Gamma P_{\Gamma^c}\rmk\Omega
}
\subset P_{\Gamma}P_{\Gamma^c}\caH.
\end{split}
\end{align}
is trivial.
In order to show the other direction,
from \cref{nias}, it suffices to show that
\begin{align}
P_\Gamma \pi\lmk\caA_{\Gamma,\mathrm{loc}}\rmk\Omega
\subset 
\overline{
\lmk P_\Gamma P_{\Gamma^c} \pi(\caA_{\Gamma})''_\mathrm{h} P_{\Gamma^c} P_{\Gamma}+
iP_{\Gamma}P_{\Gamma^c}
\pi\lmk \caA_{\Gamma^c}\rmk''_\mathrm{h} P_\Gamma P_{\Gamma^c}\rmk\Omega
}.
\end{align}
Because we clearly have
\begin{align}
P_\Gamma \pi(\caA_{\Gamma,\mathrm{loc}})_{\mathrm{h}}\Omega
\subset 
\overline{
\lmk P_\Gamma P_{\Gamma^c} \pi(\caA_{\Gamma})''_\mathrm{h} P_{\Gamma^c} P_{\Gamma}+
iP_{\Gamma}P_{\Gamma^c}
\pi\lmk \caA_{\Gamma^c}\rmk''_\mathrm{h} P_\Gamma P_{\Gamma^c}\rmk\Omega
},
\end{align}
it suffices to show
\begin{align}\label{ushi}
iP_\Gamma \pi(\caA_{\Gamma,\mathrm{loc}})_\mathrm{h}\Omega
\subset 
\overline{
\lmk P_\Gamma P_{\Gamma^c} \pi(\caA_{\Gamma})''_\mathrm{h} P_{\Gamma^c} P_{\Gamma}+
iP_{\Gamma}P_{\Gamma^c}
\pi\lmk \caA_{\Gamma^c}\rmk''_\mathrm{h} P_\Gamma P_{\Gamma^c}\rmk\Omega
}.
\end{align}
Take an arbitrary $A\in \caA_{\Gamma,\mathrm{loc},\mathrm{h}}$.
Then 
for any $\ell\in \bbN$ large enough so that 
\begin{align}
A\in \caA_{ \smash{ \Gamma_\ell^{(R)}}  },
\end{align}
we have
\begin{align*}
 & \lV
 iP_{\Gamma}\pi(A)\Omega
 - \lmk
 P_\Gamma \pi\big( S_{\ell}^{(1)} (A)\big)\Omega
 +i {P_{\Gamma^c}}\pi \big( S_{\ell}^{(2)}(A) \big)\Omega
 \rmk
\rV\\
&=\lim_{N\to\infty}
\lV
 i\pi\big( Q_{\smash{\Gamma_N^{(R)}}}\big)\pi(A)\Omega
 - \Big(
 \pi \big( Q_{\smash{\Gamma_N^{(R)}}}\big) \pi\big( S_{\ell}^{(1)} (A)\big)\Omega
 +i \pi\big( Q_{\smash{\Gamma_N^{(L)}}}\big)\pi\big( S_{\ell}^{(2)}(A)\big)\Omega
 \Big)
\rV\\
&\leq\limsup_{N\to\infty} \bigg(
\lV {Q_{\smash{\Gamma_{N}^{(R)}}}}- {Q_{\smash{\Gamma_{3\ell}^{(R)}}}}
Q_{\smash{ ( \Gamma_{2\ell}^{(R)} )^c\cap \Gamma_{N}^{(R)}} }
\rV\cdot \lmk 1+\| S_\ell^{(1)}\|\rmk\\
& \hphantom{\le\limsup_{N\to\infty}\Big(\; }\qquad 
+\lV  Q_{ \smash{ \Gamma_{N}^{(L)}}}  - {Q_{ \smash{ \Gamma_{3\ell}^{(L)}}  }}
Q_{ \smash{ ( \Gamma_{2\ell}^{(L)} )^c\cap \Gamma_{N}^{(L)}} }
\rV \cdot \lmk 1+\lV S_\ell^{(2)}\rV\rmk
\bigg) \lV A \rV
\\
&+\limsup_{N\to\infty}
\Big\|
i \pi\Big( {Q_{\Gamma_{3\ell}^{(R)}}}
Q_{ ( \Gamma_{2\ell}^{(R)} )^c\cap \Gamma_{N}^{(R)}} \Big)
 \pi(A)\Omega
\\& \hphantom{\le\limsup_{N\to\infty}\Big(\; }\qquad
- \bigg(
 \pi\big( {Q_{\Gamma_{3\ell}^{(R)}}}
Q_{( \Gamma_{2\ell}^{(R)})^c\cap \Gamma_{N}^{(R)}}\big)
 \pi \lmk S_{\ell}^{(1)} (A)\rmk \Omega
\\& \hphantom{\le\limsup_{N\to\infty}\; }\qquad\qquad\qquad
+i \pi\Big( Q_{\Gamma_{3\ell}^{(L)}}
Q_{ ( \Gamma_{2\ell}^{(L)} )^c\cap \Gamma_{N}^{(L)}}\Big)
\pi\Big( S_{\ell}^{(2)}(A) \Big)\Omega
 \bigg)
\Big\|\\
&\le\limsup_{N\to\infty}
2\lmk f(N,\ell)\rmk\lV A\rV\\
&+
\lV
\begin{gathered}
i\pi\lmk {Q_{\Gamma_{3\ell}^{(R)}}}
\rmk
 \pi(A)\Omega\\
 - \lmk
 \pi\lmk {Q_{\Gamma_{3\ell}^{(R)}}}\rmk
 \pi\lmk S_{\ell}^{(1)} (A)\rmk\Omega
 +i \pi\lmk {Q_{\Gamma_{3\ell}^{(L)}}}\rmk
\pi\lmk S_{\ell}^{(2)}(A)\rmk\Omega
 \rmk
\end{gathered}
\rV\\
&\le 2g(\ell)\lV A\rV \\
&\quad\quad +\omega_0
\bigg( \lmk i {Q_{\Gamma_{3\ell}^{(R)}}}A-
\lmk {Q_{\Gamma_{3\ell}^{(R)}}}S_{\ell}^{(1)} (A)
+i{Q_{\Gamma_{3\ell}^{(L)}}} S_{\ell}^{(2)}(A)\rmk
\rmk^* \cdot 
\\
&\qquad\qquad\qquad\cdot
\lmk i {Q_{\Gamma_{3\ell}^{(R)}}}A-
\lmk {Q_{\Gamma_{3\ell}^{(R)}}}S_{\ell}^{(1)} (A)
+i{Q_{\Gamma_{3\ell}^{(L)}}} S_{\ell}^{(2)}(A)\rmk
\rmk\bigg)^{\frac 12}\\
&=2g(\ell)\lV A\rV +\lim_{N\to\infty}
\lV\lmk
 i {Q_{\Gamma_{3\ell}^{(R)}}}A-
\lmk {Q_{\Gamma_{3\ell}^{(R)}}}S_{\ell}^{(1)} (A)
+i{Q_{\Gamma_{3\ell}^{(L)}}} S_{\ell}^{(2)}(A)\rmk
\rmk
\Omega_N\rV\\
&=
2g(\ell)\lV A\rV +\lim_{N\to\infty}
\bigg\|
 i  \lmk {Q_{\Gamma_{3\ell}^{(R)}}}
Q_{\lmk \Gamma_{2\ell}^{(R)}\rmk^c\cap \Gamma_{N}^{(R)}}\rmk
A\Omega_N\\
&\hspace{4cm}-\bigg(  \lmk {Q_{\Gamma_{3\ell}^{(R)}}}
Q_{\lmk \Gamma_{2\ell}^{(R)}\rmk^c\cap \Gamma_{N}^{(R)}}\rmk
S_{\ell}^{(1)} (A)
\\
&\hspace{5cm}+i \lmk {Q_{\Gamma_{3\ell}^{(L)}}}
Q_{\lmk \Gamma_{2\ell}^{(L)}\rmk^c\cap \Gamma_{N}^{(L)}}\rmk
S_{\ell}^{(2)}(A)\bigg)
\Omega_N
\bigg\|
\\
&\le
4g(\ell)\lV A\rV
+\limsup_{N\to\infty}
\Big\|
 i  \lmk {Q_{\smash{\Gamma_{N}^{(R)}}}}\rmk
A\Omega_N
-
\lmk  \lmk {Q_{\smash{\Gamma_{N}^{(R)}}}}\rmk
S_{\ell}^{(1)} (A)
+i \lmk {Q_{\smash{\Gamma_{N}^{(L)}}}}\rmk
S_{\ell}^{(2)}(A)\rmk
\Omega_N
\Big\|\\
&\le 5 g(\ell)\lV A\rV.
\end{align*}
Therefore,
\begin{align}
\begin{split}
iP_{\Gamma}\pi(A)\Omega & =\lim_{\ell\to\infty} \lmk
 P_\Gamma \pi\lmk S_{\ell}^{(1)} (A)\rmk\Omega
 +i {P_{\Gamma^c}}\pi\lmk S_{\ell}^{(2)}(A)\rmk\Omega
 \rmk\\
&\in 
\overline{
\lmk P_\Gamma P_{\Gamma^c} \pi(\caA_{\Gamma})''_\mathrm{h} P_{\Gamma^c} P_{\Gamma}+
iP_{\Gamma}P_{\Gamma^c}
\pi\lmk \caA_{\Gamma^c}\rmk''_\mathrm{h} P_\Gamma P_{\Gamma^c}\rmk\Omega
}.
\end{split}
\end{align}
As this holds for any $A\in \caA_{\Gamma,\mathrm{loc},\mathrm{h}}$,
we obtain \cref{ushi}, and hence \cref{asuit}.
\end{proof}

\section{Construction and properties of topologically ordered 2D states}
\label{Section:Construction}

In this section, we revisit the construction of two-dimensional tensor network states exhibiting topological order from an algebraic perspective, as recently formulated by Molnár et al.~in \cite{molnar_matrix_2022}. Those states are expected to  coincide exactly with the family of MPO-injective PEPS based on bimodule categories constructed in \cite{lootens_matrix_2021}. Along the way, we establish new results concerning these models, which are of independent interest.

We commence \cref{Subsection:Algebra} by introducing the framework for describing the symmetries exhibited by these systems, based on $C^*$-weak Hopf algebras \cite{bohm_coassociativec_1996,bohm_weak_1999,bohm_weak_2000,nikshych_semisimple_2004,etingof_tensor_2015}, whose representation categories encompass unitary multifusion categories \cite{etingof_fusion_2005}.
These algebraic structures extend the scope of symmetries in quantum systems induced by finite groups, describing models of all known topologically ordered phases of matter, e.g. Kitaev's quantum double models based on $C^*$-weak Hopf algebras \cite{kitaev_fault_2003,buerschaper_hierarchy_2013,molnar_matrix_2022,chang_kitaev_2014}, or string-net models based on unitary (multi)fusion categories \cite{levin_string-net_2005,jia_weak_2024}.
In \cref{Subsection:MPOs} we review how representations of $C^*$-weak Hopf algebras give rise to families of $C^*$-algebras of matrix product operators, which are coherent for all system sizes \cite{molnar_matrix_2022}, and in \cref{Subsection:Topo} we reformulate the algebraic notions related to the topological properties of the corresponding algebras, including the \emph{pulling-through identities}.
Building on prior work on algebras of matrix product operators, in \cref{Subsection:PEPS} we discuss the construction of two-dimensional tensor network states \cite{sahinoglu_characterizing_2021,bultinck_anyons_2017,molnar_matrix_2022},
which inherently display features of renormalization fixed points \cite{molnar_matrix_2022,ruiz_matrix_2024,cirac_matrix_2021}. These exhibit their topological properties via pulling-through movements of the string-like symmetry operators \cite{sahinoglu_characterizing_2021,molnar_matrix_2022}, such as the degeneracy of the ground state space for non-trivial geometries, the non-local nature of the symmetries, and, as we present in \cref{Subsection:BulkBoundary}, a rigorous bulk-boundary correspondence.
In \cref{Subsection:PH} we provide explicit constructions of commuting local parent Hamiltonians for these models.
Finally, \cref{Subsection:TQO} is devoted to prove that a tight subfamily of the states constructed above, induced by \emph{biconnected} $C^*$-weak Hopf algebras, satisfy the \emph{local topological quantum order} condition.

\subsection{Algebraic framework}
\label{Subsection:Algebra}

From now on, we assume that all vector spaces, tensor products, and spaces of homomorphisms considered here are over the field of complex numbers, unless explicitly specified. For a vector space $V$, let $V^* \coloneq \operatorname{Hom}(V,\mathbb{C})$ be the dual vector space, and let $\eval{\mathbin{\cdot}}{\mathbin{\cdot}}:V^*\times V\to\mathbb{C}$ stand for the corresponding natural pairing. Given a linear map $F:V\to W$ between two vector spaces $V$ and $W$, we let $F^t:W^*\to V^*$ denote the transpose map. In the case of two Hilbert spaces, we let $F^*:W\to V$ stand for the adjoint map. Furthermore, for a unital $C^*$-algebra $A$ the multiplication is denoted by juxtaposition of the factors, $1 \in A$ stands for the unit element, and let $\operatorname{Ad} x(y) = x y x^{-1}$ for any two elements $x,y\in A$, provided that $x$ is invertible.

The following provides a general notion of extensivity for quantum spin chains.
\begin{definition}
    \label{Def:Coalgebra}
    A  \emph{coalgebra} is a finite-dimensional vector space $C$ with 
    \begin{enumerate}
        \item\label{Item:Coassociativity} a linear map $\Delta:C\to C\otimes C$, called \emph{comultiplication}, satisfying 
    \begin{equation*}
        (\Delta \otimes \mathrm{Id}) \circ \Delta = (\mathrm{Id}\otimes \Delta) \circ \Delta;
    \end{equation*}
    \item\label{Item:Counit} a linear functional $\varepsilon:C\to \mathbb{C}$, calle \emph{counit}, compatible in the sense that
    \begin{equation*}
        (\varepsilon \otimes \mathrm{Id}) \circ \Delta = \mathrm{Id} = (\mathrm{Id}\otimes \varepsilon) \circ \Delta.
    \end{equation*}
    \end{enumerate}
\end{definition}
Namely, a coalgebra is a monoid in the opposite category of finite-dimensional complex vector spaces. Let
\begin{equation*}
    \Delta^{(1)} \coloneq \Delta, \quad \Delta^{(n+1)}\coloneq (\Delta\otimes \mathrm{Id}^{\otimes n})\circ \Delta^{(n)},
\end{equation*}
denote the successive applications of the comultiplication, for all $n\in\mathbb{N}$.

\begin{notation}
\label{Notation:Sweedler}
The coassociativity property in \cref{Item:Coassociativity} of \cref{Def:Coalgebra} is not expli\-ci\-tly taken into account in this formulation; this makes preferable to adopt a complementary notation simplifying the expressions where e.g.~products or evaluation of linear functionals on the tensor factors of a coproduct are involved. Here, it is particularly convenient to adopt Sweedler notation: for any element $x\in C$ and $n\in \mathbb{N}$, we let
\begin{equation*}
    x_{(1)}\otimes x_{(2)} \otimes \cdots \otimes x_{(n+1)} \coloneq \Delta^{(n)}(x),
\end{equation*}
a shorthand notation for sums of the form $\sum_{i} x_{1,i} \otimes \cdots \otimes x_{n+1,i}$; i.e.~note that the terms $x_{(i)}$ above cannot be treated as individual tensor factors.
\end{notation}

The following definitions are fundamental in the literature of coalgebras.
\begin{definition}
    \label{Def:BasicsCoalgebra}
    Let $C$ be a coalgebra. Then,
    \begin{enumerate}
    \item\label{Item:Cocentral}
        an element $x\in C$ is said to be \emph{cocentral} if it satisfies
        \begin{equation*}
            x_{(1)}\otimes x_{(2)} = x_{(2)}\otimes x_{(1)};
        \end{equation*}
    \item\label{Item:Nondegenerate}%
         $x\in C$ is \emph{non-degenerate} if for all $y\in C$ there exist $\phi,\psi\in C^\ast$ such that
        \begin{equation*}
            \eval{ \phi }{ x_{(1)} } x_{(2)} = y = x_{(1)} \eval{ \psi }{ x_{(2)} };
        \end{equation*}
    \item\label{Item:Comultiplicative}
        a linear map $F:C\to C$ is called \emph{comultiplicative} if, for all $x\in C$,
        \begin{equation*}
            F(x)_{(1)} \otimes F(x)_{(2)} = F(x_{(1)}) \otimes F(x_{(2)});
        \end{equation*}
    \item\label{Item:Anticomultiplicative}
        a linear map $F:C\to C$ is called \emph{anticomultiplicative} if, for all $x\in C$,
        \begin{equation*}
            F(x)_{(1)} \otimes F(x)_{(2)} = F(x_{(2)}) \otimes F(x_{(1)}).
        \end{equation*}
    \end{enumerate}
\end{definition}

The following structure, introduced by B\"{o}hm and Szlach\'{a}nyi in \cite{bohm_coassociativec_1996}, formalizes the minimal set of requirements expected of a $C^*$-algebraic framework capable of describing general symmetries in topologically ordered systems. In particular, it incorporates a coalgebra structure to capture extensivity, as well as a notion of invertibility or duality.

\begin{definition}

\label{Def:CWHA}

A \emph{$C^*$-weak Hopf algebra} is a finite-dimensional unital $C^*$-algebra $A$ equipped with the structure of a coalgebra for which the following holds:
\begin{enumerate}
\item\label{Item:MultiplicativeComult}
the comultiplication is multiplicative, i.e.~for all $x,y\in A$:
\begin{equation*}
    (xy)_{(1)} \otimes (xy)_{(2)} = x_{(1)} y_{(1)} \otimes x_{(2)} y_{(2)};
\end{equation*}
\item\label{Item:StarComult}
the $\ast$-operation is comultiplicative, i.e.~for all $x\in A$:
\begin{equation*}
    (x^\ast)_{(1)} \otimes (x^\ast)_{(2)} = (x_{(1)})^\ast \otimes (x_{(2)})^\ast;
\end{equation*}
\item\label{Item:UnitWeaklyComult}
the unit is \emph{weakly comultiplicative}, i.e.:
\begin{equation*}
    1_{(1)} \otimes 1_{(2)} \otimes 1_{(3)}
    =
    1_{(1)} \otimes 1_{(2)} 1_{(1')} \otimes 1_{(2')} 
    =
    1_{(1)} \otimes 1_{(1')} 1_{(2)} \otimes 1_{(2')};
\end{equation*}
\item\label{Item:CounitWeaklyMult}
the counit is \emph{weakly multiplicative}, i.e.~for all $x,y,z\in A$:
\begin{equation*}
    \varepsilon( xyz ) = \varepsilon( x y_{(1)} ) \varepsilon( y_{(2)}  z ) = \varepsilon( x y_{(2)} ) \varepsilon( y_{(1)} z );
\end{equation*}
\item \label{Item:AntipodeAxioms}
there exists a linear map $S:A\to A$, called \emph{antipode}, such that, 
for all $x\in A$:
\begin{equation*}
    x_{(1)} S( x_{(2)} ) = \eval{\varepsilon}{ 1_{(1)} x }  1_{(2)},\quad
    S( x_{(1)} ) x_{(2)} = 1_{(1)} \eval{\varepsilon}{x 1_{(2)}},
\end{equation*}
and it is both an antimultiplicative and anticomultiplicative map.
\end{enumerate}
\end{definition}

For further details, we refer the reader to Refs.~\cite{bohm_weak_1999, bohm_weak_2000, nikshych_semisimple_2004, etingof_fusion_2005}.

\begin{remark}
\label{Rem:DualCWHA}
The dual vector space $A^*$ is canonically endowed with
\begin{equation*}
    \eval{\phi\psi}{x}
        \coloneq \eval{\phi}{x_{(1)}}\eval{\psi}{x_{(2)}},
    \quad
    \eval{\phi^*}{x}
        \coloneq \overline{ \eval{\phi}{S(x^*)} }
\end{equation*}
for all $x\in A$, for which the unit element of $A^*$ is the counit of $A$, and
\begin{equation*}
    \eval{\phi_{(1)} \otimes \phi_{(2)}}{x\otimes y}\coloneq \eval{\phi}{x y},
    \quad
    \eval{\varepsilon}{\phi} \coloneq \eval{\phi}{1},
    \quad
    \eval{\hat S(\phi)}{x} \coloneq \eval{\phi}{S(x)},
\end{equation*}
for all $x,y\in A$ and $\phi,\psi\in A^*$, providing the structure of a $C^*$-weak Hopf algebra.
\end{remark}

In this context, there exist two relevant linear maps $\varepsilon^L,\varepsilon^R:A\to A$, defined by
\begin{equation}
    \label{Eq:DefEpsilonLR}
    \varepsilon^L(x) \coloneq x_{(1)} S(x_{(2)}),
    \quad
    \varepsilon^R(x) \coloneq S(x_{(1)})  x_{(2)},
\end{equation}
for all $x\in A$, known as \emph{source} and \emph{target counital maps}, respectively. Their images
\begin{equation}
    \label{Eq:DefAlgebrasLR}
    A^L \coloneq \varepsilon^L(A),
    \quad\text{ and }\quad 
    A^R\coloneq \varepsilon^R(A),
\end{equation}
constitute two commuting isomorphic separable $*$-subalgebras of $A$, known as the \emph{source} and \emph{target counital $*$-subalgebras} of $A$, respectively.

\begin{remark}\label{Rem:CharALAR}
\begin{enumerate}
    \item  $A^L$ is characterized as the $*$-subalgebra of $x\in A$ satisfying
\begin{equation*}
    x_{(1)} \otimes x_{(2)} = x 1_{(1)} \otimes 1_{(2)} = 1_{(1)} x \otimes 1_{(2)};
\end{equation*}
\item $A^R$ is characterized as the $*$-subalgebra of $y\in A$ satisfying
\begin{equation*}
    y_{(1)} \otimes y_{(2)} =  1_{(1)} \otimes y 1_{(2)} = 1_{(1)}  \otimes 1_{(2)} y.
\end{equation*}
\item\label{Item:Delta1inARAL} $1_{(1)}\otimes 1_{(2)}\otimes\cdots\otimes 1_{(n-1)}\otimes 1_{(n)}\in A^R\otimes A\otimes \cdots \otimes A \otimes A^L$.
\end{enumerate}
\end{remark}

We refer the reader to Subsection~2.2 in \cite{bohm_weak_1999}

Let us now review the basic notions of their representation theory.

\begin{definition}
\label{Def:Representation}

A \emph{$*$-representation} of a $C^*$-weak Hopf algebra $A$ is any couple $(\mathscr{H},\Phi)$, for which $\mathscr{H}$ is finite-di\-men\-sional Hilbert space and $\Phi:A\to\operatorname{End}\mathscr{H}$ is a $*$-al\-ge\-bra homomorphism.
The $*$-re\-pre\-sen\-ta\-tion is called \emph{faithful} if $\Phi$ is injective. 
\end{definition}

Let us note that the category of $*$-representations of $A$ is a unitary multifusion category \cite{etingof_fusion_2005}. In fact, every unitary multifusion category arises as the representation category of a $C^*$-weak Hopf algebra \cite{etingof_tensor_2015}.
In this context, the monoidal product of two given $*$-representations $(\mathscr{H},\Phi)$ and $(\mathscr{K},\Psi)$ of $A$ is given by the expression
\begin{equation}
    \label{Eq:MonoidalProduct}
    \quad \Phi\boxtimes \Psi \coloneq (\Phi\otimes \Psi)\circ \Delta,
\end{equation}
acting on the subspace of $\mathscr{H}_1\otimes \mathscr{H}_2$ defined by
\[
    \mathscr{H}_1 \boxtimes \mathscr{H}_2 \coloneq \Phi(1_{(1)})(\mathscr{H}_1) \otimes \Psi(1_{(2)})(\mathscr{H}_2).
\]
The monoidal unit is then given by the \emph{trivial $*$-representation}, defined by
\begin{equation}
    \label{Eq:MonoidalUnit}
    \mathscr{H}_\varepsilon \coloneq (A^*)^R,\quad \Phi_{\varepsilon}:A\to \operatorname{End}\mathscr{H}_{\varepsilon}, \quad \Phi_\varepsilon(x)(f) \coloneq \eval{f_{(2)}}{x}f_{(1)},
\end{equation}
in the sense of relaxed monoidal categories.
Two $*$-representations $(\mathscr{H}_1,\Phi_1)$ and $(\mathscr{H}_2,\Phi_2)$ of $A$ are said to be \emph{equivalent} if there exists a unitary intertwiner, i.e. 
\[ U \circ \Phi_1(x) = \Phi_2(x)\circ U \] for all $x\in A$, for some unitary operator $U:\mathscr{H}\to\mathscr{K}$.
In this setting, we also note that the category of $*$-representations is semisimple and the set $\operatorname{Irr}A$ of equivalence classes of irreducible representations is finite. This set is also known as the set of \emph{sectors} of $A$.
We note that the unusual feature of the trivial $*$-re\-pre\-sen\-ta\-tion is that it can be decomposable. 

\begin{definition}
\label{def:biconnected}
    A $C^*$-weak Hopf algebra $A$ is called \emph{biconnected} if both trivial $*$-re\-pre\-sen\-ta\-tions of $A$ and $A^*$ are indecomposable.
\end{definition}

\begin{proposition}
\label{Prop:J}
There exists a linear map $J:A\to A$ satisfying:
\begin{enumerate}
    \item\label{Item:Jantimult}
    it is a $*$-algebra antihomomorphism, i.e.~for all $x,y\in A$:
    \[
        J(xy) = J(y)J(x),\quad J(x^*) = J(x)^* \quad\text{ and }\quad J(1) =1;
    \]
    \item\label{Item:Janticomult}
    it is anticomultiplicative, i.e.~for all $x\in A$:
    \[  
        J(x)_{(1)} \otimes J(x)_{(2)} = J(x_{(2)}) \otimes J(x_{(1)});
    \]
    \item\label{Item:Jinvolutive} it is involutive, i.e.~$J\circ J = \mathrm{Id}$.
\end{enumerate}
In particular, if $S^2 = \mathrm{Id}$, then $J = S$.
\end{proposition}

See \cref{Subsection:Integrals_and_g} for a proof.

Let $\mathscr{H}$ be a finite-dimensional Hilbert space and consider the anti-iso\-mor\-phism that assigns to each vector $v\in \mathscr{H}$ the linear functional $f_v=(v,\,\cdot\,)\in\mathscr{H}^*$ induced by the inner product in $\mathscr{H}$.
Let
\[
    (f_v,f_w) \coloneq (w,v),
\]
be the inner product on the dual vector space $\mathscr{H}^*$, for all $v,w\in\mathscr{H}$.
Then, for every $*$-representation $(\mathscr{H},\Phi)$ of $A$, the linear map
\begin{equation}
\label{Eq:DualRepresentation}
\dualPhi:A\to\operatorname{End}(\mathscr{H}^*),
\quad
\dualPhi(x) \coloneq (\Phi\circ J(x))^\mathrm{t},
\end{equation}
defines a $*$-representation of $A$, known as the \emph{dual $*$-re\-pre\-sen\-ta\-tion} of $\Phi$.
We refer the reader to Sections~2 and 3.3, and Proposition~3.5 in \cite{bohm_weak_2000} for more details; see also \cref{Eq:ExplicitExpressionJ} in \cref{Appendix:Omega}.

\subsection{Tensor network representations}
\label{Subsection:MPOs}

This subsection is devoted to reformulating the construction of $*$-algebras of matrix product operators from representations of $C^*$-weak Hopf algebras introduced in  \cite{molnar_matrix_2022,ruiz_matrix_2024}. For that purpose, we first consider a finite quantum spin chain for which $\mathscr{H}_\nu \cong \mathscr{H}$ denotes a finite-dimensional  Hilbert space characterizing the physical degrees of freedom at site $\nu$. For any given $C^*$-weak Hopf algebra $A$, and a  $*$-representation $(\mathscr{H},\Phi)$ of $A$, one can fix any faithful $*$-representation $(\mathscr{K},\Psi)$ of $A^*$,  and define the rank-four tensor
\begin{equation}
    \label{Eq:DefTensorFromRep}
    T_{\Phi,\Psi} \coloneq \sum_{i=1}^d \Phi(e_i) \otimes \Psi(e^i)\in\operatorname{Im}\Phi\otimes\operatorname{Im}\Psi\subseteq\operatorname{End}\mathscr{H}\otimes\operatorname{End}\mathscr{K},
\end{equation}
where $e_1,\ldots, e_d\in A$ is any basis for $A$ and $e^1,\ldots, e^d\in A^\ast$ stands for the corresponding dual basis, i.e.~such that $\eval{e^i}{e_j} = \delta_{ij}$ for all $i,j=1,\ldots,d$.
Note that the first tensor factor in \cref{Eq:DefTensorFromRep} corresponds to an endomorphism on the physical degrees of freedom, and the second tensor factor corresponds to an endomorphism on the virtual degrees of freedom. Then, for any family of representations $(\mathscr{H}_\nu,\Phi_\nu)$, $\nu\in\mathbb{N}$, the linear maps
\begin{equation}
    \Phi_{ ( 1 , \ldots , n ) } : A\to \operatorname{End}\bigotimes_{\nu=1}^n \mathscr{H}_{\nu},
    \quad
    \Phi_{ ( 1 , \ldots , n ) } \coloneq \bigotimes_{\nu=1}^n \Phi_\nu \circ \Delta^{(n-1)}
\end{equation}
define $*$-representations of $A$ for all system sizes $n\in\mathbb{N}$. Note that, in general, it does not hold that $\Phi_{(1,\ldots,n)}(1) = \mathrm{Id}$, since the comultiplication need not be unit-preserving. Moreover, for every element $x\in A$, the corresponding images constitute matrix product operators since
\begin{equation}
    \Phi_{ ( 1 , \ldots , n ) }(x) =
    \sum_{ i_1, \ldots, i_n=1}^{d}
        \mathrm{Tr}(b(x) \Psi(e^{i_1}) \cdots \Psi(e^{i_n}))
        \,
        \Phi_1(e_{i_1})\otimes\cdots \otimes \Phi_n(e_{i_n})
\end{equation}
for some operator $b(x) \in \operatorname{End}\mathscr{K}$ associated to the element $x\in A$ via the $*$-re\-pre\-sen\-ta\-tion $\Psi$; more concretely, $b(x)$ is any endomorphism characterized by the property
\begin{equation}
    \label{Eq:DefiningProperty_bx}
    \operatorname{Tr}( b(x) \Psi( f ) )
    =
    \eval{ f }{ x }
\end{equation} for all linear functionals $f\in A^\ast$. In particular, we note that this choice is unique if we require $b(x)\in\operatorname{Im}\Psi$; here we always assume this condition. We refer the reader to Section~3 in \cite{molnar_matrix_2022} for a detailed discussion.
Let us note that, up to conjugation by a unitary operator,
\begin{equation}
    \label{Eq:PsiIrreps}
    \Psi(f) = \bigoplus_{a=1}^r \Psi_a(f)\otimes \mathrm{Id}_{\ell_a}
\end{equation}
for every linear functional $f\in A^*$, where $\Psi_a$ denotes the $a$-th irreducible $*$-re\-pre\-sen\-ta\-tion of $A^*$, and $\ell_a$ denotes the multiplicity of $\Psi_a$ within $\Psi$. In turn, the boundary terms are decomposed as follows:
\begin{equation}
    \label{Eq:bxIrreps}
    b(x) = \bigoplus_{a=1}^r b_a(x)\otimes \frac{\mathrm{Id}_{\ell_a}}{\ell_a},
\end{equation}
where $b_a(x)\in\operatorname{End}\mathscr{K}_a$ stands for the operator such that $\operatorname{Tr}(b_a(x)\Psi(f)) = \eval{fe_a}{x}$ for all linear functionals $f\in A^*$, where $e_a\in A^*$ is the corresponding minimal central idempotent of $A^*$.
\\[5pt]
We now reinterpret these expressions using tensor network notation, which is often more convenient than traditional algebraic formulas for describing quantum many-body systems.

\begin{notation}
A tensor $T\in V_1^*\otimes \cdots \otimes V_m^* \otimes W_1^{\vphantom{*}}\otimes \cdots\otimes W_n^{\vphantom{*}}$ with covariant in\-di\-ces $i_1,\ldots i_m$ and contravariant indices $j_{1},\ldots, j_{n}$ is represented by a shape, such as a circle, from which lines extend, each of them identifying a corresponding index:
\begin{equation*}
T =
\vcenter{\hbox{\includegraphics[page=001]{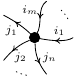}}}
\ ;
\end{equation*}
more concretely, each of the lines with ingoing arrows are identified with the degrees of freedom described by the duals of the vector spaces $V_1,\ldots,V_m$, while the lines with outgoing arrows are identified with the degrees of freedom described by the vector spaces $W_{1},\ldots, W_n$.  The tensor product corresponds to representing the tensor factors within the same diagram. Note also that the direction of an arrow can be reversed by considering that it represents the dual vector space, since a vector space and its double dual are canonically isomorphic. Additionally, one can group certain indices by considering that they reside in the tensor product, reshaping the tensor in this manner. The evaluation of two indices corresponds to connecting the respective lines, as illustrated by the following diagram:
\begin{equation*}
\vcenter{\hbox{\includegraphics[page=002]{haagduality_diagrams.pdf}}}
    \ \longmapsto  \ 
\vcenter{\hbox{\includegraphics[page=003]{haagduality_diagrams.pdf}}}
\ .
\end{equation*}
Here, we will not explicitly indicate the vector spaces of the indices or the lines in the diagrams for the sake of clarity. Instead, we will use colors, given the small number of distinct spaces that arise.
We refer the reader to Section~2 in \cite{molnar_matrix_2022} for a more detailed discussion.
\end{notation}

Based on \cref{Eq:DefTensorFromRep,Eq:DualRepresentation} and the prior considerations, we let
\begin{equation}
\label{Eq:BlackTensor}
T_{\Phi,\Psi} =
\vcenter{\hbox{\includegraphics[page=004]{haagduality_diagrams.pdf}}}
\quad\text{ and }\quad
T_{\Phi\circ J,\Psi} =
\vcenter{\hbox{\includegraphics[page=005]{haagduality_diagrams.pdf}}}
\ ,
\end{equation}
where we convey the red horizontal lines to correspond to an endomorphism on $\mathscr{K}$, while the black vertical lines correspond to an endomorphism on $\mathscr{H}$. More specifically, the left line corresponds to the index associated to $\mathscr{K}$ and the right line corresponds to the index associated to $\mathscr{K}^*$.
Note that the above diagrams are unambiguous, as each index is clearly identifiable by both the color and the direction of the arrow.
\\[3pt]
Regarding the second tensor in \cref{Eq:BlackTensor}, note that we consider  $\Phi\circ J$ instead of $\dualPhi$ the dual $*$-representation $\dualPhi$, which is originally defined on the dual Hilbert space, $\mathscr{H}^*$; in that sense, in the diagrams we will use the identification $\mathscr{H}^{**}\cong\mathscr{H}$ to consider it on $\mathscr{H}$, by transposing the physical indices in its expression; thus, black lines are always associated to the Hilbert space $\mathscr{H}$.

Regarding the decomposition into irreducible $*$-representations presented in \cref{Eq:PsiIrreps,Eq:bxIrreps},
\begin{equation}
\label{Eq:MultiplPsibx}
\vcenter{\hbox{\includegraphics[page=006]{haagduality_diagrams.pdf}}}
    =
    \bigoplus_{a=1}^r
    \ 
\vcenter{\hbox{\includegraphics[page=007]{haagduality_diagrams.pdf}}}
    \qquad\text{ and }\qquad
\vcenter{\hbox{\includegraphics[page=008]{haagduality_diagrams.pdf}}}
    =
    \bigoplus_{a=1}^r 
    \frac{1}{\ell_a} \ 
\vcenter{\hbox{\includegraphics[page=009]{haagduality_diagrams.pdf}}}
 \ .
\end{equation}

\subsection{Pulling-through identities}
\label{Subsection:Topo}

The following result encompasses the fundamental properties for the description of topologically ordered states in this framework \cite{molnar_matrix_2022,ruiz_matrix_2024}. For clarity, we postpone the proof to \cref{Appendix:Omega}.

\begin{theorem}\label{Thm:PT}
    Let $A$ be a $C^*$-weak Hopf algebra. Then:
    \begin{enumerate}
        \item\label{Item:Omega-And-PT}
        There exists a cocentral non-degenerate element $\Omega\in A$, which defines a positive functional on $A^*$,  such that
        \[
        \Omega^2 = \Omega^* = S(\Omega) = J(\Omega) = \Omega,
        \]
        and there exist two linear maps $J^L,J^R:A\to A$ satisfying, for all $x,y\in A$,
    \begin{align*}
        J^L(x) \Omega_{(1)}\otimes\Omega_{(2)} &=
        \Omega_{(1)} \otimes x \Omega_{(2)}
        ,\\
        \Omega_{(1)} J^R(y) \otimes\Omega_{(2)} &=
        \Omega_{(1)} \otimes  \Omega_{(2)}y,
    \end{align*}
    called pulling-through identities. The element $\Omega$ is known as a pulling-through element of $A$. Let $\omega\in A^*$ denote the dual analogue of $\Omega$ from now on.\vspace{3pt}
    \item\label{Item:propsJLJR} $J^L$ and $J^R$ are involutive algebra anti-homomorphisms and
    \[
        \omega \circ J = \omega \circ J^L = \omega \circ J^R = \omega \circ S = \omega.
    \]
    \item\label{Item:JLJRXi} There exists an invertible positive element $\xi \in A$ such that, for all $x\in A$,
    \begin{equation*}
        \eval{\omega}{\xi^{ \frac{1}{2} } J(\Omega_{(1)}) \xi^{ \frac{1}{2} } x } \Omega_{(2)} = x;
    \end{equation*}
    and it relates $J$, $J^L$ and $J^R$, for all $x\in A$, via
    \begin{equation*}
        J^L(x) \coloneq \xi^{-\frac{1}{2}} J(x)  \xi^{\frac{1}{2}} 
        ,\quad
        J^R(x)\coloneq \xi^{ \frac{1}{2} } J(x) \xi^{-\frac{1}{2} },
    \end{equation*}
    by virtue of the pulling-through identities. It is uniquely determined by
    \[
        \eval{\omega}{\Omega_{(1)}} \Omega_{(2)} = \xi^{-1}.
    \]
    \item\label{Item:XiFactor} There exist positive invertible elements $\xi_L\in A^L$, $\xi_R = J(\xi_L)\in A^R$ such that 
    \[ \xi = \xi_L \xi_R; \]
    in particular, $J(\xi) = \xi$, and the coproducts of $\xi$ are of the form
    \[ \Delta^{(n)} (\xi) = \xi_L 1_{(1)} \otimes 1_{(2)}\otimes \cdots \otimes 1_{(n)} \otimes \xi_R 1_{(n+1)}. \]
    \item\label{Item:eqPT} The pulling-through identities can be then rewritten, for all $x,y\in A$, as:
    \begin{align*}
        x \xi^\frac{1}{2} J(\Omega_{(1)}) \otimes \Omega_{(2)}
        = \xi^\frac{1}{2} J(\Omega_{(1)})  \otimes \Omega_{(2)} x,\\
        J(\Omega_{(1)}) \xi^\frac{1}{2}  y \otimes \Omega_{(2)}
        = J(\Omega_{(1)}) \xi^\frac{1}{2} \otimes y \Omega_{(2)}.
    \end{align*}
    \item\label{Item:gRgLexch} Let $\hat\xi_L$, $\hat\xi_L$ and $\hat\xi\in A^*$ stand for their dual analogues; then, for all $x\in A$,
    \begin{align*}
        \xi_R x &= \eval{\hat{\xi}_L}{x_{(2)}} x_{(1)},
        \quad 
        \xi_L x = \eval{\hat{\xi}_L}{x_{(1)}}  x_{(2)} ,
        \\ 
        x \xi_L &= \eval{\hat{\xi}_R}{x_{(1)}} x_{(2)},
        \quad 
        x \xi_R = \eval{\hat{\xi}_R}{x_{(2)}}x_{(1)}.
    \end{align*}
    \item \label{Item:ClosingXiLXiR}
    Let $g\coloneq \xi_L\xi_R^{-1}$, then, the following identities hold 
    for all $x\in A$:
        \begin{align*}
            \eval{\omega}{  g^{\frac{1}{2}} x_{(2)} g^{\frac{1}{2}} J(x_{(3)})  } x_{(1)}& = x,
            \\
            \eval{\omega}{ g^{-\frac{1}{2}} J(x_{(1)}) g^{-\frac{1}{2}} x_{(2)}  } x_{(3)} & = x.
        \end{align*}
    \end{enumerate}
\end{theorem}

Let us elaborate now further on the identities introduced above from a tensor network perspective.
In first place, the fact that $\Omega$ is cocentral and defines a po\-si\-ti\-ve linear functional implies that $b(\Omega)$ is a positive central element of the $*$-sub\-al\-ge\-bra $\operatorname{Im}\Psi$ of $\operatorname{End}\mathscr{K}$. 
Regarding the first pulling-through identity in \cref{Item:eqPT} of \cref{Thm:PT}, this statement can be in turn represented as follows:
\begin{equation}\label{Eq:PT_2sites}
\vcenter{\hbox{\includegraphics[page=010]{haagduality_diagrams.pdf}}}
    \ = \ 
\vcenter{\hbox{\includegraphics[page=011]{haagduality_diagrams.pdf}}}
    \ ;
\end{equation}
here, the matrix product operator wrapped in a circular shape corresponds to the endomorphism $(\Phi\circ J)(\Omega_{(1)})\otimes \Phi(\Omega_{(2)})$ on $\mathscr{H}\otimes \mathscr{H}$, while the rank-four tensor with open virtual indices gives rise to any operator $\Phi(x)$, as the preceding identities hold for any element $x\in A$ and the $*$-representation $\Psi$ at the virtual level is faithful. Note that it is possible to recover the exact expression by simply closing at the virtual level with the operator $b(x)$.
\\[5pt]
The second pulling-through identity in \cref{Item:eqPT} of \cref{Thm:PT} can be interpreted analogously by simply considering the opposite orientation for the arrows:
\begin{equation}\label{Eq:PT_2sites_reversed}
\vcenter{\hbox{\includegraphics[page=012]{haagduality_diagrams.pdf}}}
    \ = \ 
\vcenter{\hbox{\includegraphics[page=013]{haagduality_diagrams.pdf}}}
    \ .
\end{equation}
In the present work we will need to consider identities involving matrix product operators acting on more than two sites. The following examples are provided to offer further insight into how such identities can be constructed in these cases.
\\[5pt]
As a first example, let us apply $\mathrm{Id}\otimes \Delta$ on both sides of the first pulling-through equation. Then,
\[
    x \xi^\frac{1}{2} J(\Omega_{(1)}) \otimes \Omega_{(2)}\otimes \Omega_{(3)}
    = \xi^\frac{1}{2} J(\Omega_{(1)}) \otimes \Omega_{(2)} x_{(1)} \otimes \Omega_{(3)} x_{(2)},
\]
for all $x\in A$, which in graphical notation can be reinterpreted as follows:
\begin{equation}\label{Eq:PT_3sites}
\vcenter{\hbox{\includegraphics[page=014]{haagduality_diagrams.pdf}}}
    \ = \ 
\vcenter{\hbox{\includegraphics[page=015]{haagduality_diagrams.pdf}}}
    \ .
\end{equation}
Another possibility is to apply $(\mathrm{Id}\otimes J\otimes \mathrm{Id}\otimes\mathrm{Id})\circ (\mathrm{Id}\otimes \Delta^{(2)})$, obtaining the identity
\begin{align*}
    x \xi^\frac{1}{2} J(\Omega_{(1)}) \otimes { } & { } J(\Omega_{(2)})\otimes \Omega_{(3)} \otimes \Omega_{(4)}
    \\ &= \xi^\frac{1}{2} J(\Omega_{(1)}) \otimes J(x_{(1)}) J(\Omega_{(2)})  \otimes \Omega_{(3)} x_{(2)} \otimes \Omega_{(4)} x_{(2)},
\end{align*}
for all $x\in A$, which can be interpreted as follows:
\begin{equation}
\label{Eq:PT_4sites_vs1}
\vcenter{\hbox{\includegraphics[page=016]{haagduality_diagrams.pdf}}}
    \ = \ 
\vcenter{\hbox{\includegraphics[page=017]{haagduality_diagrams.pdf}}}
    \ .
\end{equation}
Finally, apply $\Delta\otimes \Delta$ on both sides of the pulling-through identity, then
\begin{align*}
    x_{(1)} \xi_L^{\frac{1}{2}} J(\Omega_{(2)}) \otimes {} & {} x_{(2)}\xi_R^{\frac{1}{2}} J(\Omega_{(1)}) \otimes \Omega_{(3)}  \otimes \Omega_{(4)} 
    \\
    &= 
    \xi_L^{\frac{1}{2}} J(\Omega_{(2)}) \otimes \xi_R^{\frac{1}{2}} J(\Omega_{(1)}) \otimes \Omega_{(3)} x_{(1)} \otimes \Omega_{(4)} x_{(2)},
\end{align*}
by virtue of \cref{Item:XiFactor} of \cref{Thm:PT} and the anti-multiplicativity of $J$, i.e.
\begin{equation}\label{Eq:PT_4sites_vs2}
\vcenter{\hbox{\includegraphics[page=018]{haagduality_diagrams.pdf}}}
    \ = \ 
\vcenter{\hbox{\includegraphics[page=019]{haagduality_diagrams.pdf}}}
    \ .
\end{equation}
In addition, the first identity in \cref{Item:JLJRXi} of \cref{Thm:PT} implies
\begin{equation}
\label{Eq:IdentityBlackVsWhite}
\vcenter{\hbox{\includegraphics[page=020]{haagduality_diagrams.pdf}}}
    \ = \ 
    \delta_{a,b}
    \ 
\vcenter{\hbox{\includegraphics[page=021]{haagduality_diagrams.pdf}}}
\end{equation}
for all sectors $a,b = 1,\ldots, r$. Here, the label $a$ next to the arrow stands for the restriction to the $a$-th irreducible representation. Moreover, once all sectors and their associated multiplicities are taken into account, the following identity holds for the full rank-four tensors:
\begin{equation}
    \label{Eq:IdentityBlackVsWhiteFull}
\vcenter{\hbox{\includegraphics[page=022]{haagduality_diagrams.pdf}}}
    = \ 
    \sum_{a=1}^r
    \frac{1}{\ell_a}\ 
\vcenter{\hbox{\includegraphics[page=023]{haagduality_diagrams.pdf}}}
    \ .
\end{equation}
The identities in \cref{Item:ClosingXiLXiR} of \cref{Thm:PT} are represented as follows:
\begin{equation}
\label{Eq:TransferEigenvalue}
\vcenter{\hbox{\includegraphics[page=024]{haagduality_diagrams.pdf}}}
    = \ 
\vcenter{\hbox{\includegraphics[page=025]{haagduality_diagrams.pdf}}}
    \quad\text{ and }\quad
\vcenter{\hbox{\includegraphics[page=026]{haagduality_diagrams.pdf}}}
    = \ 
\vcenter{\hbox{\includegraphics[page=027]{haagduality_diagrams.pdf}}}
    \ .
\end{equation}
The algebraic conditions in \cref{Item:gRgLexch} of \cref{Thm:PT} are equivalent to the following:
\begin{align}
\label{Eqs:xiExchGraphBlack}
\vcenter{\hbox{\includegraphics[page=028]{haagduality_diagrams.pdf}}}
    \ = \ 
\vcenter{\hbox{\includegraphics[page=029]{haagduality_diagrams.pdf}}}
    \ , \qquad
\vcenter{\hbox{\includegraphics[page=030]{haagduality_diagrams.pdf}}}
    \ = \ 
\vcenter{\hbox{\includegraphics[page=031]{haagduality_diagrams.pdf}}}
    ,
\\[5pt]
\label{Eqs:xiExchGraphBlack2}
\vcenter{\hbox{\includegraphics[page=032]{haagduality_diagrams.pdf}}}
    \ = \ 
\vcenter{\hbox{\includegraphics[page=033]{haagduality_diagrams.pdf}}}
    \ , \qquad
\vcenter{\hbox{\includegraphics[page=034]{haagduality_diagrams.pdf}}}
    \ = \ 
\vcenter{\hbox{\includegraphics[page=035]{haagduality_diagrams.pdf}}}
    \ .
\end{align}
Similar identities apply for the tensor arising from the dual $*$-representation. It is straightforward to check that, due to the fact that $\xi_R = J(\xi_L)$, the identities in \cref{Eqs:xiExchGraphBlack,Eqs:xiExchGraphBlack2} are equivalent to the following:
\begin{align}
\label{Eqs:xiExchGraphWhite}
\vcenter{\hbox{\includegraphics[page=036]{haagduality_diagrams.pdf}}}
    ~=~
\vcenter{\hbox{\includegraphics[page=037]{haagduality_diagrams.pdf}}}
    ~, \qquad 
\vcenter{\hbox{\includegraphics[page=038]{haagduality_diagrams.pdf}}}
    ~=~
\vcenter{\hbox{\includegraphics[page=039]{haagduality_diagrams.pdf}}}
    ~,\\[5pt]
\label{Eqs:xiExchGraphWhite2}
\vcenter{\hbox{\includegraphics[page=040]{haagduality_diagrams.pdf}}}
    ~=~
\vcenter{\hbox{\includegraphics[page=041]{haagduality_diagrams.pdf}}}
    ~, \qquad  
\vcenter{\hbox{\includegraphics[page=042]{haagduality_diagrams.pdf}}}
    ~=~
\vcenter{\hbox{\includegraphics[page=043]{haagduality_diagrams.pdf}}}
    ~.
\end{align}

\begin{remark}\label{Rem:ExamplBic}
The description and properties of $\Omega$ become simpler in the particular case of biconnected $C^*$-weak Hopf algebras, see \cref{def:biconnected}.
For instance, for each sector $a=1,\ldots, r$ of $A^*$ with irreducible character $\mathrm{x}_a\in A^{**}\cong A$, there exists a positive real number $d_a$, called Frobenius-Perron dimension or quantum dimension of sector $a$, and these are the unique positive real numbers such that the element $\Omega = D^{-2} (d_1 \mathrm{x}_1+\cdots + d_r\mathrm{x_r})$, where we let $D^2 \coloneq d_1^2 + \cdots + d_r^2$ stand for the total quantum dimension of $A^*$, satisfies
\begin{equation}\label{Eq:OmegaEigBiconn}
    \Omega \mathrm{x}_a = \mathrm{x}_a \Omega = d_a \Omega.
\end{equation}
We refer the reader to Section 3, Propositions 3.3.6 and 3.3.11 in \cite{etingof_tensor_2015} or Appendix B in \cite{ruiz_matrix_2024}.
Furthermore, for $C^*$-weak Hopf algebras $A$ with $S^2 = \mathrm{Id}$, the element  $\Omega$ coincides with the Haar integral of $A$, and
\(
    J^L = J^R = J = S = S^{-1}.
\)
Finally, if $G$ is a finite group this is the case of group Hopf-algebras $\mathbb{C}G$, where $G$ is a finite group, for which $\Omega= |G|^{-1}\sum g$, where the sum is over all $g\in G$, and the antipode is given by the linear extension of the group inverse $S(g) = g^{-1}$. This is also the case for their dual counterparts $(\mathbb{C}G)^*\cong\mathbb{C}^G$, for which $\Omega = \delta_e$, the dual basis element of the unit element $e\in G$.
\end{remark}

\subsection{Construction of tensor network states}
\label{Subsection:PEPS}
We begin by introducing the relevant graph-theoretic framework. Let $(\mathcal{L},\mathcal{E})$ a directed rectangular lattice, where the vertex set is given by $\mathcal{L}\subset \mathbb{Z}^2$, and the edge set $E\subset \mathcal{L}\times \mathcal{L}$ consists of directed nearest-neighbor pairs. We adopt the convention that edges are oriented from right to left and from top to bottom, as illustrated below.
\[
\vcenter{\hbox{\includegraphics[page=044]{haagduality_diagrams.pdf}}}
\]

\begin{definitions}
    Let $\Lambda$ be a subset of $\mathbb{Z}^2$. Then,
    \begin{enumerate}
        \item the \emph{sets of edges with only their heads} (resp. \emph{tails}) on $\Lambda$ are denoted
\begin{equation*}
    \partial_{\rightarrow}{\Lambda} \coloneq \mathcal{E} \cap ((\mathbb{Z}^2\setminus\Lambda) \times \Lambda)
    ,
    \qquad
    \partial_{\leftarrow}{\Lambda} \coloneq \mathcal{E} \cap (\Lambda \times (\mathbb{Z}^2\setminus\Lambda))
    ;
\end{equation*}
\item  the \emph{set of edges with only one endpoint on $\Lambda$} is denoted
\begin{equation*}
    \partial{\Lambda} \coloneq \partial_{\leftarrow}{\Lambda} \cup \partial_{\rightarrow}{\Lambda};
\end{equation*}
\item the set of \emph{interior edges}, for which both endpoints are on $\Lambda$, is denoted
\begin{equation*}
    \Interior{\Lambda} \coloneq \mathcal{E} \cap (\Lambda\times \Lambda).
\end{equation*}
    \end{enumerate}
\end{definitions}

The following figures exemplify these definitions:
\[
\vcenter{\hbox{\includegraphics[page=045]{haagduality_diagrams.pdf}}}
\]
\begin{definitions}
A \emph{row} (resp.~a \emph{column}) is a non-empty subset of $\mathbb{Z}^2$ of the form 
\[
\{i,\ldots,j\}\times\{k\},\qquad (\text{resp.}~\{i\}\times\{j,\ldots,k\}),\] 
for some $i, j, k \in \mathbb{Z}$.
A subset that is either a row or a column is called a \emph{segment}.
\end{definitions}

\begin{definition}\label{Def:Comblike}
A \emph{comb-like region} is any subset $\Lambda = \Lambda_1 \cup \Lambda_2 \cup \cdots \cup \Lambda_\kappa
\subset \mathbb{Z}^2$, for some $\kappa\geq 1$, constructible by the following iterative procedure:
\begin{itemize}
\setlength\itemsep{3pt}
\item[Step 0:] Fix a direction of growth \(\vec{d} \in \{(\pm 1,0),(0,\pm 1)\}\).
We adopt the convention that for a vertical direction $\vec{d}=(0,\pm 1)$ all segments below are rows, whereas for a horizontal direction $\vec{d}=(\pm 1,0)$, all segments below are columns.
\item[] For simplicity, let $\vec{d}=(0,1)$, i.e.~the direction is from bottom to top.
\item[Step 1:] Let \(\Lambda_1\) be a segment of \(\mathbb{Z}^2\). Assume that $\Lambda_1\subset \mathbb{Z}\times\{k\}$ for some $k\in\mathbb{Z}$.
\item[Step $s$:]  For each step $s= 2,\ldots, \kappa$, let $\Lambda_s := \Lambda_{s,1}\cup\cdots \cup \Lambda_{s,\ell_s}$
denote a union of segments satisfying the following properties:
\begin{enumerate}
\setlength\itemsep{3pt}
  \item for each $i\in\{1,\ldots, \ell_s\}$, the segment \(\Lambda_{s,i}\) lies immediately ahead of the segments in  \(\Lambda_{s-1}\) in the designated direction $\vec{d}$, i.e.~$\Lambda_{s,i}\in\mathbb{Z}\times \{k+s-1\}$;
  \item for any $i,j\in \{1,\ldots, \ell_s\}$, $i\neq j$, the union $\Lambda_{s,i}\cup \Lambda_{s,j}$ is not a segment;
  \item for each $i\in\{1,\ldots,\ell_s\}$ there exists a unique $j_i\in\{1,\ldots,\ell_{s-1}\}$ such that the segment \(\Lambda_{s,i}\) is adjacent to \(\Lambda_{s-1,j_i}\), called its \emph{predecessor};
  \item for each $i\in\{1,\ldots,\ell_s\}$ and $j\in\{1,\ldots,\ell_{s-1}\}\setminus\{j_i\}$, the segment  \(\Lambda_{s,i}\) is at lattice distance at least three from the segment $\Lambda_{s-1,j}$.
\end{enumerate}
\end{itemize}
\end{definition}

\begin{example}
Let us consider the following examples:
\[
\vcenter{\hbox{\includegraphics[page=046]{haagduality_diagrams.pdf}}}
\]
In Example~(a), we consider a comb-like region constructed by joining four segments over three iterations in the vertical direction \(\vec{d} = (0,1)\), proceeding from bottom to top. In Example~(b), we present a comb-like region formed by the union of three segments, also over three iterations, but in the horizontal direction \(\vec{d} = (1,0)\), from left to right. While such regions can, in principle, be constructed using any of the four allowed directions of growth, this may not always possible in practice. The region depicted in Example~(c) is not a comb-like region, as the segment \(\Lambda_3\) does not satisfy condition (4) in \cref{Def:Comblike}.
\end{example}

The use of comb-like subregions will play a central role in the subsequent analysis of Haag duality, as such configurations naturally arise as intersections of (possibly non-convex) cones and finite rectangular regions. We remark that this choice is motivated by the context and the fact that it is technically convenient, as it facilitates a transparent formulation using iterative constructions. However, it is natural to consider broader classes of subregions, provided they satisfy a suitable notion of contractibility for lattice graphs.
An important remark is the following:
\begin{remark}\label{Rem:BdryComblike}
    Let $\Lambda$ be a comb-like subregion of $\mathbb{Z}^2$. Then, the set of boundary edges $\partial\Lambda$ induces a simple closed curve in the plane, in the sense that it forms a connected, non-self-intersecting loop separating the interior of $\Lambda$ from its exterior. This clearly follows from the spatial separation conditions (3) and (4) involving different branches, assuming they exist. In particular, $\partial\Lambda$ intersects the interior edges of any plaquette of $\mathbb{Z}^2$ at either zero or two edges.
\end{remark}
The construction of two-dimensional tensor network states that exhibit topological order is described as follows.
Let us consider a fixed $C^*$-weak Hopf algebra $A$, and let $(\mathscr{H}, \Phi)$ be any given faithful $*$-representation of $A$. First, for each edge $e \in \mathcal{E}$, we associate a Hilbert space \[ \mathscr{H}_e \coloneq \mathscr{H},\] representing the virtual or bond Hilbert space that models correlations between sites. Second, at each vertex ${\boldsymbol{x}} \in \mathcal{L}$, the physical degrees of freedom are described by the tensor product Hilbert space 
\begin{equation}
    \mathscr{H}_{\boldsymbol{x}}
    \coloneq  \bigotimes_{ \partial_{\leftarrow}{ \{ \boldsymbol{x} \} } } \mathscr{H}^\ast \otimes \bigotimes_{ \partial_{\rightarrow}{ \{\boldsymbol{x}\}} } \mathscr{H}.
\end{equation}
As we consider rectangular lattices, $\mathscr{H}_{\boldsymbol{x}} \cong \mathscr{H}^*\otimes \mathscr{H}^*\otimes \mathscr{H}\otimes \mathscr{H}$, since the set $\partial_{\leftarrow} {\boldsymbol{x}}$ consists of both left and bottom edges, and $\partial_{\rightarrow}{\boldsymbol{x}}$ consists of both right and top edges.
Furthermore, for any finite subset $\Lambda\subset\mathbb{Z}^2$ of vertices, let
\begin{equation}
    \mathscr{H}_{\Lambda} \coloneq \bigotimes_{{\boldsymbol{x}}\in \Lambda}^{\vphantom{.}} \mathscr{H}_{\boldsymbol{x}}
    ,\qquad 
    \mathscr{H}_{\partial\Lambda} \coloneq  \bigotimes_{\partial_{\leftarrow}\Lambda} \mathscr{H}^* \otimes \bigotimes_{ \partial_{\rightarrow}\Lambda} \mathscr{H},
\end{equation}
stand for the total physical Hilbert space associated to the subregion $\Lambda$ and the corresponding Hilbert space of its boundary $\partial\Lambda$, respectively.
\\[5pt]
Due to the formulation of the model, a vertex ${\boldsymbol{x}}$ can be also interpreted as a union of $|\partial {\boldsymbol{x}}|$ physical \emph{subsites}. We shall refer to the boundary $\partial\Lambda$ interchangeably as either the edges on the boundary of $\Lambda$
or as a collection of physical \emph{subsites} of $\Lambda$.
\\[5pt]
Then, the single-vertex map
defining the tensor network states is the linear map
\begin{equation}
    \label{Eq:DefLocalTensorMap}
    \Gamma_{ {\boldsymbol{x}} } : \mathscr{H}_{\partial{{\boldsymbol{x}}}} \to \mathscr{H}_{{\boldsymbol{x}}},
    \qquad
    \Gamma_{ {\boldsymbol{x}} } \coloneq \Phi_{ \partial{{\boldsymbol{x}}} }(\Omega) \circ  \kappa_{ \partial{{\boldsymbol{x}}} },
\end{equation}
where $\Phi_{\partial{{\boldsymbol{x}}}}:A\to\operatorname{End}\mathscr{H}_{\boldsymbol{x}}$ is the $*$-representation defined by the expression
\begin{equation}
    \label{Eq:DefPhiBdX}
    \Phi_{ \partial\boldsymbol{x}} \coloneq
        (
            \bigotimes_{ \partial_{\leftarrow}{{\boldsymbol{x}}} }^{\vphantom{.}} \dualPhi
            \otimes \bigotimes_{ \partial_{\rightarrow}{{\boldsymbol{x}}} } \Phi
        )
        \circ \Delta^{ ( | \partial{\boldsymbol{x}} | -1 ) },
\end{equation}
and $\kappa_{ \partial{{\boldsymbol{x}}} }:\mathscr{H}_{\partial {\boldsymbol{x}}}\to\mathscr{H}_{\partial {\boldsymbol{x}}}$ is a positive endomorphism defined by the tensor product 
\begin{equation}
    \label{Eq:DefKappaBdX}
    \kappa_{\partial{{\boldsymbol{x}}}}
        \coloneq
            (b(\omega)^\frac{1}{4}\Phi(\xi)^{\frac{1}{4}})^\mathrm{t}
            \otimes
            (b(\omega)^\frac{1}{4}\Phi(g)^{-\frac{1}{4}})^\mathrm{t}
            \otimes
            b(\omega)^\frac{1}{4}\Phi(g)^{-\frac{1}{4}}
            \otimes
            b(\omega)^\frac{1}{4}\Phi(\xi)^{\frac{1}{4}},
\end{equation}
where the tensor factors are ordered starting from the left index and in counterclockwise direction, and $g:= \xi_L\xi_R^{-1}$.
In the diagrammatic formalism,
\begin{equation*}
    \Phi_{\partial{\boldsymbol{x}}}(x) \ = \ 
\vcenter{\hbox{\includegraphics[page=047]{haagduality_diagrams.pdf}}}
\end{equation*}
here, the outer lines correspond to the input of the maps, while the inner legs correspond to the output. In order to ensure consistency with respect to the choice of lattice orientation, the black tensors $T_{\Phi, \Psi}$ are selected for the top and right positions, while the white tensors $T_{\dualPhi, \Psi}$ are consequently relegated to the bottom and left positions.
\\[5pt]
The tensor network map defined in \cref{Eq:DefLocalTensorMap} is then respresented as follows:
\[
    \Gamma_{{\boldsymbol{x}}} 
    =
\vcenter{\hbox{\includegraphics[page=048]{haagduality_diagrams.pdf}}}
\]
and, by virtue of \cref{Eqs:xiExchGraphBlack,Eqs:xiExchGraphBlack2,Eqs:xiExchGraphWhite,Eqs:xiExchGraphWhite2}, the following is an equivalent description:
\[
    \Gamma_{{\boldsymbol{x}}} 
    =
\vcenter{\hbox{\includegraphics[page=049]{haagduality_diagrams.pdf}}}
\ .
\]
For any subset $\Lambda\subset \mathcal{L}$ of vertices, the linear map $\Gamma_{ \Lambda } : \mathscr{H}_{ \partial{\Lambda} } \to \mathscr{H}_{ \Lambda }$ that defines the tensor network on $\Lambda$ with open boundary conditions is given by the expression
\begin{equation}\label{Eq:DefGammaLambda}
    \Gamma_\Lambda =
    \bigotimes_{{\boldsymbol{x}}\in\Lambda}
    \Gamma_{\boldsymbol{x}}\circ 
       (\operatorname{Id}_{\partial\Lambda}\otimes \bigotimes_{e\in \Interior{\Lambda}}^{\vphantom{.}} \myket{ \mathrm{Id}_e }),
\end{equation}
i.e.~it is a map from the outer virtual indices to the physical indices.

For further results, it is useful to distinguish between the weights on the inner virtual edges of the tensor network state and those on the boundary virtual edges. To this end, we first introduce the following technical observation on the properties of these weights.

\begin{lemma}\label{Prop:FactKappaBd}
On a single plaquette, the following expressions are equivalent:
\[
\begin{array}{c@{\,}c@{\,}c@{\,}c@{\,}c@{\,}c@{\,}c@{\,}c}
\vcenter{\hbox{\includegraphics[page=050]{haagduality_diagrams.pdf}}}
    &=&
\vcenter{\hbox{\includegraphics[page=051]{haagduality_diagrams.pdf}}}
    &=&
\vcenter{\hbox{\includegraphics[page=052]{haagduality_diagrams.pdf}}}
    &=&
\vcenter{\hbox{\includegraphics[page=053]{haagduality_diagrams.pdf}}}
    &=\vspace{7pt}
    \\
\vcenter{\hbox{\includegraphics[page=054]{haagduality_diagrams.pdf}}}
    &=&
\vcenter{\hbox{\includegraphics[page=055]{haagduality_diagrams.pdf}}}
    &=&  
\vcenter{\hbox{\includegraphics[page=056]{haagduality_diagrams.pdf}}}
    &=&
\vcenter{\hbox{\includegraphics[page=057]{haagduality_diagrams.pdf}}}
    &
\end{array}
\]
for all $\theta\in\mathbb{R}$.
In particular,
\begin{equation*}
\begin{array}{c@{}c@{}c@{}c@{}c@{}c@{}c@{}c@{}c}
\vcenter{\hbox{\includegraphics[page=058]{haagduality_diagrams.pdf}}}
&=&
\vcenter{\hbox{\includegraphics[page=059]{haagduality_diagrams.pdf}}}
&=&
\vcenter{\hbox{\includegraphics[page=060]{haagduality_diagrams.pdf}}}
&=&
\vcenter{\hbox{\includegraphics[page=061]{haagduality_diagrams.pdf}}}
&=&
\vcenter{\hbox{\includegraphics[page=062]{haagduality_diagrams.pdf}}}
\\
\vcenter{\hbox{\includegraphics[page=063]{haagduality_diagrams.pdf}}}
&=&
\vcenter{\hbox{\includegraphics[page=064]{haagduality_diagrams.pdf}}}
&=&
\vcenter{\hbox{\includegraphics[page=065]{haagduality_diagrams.pdf}}}
&=&
\cdots
&=&
\vcenter{\hbox{\includegraphics[page=066]{haagduality_diagrams.pdf}}}
\end{array}
\end{equation*}
\end{lemma}

\begin{proof}
The first set of equalities is an straightforward consequence of \cref{Eqs:xiExchGraphBlack,Eqs:xiExchGraphBlack2,Eqs:xiExchGraphWhite,Eqs:xiExchGraphWhite2}. The second set of equalities is immediate by virtue of the former ones, together with the fact that $\xi^\theta = \xi_L^\theta \xi_R^\theta $ and both factors commute.
\end{proof}

We will use the previous expressions interchangeably throughout, usually without explicitly stating it. These identities primarily apply to the expression of tensor network states, as the following result shows.

\begin{remark}
\label{Prop:RewrittenStatesKappa}
The tensor network states above are given by
\begin{equation}
    \Gamma_\Lambda =
       \bigotimes_{{\boldsymbol{x}}\in \Lambda }
            \Phi_{\partial {\boldsymbol{x}}}(\Omega)
        \circ
        \bigotimes_{{\boldsymbol{x}}\in \Lambda}
            \kappa_{\partial {\boldsymbol{x}}}
        \bigotimes_{e\in \Interior{\Lambda}}
            \myket{ \mathrm{Id}_e },
\end{equation}
and it is possible to replace, using \cref{Prop:FactKappaBd},
\begin{equation}
    \bigotimes_{{\boldsymbol{x}}\in \Lambda}
        \kappa_{\partial{{\boldsymbol{x}}}}
    \quad\text{ with }\quad
    \bigotimes_{e\in \partial\Lambda}
        \tilde\kappa_{e}
    \otimes
    \bigotimes_{e\in \Interior{\Lambda}}
        \tilde\kappa_{e}
    \eqcolon
        \tilde\kappa_{\partial{\Lambda}}\otimes \tilde\kappa_{\Interior{\Lambda}};
\end{equation}
we may choose $\tilde\kappa_{\Interior{\Lambda}}$ such that
\begin{equation}
    \label{Eq:ChoiceRightmostTildeKappaE}
    \tilde\kappa_e
    =
    \begin{cases}
        b(\omega)^{1/2}\Phi(g)^{1/2}
            &\text{for every } e\in \Interior{\Lambda} \text{ being the right-most}
        \\
            & \text{edge of some plaquette inside } \Lambda,\\
        b(\omega)^{1/2}\Phi(\xi)^{1/2} &\text{for every other edge } e\in \Interior{\Lambda}.
    \end{cases}
\end{equation}
Note that the operator $\tilde\kappa_{\partial\Lambda}$ is not uniquely determined because of \cref{Eqs:xiExchGraphBlack,Eqs:xiExchGraphBlack2,Eqs:xiExchGraphWhite,Eqs:xiExchGraphWhite2}, but any valid choice is sufficient for all subsequent results.
\end{remark}

The following result establishes the existence of a weight assignment on the virtual edges that will be useful in the subsequent sections.

\begin{proposition}\label{Rem:DistribW}
Let $\Sigma$ be a comb-like subregion of a finite rectangular lattice $\mathcal{L}\subset\mathbb{Z}^2$, sufficiently large such that $\partial\Sigma\subseteq\Interior{\mathcal{L}}$.
Then, 
\[
\tilde\kappa_e := b(\omega)^{1/2} \quad \text{ for all } e \in \partial\Sigma,
\]
is a valid choice for $\Gamma_{\mathcal{L}}$ as in \cref{Prop:RewrittenStatesKappa}.
\end{proposition}

\begin{proof}
The proof consists on distributing the virtual weights $\Psi(\hat{\xi}_R^{-1})$, which are intrinsic to the plaquettes, along the edges of $\partial\Sigma$. As mentioned in \cref{Rem:BdryComblike}, the boundary of a comb-like region induces a simple closed curve on the plane and, in particular, it intersects the edges of each plaquette at either zero or two edges. 
Let us illustrate this in the case in which $\mathcal{L}$ is a $3\times 5$ rectangular lattice and $\Sigma$ stands for two contiguous sites, and $\Gamma_{\mathcal{L}}$ takes the form:
\begin{equation}
\label{Eq:TNgammaDistrib1}
\vcenter{\hbox{\includegraphics[page=067]{haagduality_diagrams.pdf}}}
\end{equation}
Dotted arrows indicate how the factors of $\Psi(\hat\xi_R^{-1}) = \Psi(\hat\xi_R^{-1/2})\Psi(\hat\xi_R^{-1/2})$ can be distributed using the rules from \cref{Prop:FactKappaBd}. 
Note that $\partial\Sigma$ consists of two edges at each plaquette. For each edge, the weight arising from one plaquette will be $\Phi(\xi_L^{-1/2})$, and hence from the other plaquette it will be $\Phi(\xi_R^{-1/2})$, both cancelling the weight $\Phi(\xi^{1/2})$ in the edge.
Therefore, \cref{Eq:TNgammaDistrib1} is equal to
\begin{equation}
\vcenter{\hbox{\includegraphics[page=068]{haagduality_diagrams.pdf}}}
\end{equation}
This reasoning can be extended to the case of an arbitrary comb-like subregion $\Sigma$, as commented above.
\end{proof}

\subsection{Bulk-boundary correspondence}
\label{Subsection:BulkBoundary}

\begin{definition}
\label{Eq:DefRhoLambda}
Let $\Lambda \subset \mathbb{Z}^2$ be a finite subset of vertices. The map
\begin{equation*}
    \rho_{\partial \Lambda} : \mathscr{H}_{\partial \Lambda} \to \mathscr{H}_{\partial \Lambda}
    ,\qquad
    \rho_{\partial \Lambda} \coloneq ( \Gamma_{\Lambda} )^\dagger \circ \Gamma_{\Lambda},
\end{equation*}
is called the \emph{transfer operator} associated with the tensor network state on $\Lambda$.
\end{definition}

\begin{remark}
    \label{Rem:BdOpSimpl1site}
    For a single vertex ${\boldsymbol{x}}\in \mathbb{Z}^2$,
    \begin{equation*}
        \rho_{\partial {\boldsymbol{x}}} = \kappa_{\partial {\boldsymbol{x}}}\circ \Phi_{\partial {\boldsymbol{x}}}(\Omega) \circ \kappa_{\partial {\boldsymbol{x}}},
    \end{equation*}
    since $\Phi_{\partial {\boldsymbol{x}}}$ is a $*$-representation and $\Omega$ is a self-adjoint idempotent.
\end{remark}
In the case of a comb-like subregion $\Lambda\subset \mathbb{Z}^2$, the bulk of the transfer operator can be further simplified as we prove below, essentially resulting on a $*$-representation
\( \Phi_{\partial\Lambda}:A\to \operatorname{End}\mathscr{H}_{\partial\Lambda} \) along $\partial\Lambda\subset\mathcal{E}$, given by the expression
\begin{equation}
    \label{eq:defPhiBdLambda}
    \Phi_{\partial\Lambda} \coloneq
        (
            \bigotimes_{ \partial_{\leftarrow}{\Lambda} }^{\vphantom{.}} \dualPhi
            \otimes
            \bigotimes_{ \partial_{\rightarrow}{\Lambda} } \Phi
        )
        \circ \Delta^{ ( | \partial\Lambda| -1  ) }.
\end{equation}
This extends the $*$-representation defined on a single-site as in \cref{Eq:DefPhiBdX}, where $\Phi$ denotes the given $*$-representation and $\dualPhi$ its dual. The order of the coproducts is taken to be anticlockwise; that is, the arrow on the virtual level of the matrix product operators points clockwise. However, it should be noted that, in general, it does not necessarily hold that $\Phi_{\partial\Lambda}(1) = \mathrm{Id}$, since $\Delta$ may not be unit-preserving.
\\[5pt]
For the purposes of this discussion, the choice of origin for the coproduct or, equivalently, the position of the virtual boundary condition, is not relevant, since our focus is on the representation of $\Omega\in A$, which is cocentral.
\\[5pt]
The following result is called a \emph{bulk-boundary correspondence}, and extends the result of Section~4 in \cite{ruiz_matrix_2024} beyond the case of biconnected $C^*$-weak Hopf algebras. 

\begin{theorem}\label{Thm:BulkBoundary}
    Let $\Lambda$ be a comb-like subregion of $\mathbb{Z}^2$.
    Then, 
    \begin{equation*}
        \rho_{\partial{\Lambda}} =
            \tilde\kappa_{\partial{\Lambda}}
            \circ \Phi_{\partial\Lambda}(\Omega)
            \circ \tilde\kappa_{\partial{\Lambda}},
    \end{equation*}
    for any automorphism $\tilde\kappa_{\partial\Lambda}$ as in \cref{Prop:RewrittenStatesKappa}.
\end{theorem}

\begin{proof}
Let us note that proving that
\begin{equation}\label{Eq:SuffBulkBdry}
    \mybra{ \tilde\kappa_{\Interior{\Lambda} } } 
    \bigotimes_{{\boldsymbol{x}}\in \Lambda}  \Phi_{\partial{{\boldsymbol{x}}}}(\Omega)
     \myket{ \tilde\kappa_{\Interior{\Lambda} } }
     =
     \Phi_{\partial\Lambda}(\Omega)
\end{equation}
is a sufficient condition, as the following calculation shows:
\begin{align}
    \rho_{ \partial\Lambda }
    &=
        (\bigotimes_{ e\in \Interior{\Lambda}  }^{\vphantom{{\boldsymbol{x}}}}  \mybra{ \mathrm{Id}_e } \,)
        \bigotimes_{ {\boldsymbol{x}}\in \Lambda } \kappa_{ \partial{{\boldsymbol{x}}} }\circ  \Phi_{ \partial{{\boldsymbol{x}}} } (\Omega)^\dagger \circ \Phi_{\partial{{\boldsymbol{x}}}}(\Omega) \circ \kappa_{ \partial{{\boldsymbol{x}}} }
       ( \bigotimes_{ e\in \Interior{\Lambda} } \myket{ \mathrm{Id}_e } \,)
       \\
    & =
        \bigotimes_{e\in \Interior{\Lambda}} \mybra{ \mathrm{Id}_e }
        \bigotimes_{{\boldsymbol{x}}\in \Lambda}^{\vphantom{{\boldsymbol{x}}}} \kappa_{\partial{{\boldsymbol{x}}}} \circ \Phi_{\partial{{\boldsymbol{x}}}}(\Omega)\circ \kappa_{\partial{{\boldsymbol{x}}}}
        \bigotimes_{e\in\Interior{\Lambda}} \myket{ \mathrm{Id}_e }
        \\
    & =
        (\tilde\kappa_{\partial\Lambda}\otimes \mybra{ \tilde\kappa_{\Interior{\Lambda} } } )
        \circ 
        \bigotimes_{{\boldsymbol{x}}\in \Lambda}^{\vphantom{{\boldsymbol{x}}}}  \Phi_{\partial{{\boldsymbol{x}}}}(\Omega)
        \circ (\tilde\kappa_{\partial\Lambda}\otimes \myket{ \tilde\kappa_{\Interior{\Lambda} } } ) 
    \\[5pt]
    & =
        \tilde\kappa_{\partial\Lambda} 
        \circ 
        \Phi_{\partial{\Lambda}}(\Omega)
        \circ 
        \tilde\kappa_{\partial\Lambda},
\end{align}
here, the first equality holds by the expressions of $\rho_{\partial\Lambda}$ and $\Gamma_\Lambda$ in \cref{Eq:DefRhoLambda,Eq:DefGammaLambda,Eq:DefLocalTensorMap}, respectively, the second follows from the fact that $\Omega$ is a self-adjoint idempotent and $\Phi_{\partial {\boldsymbol{x}}}$ is a $*$-representation, and the third equality is due to \cref{Prop:FactKappaBd}.
\\[5pt]
The remaining part of the proof is devoted to prove \cref{Eq:SuffBulkBdry} for any comb-like region $\Lambda$ as in \cref{Def:Comblike}. Assume without loss of generality that the direction of growth is vertical, from bottom to top. In particular, all segments considered are rows.
In first place, note that \cref{Eq:SuffBulkBdry} holds for a segment of two sites, as it can be easily simplified by virtue of \cref{Eq:IdentityBlackVsWhiteFull}. The following picture describes this situation:
\[
\vcenter{\hbox{\includegraphics[page=069]{haagduality_diagrams.pdf}}}
\ = \ 
\vcenter{\hbox{\includegraphics[page=070]{haagduality_diagrams.pdf}}}
\]
Note that the weights in the middle correspond to $\Phi(\xi)^{1/2}$ by the choice in \cref{Eq:ChoiceRightmostTildeKappaE}.
Trivially, incorporating more sites in the horizontal direction follows analogously. This proves that all segments satisfy \cref{Eq:SuffBulkBdry}. Let us now describe the aforementioned condition for a single plaquette. If this is the case, by virtue of \cref{Eq:IdentityBlackVsWhite}, the left-hand side takes the form
\begin{align*}
&
\vcenter{\hbox{\includegraphics[page=071]{haagduality_diagrams.pdf}}}
 \ = \
\vcenter{\hbox{\includegraphics[page=072]{haagduality_diagrams.pdf}}}
\\
\intertext{and as a consequence repeated applications of \cref{Eq:IdentityBlackVsWhiteFull,Eq:TransferEigenvalue},}
=\cdots = &
\vcenter{\hbox{\includegraphics[page=073]{haagduality_diagrams.pdf}}}
\ = \ 
\vcenter{\hbox{\includegraphics[page=074]{haagduality_diagrams.pdf}}}
\hphantom{=\cdots=}
\end{align*}
we recover $\Phi_{\partial\Lambda}(\Omega)$, as we wanted to prove. Finally, let us consider a general comb-like region $\Lambda = \Lambda_1\cup \Lambda_2\cup \cdots \cup \Lambda_\kappa$ and discuss the inductive step in which we incorporate $\Lambda_1\cup \cdots \cup \Lambda_{s-1}$, for which \cref{Eq:SuffBulkBdry} holds by assumption, and a segment $\Lambda_{s,1}$. Assume this corresponds to a unique branch or that there are no branches. The following figures illustrate the resulting simplification in a general configuration, obtained by applying iteratively \cref{Eq:IdentityBlackVsWhiteFull,Eq:TransferEigenvalue}:
\begin{align*}
\vcenter{\hbox{\includegraphics[page=075]{haagduality_diagrams.pdf}}}
\qquad\qquad
\\
\qquad =\cdots =
\vcenter{\hbox{\includegraphics[page=076]{haagduality_diagrams.pdf}}}
\\
\qquad =\cdots =
\vcenter{\hbox{\includegraphics[page=077]{haagduality_diagrams.pdf}}}
\end{align*}
Note that the choice of weights considered here, as in \cref{Eq:ChoiceRightmostTildeKappaE}, is relevant for the inductive step to match the specific order of applying \cref{Eq:IdentityBlackVsWhiteFull,Eq:TransferEigenvalue}. For other directions, one can distribute the intrinsic weights according to more appropriate configurations using \cref{Prop:FactKappaBd,Prop:RewrittenStatesKappa}.
This concludes the proof of \cref{Thm:BulkBoundary}.
\end{proof}
    
\subsection{Parent Hamiltonians}
\label{Subsection:PH}
Given a two-dimensional tensor network state defined on a finite region of the lattice, one can associate to it a local, frustration-free Hamiltonian known as the parent Hamiltonian. This operator is constructed in such a way that the tensor network state, with arbitrary boundary conditions, spans its ground state space. The structure of the parent Hamiltonian will serve as a central tool in our analysis.
\\[5pt]
With the notations from the previous subsection, consider a comb-like subregion $\Lambda$ of $\mathbb{Z}^2$, and assume that it is expressible as a union of plaquettes and let
\begin{equation}\label{Eq:DefGSStab}
    \Stab_{\Lambda} : \mathscr{H}_{\Lambda}\to \mathscr{H}_{\Lambda}
    ,\qquad
    \Stab_{ \Lambda }
    \coloneq
    \Gamma_{ \Lambda }
        \circ \tilde\kappa_{ \partial{\Lambda} }^{-2}
        \circ (\Gamma_{ \Lambda })^{\dagger},
\end{equation}
where $\tilde\kappa_{ \partial{\Lambda} } $ is any positive endomorphism as introduced in \cref{Thm:BulkBoundary}.
\\[5pt]
It is simple to check that, for a single plaquette $\mathcal{P}$, this map takes the form
\[
    \Stab_{\mathcal{P}} = \ 
\vcenter{\hbox{\includegraphics[page=078]{haagduality_diagrams.pdf}}}
\]
In the particular case of two contiguous plaquettes $\mathcal{P}\cup\mathcal{Q}$ along the horizontal axis,
\[
    \Stab_{\mathcal{P}\cup \mathcal{Q}} = \ 
\vcenter{\hbox{\includegraphics[page=079]{haagduality_diagrams.pdf}}}
    \ .
\]
\begin{theorem}
\label{Thm:ParentHamiltonian}
Consider the finite-range interaction $\Theta$ on $\caA$ given by
\[
    \Theta(\Sigma) := \begin{cases}
        Q_\Sigma^\perp = \mathrm{Id} - Q_\Sigma & \text{ if }\Sigma\text{ is a plaquette},\\
        0        & \text{ otherwise,}
    \end{cases}
\]
for all finite subsets $\Sigma$ of $\mathbb{Z}^2$.
For any finite subset $\Lambda$ of $\mathbb{Z}^2$, the corresponding local Hamiltonian is defined by the expression
\[
    H_{\Lambda}
    := \sum_{\Sigma\subseteq \Lambda} \Theta(\Sigma)
    =  \sum_{\mathcal{P}\subseteq \Lambda} Q_{\mathcal{P}}^\perp,
\]
where the second sum is over all plaquettes $\mathcal{P}$ contained in $\Lambda$.
In the particular case in which $\Lambda$ is a comb-like subregion expressible as union of plaquettes, $H_{\Lambda}$ is a minimal local commuting parent Hamiltonian for the tensor network states introduced above. That is, $H_\Lambda$ is self-adjoint, (moreover, positive), and satisfies
\[
    \mathcal{G}_\Lambda := \operatorname{Im}\Gamma_\Lambda = \ker{H};
\]
furthermore, it is frustation-free, i.e.~for any plaquette $\mathcal{P}\subseteq \Lambda$:
\[
    Q_{\mathcal{P}}^\perp \circ \Gamma_\Lambda = 0,
\]
and its local terms pairwise commute, i.e.~for any two plaquettes $\mathcal{P},\mathcal{Q}\subseteq\Lambda$:
\[
    Q_{\mathcal{P}}^\perp \circ Q_{\mathcal{Q}}^\perp
    =
    Q_{\mathcal{Q}}^\perp \circ Q_{\mathcal{P}}^\perp .
\]
\end{theorem}

The proof of the theorem is presented below as a consequence of two lemmas.

\begin{lemma}\label{Lemma:GammaInvarBd}
    Let $\Lambda$ be a comb-like subregion of $\mathbb{Z}^2$. Then, 
    \begin{equation*}
        \Gamma_{ \Lambda }^{\vphantom{-2}}
            \circ \tilde\kappa_{ \partial{\Lambda} }^{-2}
            \circ \rho_{ \partial{\Lambda} }^{\vphantom{-2}}
        = 
        \Gamma_{ \Lambda }.
    \end{equation*}
\end{lemma}

\begin{proof}
    By virtue of \cref{Thm:BulkBoundary}, the left-hand side of the statement is 
    \begin{align}
        \Gamma_{ \Lambda } \circ \tilde\kappa_{ \partial{\Lambda} }^{-2} \circ \rho_{ \partial{\Lambda} }
        &= 
        \bigotimes_{ {\boldsymbol{x}} \in \Lambda }^{\vphantom{.}} \Gamma_{ {\boldsymbol{x}} }
                \bigotimes_{ e\in \Interior{\Lambda} } \myket{ \mathrm{Id}_e }
            \circ \tilde\kappa_{ \partial{\Lambda} }^{-1}
            \circ \Phi_{ \partial{\Lambda} }(\Omega)
            \circ \tilde\kappa_{ \partial{\Lambda} }
        \\ &=
        \bigotimes_{ {\boldsymbol{x}} \in \Lambda } \Phi_{ \partial{{\boldsymbol{x}}} }(\Omega)
            \myket{ \tilde\kappa_{ \Interior{\Lambda} } }
            \circ  \Phi_{\partial{\Lambda}}(\Omega)
            \circ \tilde\kappa_{\partial{\Lambda}},
    \end{align}
    where $\tilde\kappa_{\Interior{\Lambda}}$ is as in \cref{Prop:FactKappaBd}.
    Therefore, it is sufficient to check that
    \begin{equation}
        \label{eq:toProveBdInv}
        \bigotimes_{ {\boldsymbol{x}} \in \Lambda} \Phi_{ \partial{{\boldsymbol{x}}} }( \Omega )
            \myket{ \tilde\kappa_{ \Interior{\Lambda} } }
            \circ \Phi_{ \partial{\Lambda} }( \Omega )
        = 
        \bigotimes_{ {\boldsymbol{x}} \in \Lambda } \Phi_{ \partial{{\boldsymbol{x}}} }( \Omega )
            \myket{ \tilde\kappa_{ \Interior{\Lambda} } }.
    \end{equation}
    First, if $\Lambda$ consists of a single segment, assume it is a row without loss of generality, the previous equation holds by virtue of repeated applications of the pulling-through identities presented in \cref{Eq:PT_4sites_vs2}:
\begin{align*}
\vcenter{\hbox{\includegraphics[page=080]{haagduality_diagrams.pdf}}}
&= 
\vcenter{\hbox{\includegraphics[page=081]{haagduality_diagrams.pdf}}}
\\
&=
\vcenter{\hbox{\includegraphics[page=082]{haagduality_diagrams.pdf}}}
\\
&= 
\vcenter{\hbox{\includegraphics[page=083]{haagduality_diagrams.pdf}}}
\end{align*}
where the last equality holds since the composition of the two corresponding operators on the bottom-left part rephrases the product $\Phi_{\partial {\boldsymbol{x}}}(\Omega)\circ \Phi_{\partial {\boldsymbol{x}}}(\Omega) = \Phi_{\partial {\boldsymbol{x}}}(\Omega^2) = \Phi_{\partial {\boldsymbol{x}}}(\Omega)$, in such a way that the outer operator is completely absorbed and we obtain the right-hand side of \cref{eq:toProveBdInv}.
Let us now consider the particular case of a plaquette $\Lambda = \mathcal{P}$ and provide a detailed proof of \cref{eq:toProveBdInv}. First, the left-hand side of the equation is then given by:
\begin{equation}
\vcenter{\hbox{\includegraphics[page=084]{haagduality_diagrams.pdf}}}
        \ = \ 
\vcenter{\hbox{\includegraphics[page=085]{haagduality_diagrams.pdf}}}
        \ = \ 
\vcenter{\hbox{\includegraphics[page=086]{haagduality_diagrams.pdf}}}
    \end{equation}
    on the left, all inner circles correspond to the operators $\Phi_{\partial {\boldsymbol{x}}}(\Omega)\in A$ for all vertices ${\boldsymbol{x}}\in\mathcal{P}$, the inner straight lines correspond to the vectorization of $\tilde\kappa_{\Interior{\Lambda}}$, while the outer operator corresponds to $\Phi_{\partial\mathcal{P}}(\Omega)$; note that for the first equality we have simply applied \cref{Prop:FactKappaBd}, while the second equality follows from the pulling-through identity in \cref{Eq:PT_4sites_vs2}.
    Moreover,
    \begin{equation}
\vcenter{\hbox{\includegraphics[page=087]{haagduality_diagrams.pdf}}}
        \ = \ 
\vcenter{\hbox{\includegraphics[page=088]{haagduality_diagrams.pdf}}}
        \ = \ 
\vcenter{\hbox{\includegraphics[page=089]{haagduality_diagrams.pdf}}}
    \end{equation}
    where in the first equality we have used the version of the pulling-through identity in \cref{Eq:PT_4sites_vs1} of the outer operator against the corresponding top-left inner operator, and for the second equality we have used an analogous argument, applying the pulling-through identity in \cref{Eq:PT_4sites_vs1} of the outer operator against the bottom-right inner operator. Finally,
        \begin{equation}
\vcenter{\hbox{\includegraphics[page=090]{haagduality_diagrams.pdf}}}
        \ = \ 
\vcenter{\hbox{\includegraphics[page=091]{haagduality_diagrams.pdf}}}
        \ = \ 
\vcenter{\hbox{\includegraphics[page=092]{haagduality_diagrams.pdf}}}
\end{equation}
where the first equality holds as in the previous cases by virtue of the fact that $\Omega$ is an idempotent, and the second equality follows after re-ordering the weights.
\\[5pt]
Now, consider comb-like region of the form $\Lambda = \Lambda_1\cup \cdots \cup \Lambda_\kappa$, assume that \cref{eq:toProveBdInv} holds for $\Lambda_1\cup\cdots \cup \Lambda_{s-1}$ and let us consider incorporating a segment in $\Lambda_s$. The following figures apply for both a single branch or when considering an isolated individual branch, and illustrate how the pulling-through identities are applied iteratively along the sites of the segment in $\Lambda_s$. Note that, as in the case of a single plaquette, the weights associated with the newly incorporated plaquettes can be distributed accordingly, in view of \cref{Prop:FactKappaBd}: 
\begin{align*}
\vcenter{\hbox{\includegraphics[page=093]{haagduality_diagrams.pdf}}}
\qquad\qquad\qquad
\\
\qquad\qquad=
\vcenter{\hbox{\includegraphics[page=094]{haagduality_diagrams.pdf}}}
\\
\qquad\qquad=
\vcenter{\hbox{\includegraphics[page=095]{haagduality_diagrams.pdf}}}
\\
\qquad\qquad=
\vcenter{\hbox{\includegraphics[page=096]{haagduality_diagrams.pdf}}}
\end{align*}
This concludes the proof of \cref{eq:toProveBdInv}, and thus establishes the result.
\end{proof}

\begin{lemma}\label{Lem:int}
    \label{Lemma:GammaIntProp}
    (1) For any two plaquettes $\mathcal{P}$ and $\mathcal{Q}$ of $\mathbb{Z}^2$,
    \[ Q_{\mathcal{P}\cup\mathcal{Q}} = Q_{\mathcal{P}} \circ Q_{\mathcal{Q}}.\] 
    In particular, the local terms of the Hamiltonian commute pairwise.
    \\[5pt]
    (2) Let $\Lambda$ be a comb-like subregion of $\mathbb{Z}^2$, expressible as union of plaquettes. Then,
    \[
        Q_{\Lambda} = \prod_{\mathcal{P}\subseteq \Lambda} Q_{\mathcal{P}}.
    \]
    In particular, the ground-state projections satisfy the intersection property.
\end{lemma}

\begin{proof}
To prove (1), let us consider the following eight possible cases:
\[
\vcenter{\hbox{\includegraphics[page=097]{haagduality_diagrams.pdf}}}
\]
We now provide detailed proofs for Cases 1 and 5. The remaining cases can be treated analogously.
\begin{stepPT}
Let us consider two plaquettes $\mathcal{P},\mathcal{Q}\subset \mathbb{Z}^2$ that intersect at two contiguous sites along the vertical axis, in such a way that the composition of the local terms $\Stab_{\mathcal{P}}\circ \Stab_{\mathcal{Q}}$ is as in the left-hand side of the following picture:
\begin{equation*}
\vcenter{\hbox{\includegraphics[page=098]{haagduality_diagrams.pdf}}}
    \ = \  
\vcenter{\hbox{\includegraphics[page=099]{haagduality_diagrams.pdf}}}
\end{equation*}
where the green weights correspond to $\Phi(\xi^{1/2})$, the blue weights correspond to $\Phi(g^{1/2})$, the yellow weights correspond to $\Phi(\xi_L^{1/2})$ and $\Phi(\xi_R^{1/2})$, and the order of the composition is read from bottom to top.
\\[5pt]
Note that in this picture we have employed the equivalent expressions presented in \cref{Prop:FactKappaBd} in order to re-express the corresponding weights. Note also that we have omitted the weights $b(\omega)^{1/2}$ on the physical legs for the sake of clarity. In the second picture we remark that the terms in the middle, represented using bold lines, can be simplified, where we have employed that $\Phi_{\partial {\boldsymbol{x}}}(\Omega)$ is idempotent for all sites ${\boldsymbol{x}}\in \mathcal{P}\cap\mathcal{Q}$.
Now, this expression becomes
\begin{equation*}
\vcenter{\hbox{\includegraphics[page=100]{haagduality_diagrams.pdf}}}
    \ = \  
\vcenter{\hbox{\includegraphics[page=101]{haagduality_diagrams.pdf}}}
\end{equation*}
for which we have used \cref{Eq:IdentityBlackVsWhite}, which provides the operator $\Phi_{\partial (\mathcal{P}\cap\mathcal{Q})}(\Omega)$, as emphasized in the middle of the picture; the picture on the right can be obtained by employing the pulling-through identity. Finally, the last expression equals
\begin{equation*}
\vcenter{\hbox{\includegraphics[page=102]{haagduality_diagrams.pdf}}}
    \ = \  
\vcenter{\hbox{\includegraphics[page=103]{haagduality_diagrams.pdf}}}
\end{equation*}
by simply applying the pulling-through identities, as in the proof of \cref{Lemma:GammaInvarBd}, while to obtain the second expression it suffices to note that on the left we have the operator $\Phi_{\partial\mathcal{P}}(\Omega)$ supported on $\partial\mathcal{P}$ followed by the tensor network state map on top, and the former vanishes by virtue of \cref{Lemma:GammaInvarBd}. Note that the expression on the right is precisely $\Stab_{\mathcal{P}\cup \mathcal{Q}}$, as we wanted to prove.
\end{stepPT}\setcounter{stepPT}{4}
\begin{stepPT}
Consider now the case in which two different plaquettes $\mathcal{P},\mathcal{Q}\subset \mathbb{Z}^2$ intersect at a single site, and $\Stab_{\mathcal{P}}\circ \Stab_{\mathcal{Q}}$ is represented as follows
\begin{equation*}
\vcenter{\hbox{\includegraphics[page=104]{haagduality_diagrams.pdf}}}
    \ = \ 
\vcenter{\hbox{\includegraphics[page=105]{haagduality_diagrams.pdf}}}
\end{equation*}
here, the expression on the left corresponds to the composition $\Stab_{\mathcal{P}}\circ \Stab_{\mathcal{Q}}$, and to obtain the second expression, we have just simplified $\Phi_{\partial {\boldsymbol{x}}}(\Omega)^\dagger \circ \Phi_{\partial {\boldsymbol{x}}}(\Omega) = \Phi_{\partial {\boldsymbol{x}}}(\Omega)$, where ${\boldsymbol{x}}\in \mathcal{P}\cap\mathcal{Q}$; furthermore, the latter diagram is in turn equal to
\begin{equation*}
\vcenter{\hbox{\includegraphics[page=106]{haagduality_diagrams.pdf}}}
    \ = \ 
\vcenter{\hbox{\includegraphics[page=107]{haagduality_diagrams.pdf}}}
    \ ,
\end{equation*}
where to obtain the left-hand side expression we have use the pulling-through identities as in Case~1, and to obtain the right-hand side expression we use that the boundary operator vanishes with the projected entangled pair map as has been shown in \cref{Lemma:GammaInvarBd}. 
\end{stepPT}%
\noindent{}Let us now prove (2). For general comb-like regions, we proceed inductively. As defined in \cref{Def:Comblike}, consider a comb-like region $\Lambda = \Lambda_1\cup \cdots \cup\Lambda_\kappa$. Let us discuss the inductive step, in which we consider $\Lambda=R\cup \mathcal{P}$, the union of a comb-like region $R$ expressible as union of plaquettes for which the statement holds, and a contiguous plaquette $\mathcal{P}\subsetneq R$.
The following figure illustrates three representative cases of this procedure.
\[
\vcenter{\hbox{\includegraphics[page=108]{haagduality_diagrams.pdf}}}
\]
In all Case~9 to 11, we consider the composition \( Q_{\mathcal{P}} \circ Q_R \), where \( R \) and \( \mathcal{P} \) overlap on two or three sites along different directions. This situation is slightly different, but as in the previous cases, the tensor network arising at the intersection \( R \cap \mathcal{P} \) can still be simplified. In particular, it yields an operator of the form \( \Phi_{\partial(R \cap \mathcal{P})}(\Omega) \), and a generalized version of the pulling-through identity along $\mathcal{P}$ can again be applied to conclude the result.
\\[5pt]
Let us finally consider the case in which $\Lambda_{s}$ is union of two disjoint segments, for some $s\in \{3, \ldots, \kappa\}$. Without loss of generality, let $s = 3$. The following picture exemplifies this situation.
\[
\vcenter{\hbox{\includegraphics[page=109]{haagduality_diagrams.pdf}}}
\]
As above, we proceed inductively by incorporating all the segments of one branch. The same reasoning applies for the composition $Q_{\mathcal{P}_{2,2}}\circ Q_{\mathcal{P}_{2,1}\cup R}$, as each segment in steps $s,\ldots,\kappa$ is separated by assumption and the previous inductive cases apply. This concludes the proof.
\end{proof}

\begin{proof}[Proof of \cref{Thm:ParentHamiltonian}]
On the one hand, \cref{Lemma:GammaInvarBd} has two straightforward consequences. First, that $\operatorname{Im}\Gamma_{\Lambda}$ is a set of ground states of $H_{\Lambda}$, since
\begin{equation}
    \label{eq:PlaqIsStab}
    \Stab_{ \mathcal{P} } \circ \Gamma_{ \mathcal{P} }
    =
    \Gamma_{\mathcal{P}}^{\vphantom{-2}}
        \circ \tilde\kappa_{ \partial{\mathcal{P}} }^{-2}
        \circ \rho_{ \partial{\mathcal{P}} }^{\vphantom{-2}}
    =
    \Gamma_{\mathcal{P}}.
\end{equation}
Second, that the local terms are positive projections,
\begin{equation}
    \label{eq:PlaqIsProj}
    \Stab_{ \mathcal{P} }^2 
    =
    \Gamma_{\mathcal{P}}^{\vphantom{-2}}
        \circ \tilde\kappa_{ \partial{\mathcal{P}} }^{-2}
        \circ \rho_{ \partial{\mathcal{P}} }^{\vphantom{-2}}
        \circ \tilde\kappa_{ \partial{\mathcal{P}} }^{-2}
        \circ (\Gamma_{ \mathcal{P} })^{\dagger} 
    =
    \Gamma_{ \mathcal{P} }^{\vphantom{-2}}
        \circ \tilde\kappa_{ \partial{\mathcal{P}} }^{-2}
        \circ (\Gamma_{ \mathcal{P} })^{\dagger} 
    =
    \Stab_{ \mathcal{P} }^{\vphantom{-2}}.
\end{equation}
On the other hand, an immediate consequence of \cref{Lemma:GammaIntProp} is that the local terms of the Hamiltonian pairwise commute and, in addition, the  manifold of ground states of the local Hamiltonian consists precisely on the tensor network states with arbitrary boundary conditions.
Indeed, let us consider an arbitrary ground state $\myket{\psi_0}\in \mathscr{H}_\Lambda $ of $H_\Lambda$. Since the local terms are positive, $\Stab_{\mathcal{P}} \myket{\psi_0} =  \myket{\psi_0}  $ for all plaquettes $\mathcal{P}\subset \Lambda$, and hence $ \Stab_{\Lambda}\myket{\psi_0} =  \myket{\psi_0} $ by virtue of \cref{Lemma:GammaIntProp}. In conjunction with \cref{Eq:DefGSStab}, this implies that $\myket{\psi_0} \in \operatorname{Im}\Stab_{\Lambda}\subset \operatorname{Im}\Gamma_\Lambda$, as we wanted to prove.
\end{proof}

\subsection{Topological Quantum Order}
\label{Subsection:TQO}
This subsection is devoted to proving that the models introduced above satisfy the \emph{local topological quantum order} condition as introduced in the work of Michalakis and Zwolak \cite{michalakis_stability_2013}; see also \cite{cirac_robustness_2013}.
\\[5pt]
We note that the $C^*$-weak Hopf algebra of symmetries in this context is, in addition, biconnected, thereby excluding states arising from proper unitary multi-fusion categories, for which this property clearly does not hold.

\begin{theorem}\label{Thm:TQO}
Let $A$ be a biconnected $C^*$-weak Hopf algebra. Then, for any fi\-ni\-te rectangular region $\Lambda$ of $\mathbb{Z}^2$ and all rectangular subregions $\Sigma\subset\Lambda$ such that $\partial\Sigma\subset \Interior{\Lambda}$, there exists a positive linear functional $c_\Sigma:\operatorname{End}\mathscr{H}_\Sigma\to\mathbb{C}$ satisfying
\begin{equation}\label{Eq:TQO1v1}
    (\Gamma_\Lambda)^\dagger \circ X_\Sigma \circ \Gamma_{\Lambda} = c_\Sigma( X_\Sigma) \cdot \rho_{\partial\Lambda},
\end{equation}
for all $X_\Sigma\in\operatorname{End}\mathscr{H}_\Sigma$. This is equivalent to the following condition:
\begin{equation}\label{Eq:TQO1v2}
    \Stab_\Lambda \circ X_\Sigma \circ \Stab_\Lambda = c_\Sigma(X_\Sigma) \cdot  \Stab_\Lambda
\end{equation}
for all $X_\Sigma\in\operatorname{End}\mathscr{H}_\Sigma$, i.e.~the states satisfy the local to\-po\-lo\-gi\-cal quan\-tum order condition.
\end{theorem}
The latter identity corresponds to Definition~4 in \cite{michalakis_stability_2013}.

\begin{proof}
First, \cref{Eq:TQO1v1} implies \cref{Eq:TQO1v2} by simply considering the property in \cref{Lemma:GammaInvarBd}:
\begin{align*}
    \Stab_\Lambda \circ X_\Sigma \circ \Stab_\Lambda
    &=
    \Gamma_\Lambda \circ \tilde\kappa_{\partial\Lambda}^{-2} \circ (\Gamma_\Lambda)^\dagger \circ X_\Sigma \circ \Gamma_\Lambda \circ \tilde\kappa_{\partial\Lambda}^{-2} \circ (\Gamma_\Lambda)^\dagger
    \\&=
    c_\Sigma(X_\Sigma) \cdot \Gamma_\Lambda \circ \kappa_{\partial\Lambda}^{-2} \circ \rho_{\partial\Lambda} \circ \tilde\kappa_{\partial\Lambda}^{-2} \circ (\Gamma_\Lambda)^\dagger
    \\&=
    c_\Sigma(X_\Sigma) \cdot \Gamma_\Lambda \circ \tilde\kappa_{\partial\Lambda}^{-2} \circ (\Gamma_\Lambda)^\dagger
    =
    c_\Sigma(X_\Sigma) \cdot \Stab_\Lambda.
\end{align*}
The converse implication holds because $\Stab_\Lambda \circ \Gamma_\Lambda  = \Gamma_\Lambda$, see e.g.~\cref{eq:PlaqIsStab}.
Now we proceed to the proof of \cref{Eq:TQO1v1}. For the sake of clarity we consider the following diagram, which corresponds to the left-hand side of the equation, where $\Lambda$ is a $3\times 3$ lattice and the subregion $\Sigma$, highlighted in yellow, corresponds to the inner site of the lattice:
\begin{equation}
\vcenter{\hbox{\includegraphics[page=110]{haagduality_diagrams.pdf}}}
\ .
\end{equation}
For simplicity, we have omitted all weights on the boundary legs, and the weights $b(\omega)$ on the physical lines. Furthermore, as discussed in \cref{Rem:DistribW}, the intrinsic weights $\Psi(\hat\xi_R^{-1})$ have been distributed along the edges of $\partial\Sigma$; hence, there are no weights along these edges. A general proof follows readily by extending this specific case. The observable $X_\Sigma$ corresponds to the stars in the middle, depicted here as a tensor-product operator for clarity.
As in the proof bulk-boundary correspondence outlined in \cref{Thm:BulkBoundary}, note that $\Omega^2 = \Omega$ and consequently one can first simplify the product of the on-site matrix product operators on the sites of $\Lambda\setminus\Sigma$, obtaining the following equivalent expression:
\begin{equation}
\vcenter{\hbox{\includegraphics[page=111]{haagduality_diagrams.pdf}}}
\ .
\end{equation}
Since $\Lambda\setminus \Sigma$ is not a comb-like region by construction, the remaining simplification of these operators, using \cref{Eq:IdentityBlackVsWhiteFull}, requires taking into account the sum over all sectors $a=1,\ldots,r$ of $A^*$ and their corresponding multiplicities. This leads to the expression:
\begin{equation}
\label{Eq:LTQOaa}
    \sum_{a=1}^r \ 
\vcenter{\hbox{\includegraphics[page=112]{haagduality_diagrams.pdf}}}
\ .
\end{equation}
In order to decouple the inner and outer regions, which are currently related by the sum over all sectors, we will multiply the inner matrix product operator associated with sector $a$ by the matrix product operator representation of $\Omega$ acting on $\Sigma$. We remark that, if $\Sigma$ consists of multiple sites, then pulling-through equations are applied first. However, this multiplication cannot be performed directly, as (i) it involves contracting black tensors with white tensors, and (ii) the virtual representation only accounts for the irreducible $*$-representations $\Psi_a$, and we need the full $*$-representation. The following claim provides a workaround for the first issue.
\begin{claim}\label{Claim:BvsW}
For any comb-like subregion $R$ of $\mathbb{Z}^2$, recall that $\Phi_{\smash{\partial R}}(\Omega)$ is given by
\[
    \Phi_{\smash{\partial{R}}}(\Omega) =
        (
            \bigotimes_{ \partial_{\leftarrow}{R} } \dualPhi
            \otimes
            \bigotimes_{ \partial_{\rightarrow}{R} } {\Phi}
        )
         (\Omega_{(1)}\otimes \Omega_{(2)} \otimes \cdots \otimes \Omega_{\smash{(|\partial R|)}});
\]
then, the following is an equivalent expression:
\[
    \Phi_{\smash{\partial R}}(\Omega) =
        (
            \bigotimes_{ \partial_{\leftarrow}R} \Phi^{\mathrm{t}}
            \otimes
            \bigotimes_{ \partial_{\rightarrow}R} {\dualPhi}^{\mathrm{t}}
        )
         (\Omega_{(1)}\otimes \Omega_{\smash{(|\partial R|)}} \otimes \cdots \otimes \Omega_{(2)}),
\]
i.e.~for the resulting matrix product operator representation, it is possible to exchange black tensors and white tensors simultaneously by reversing the order of composition at the virtual level.
\end{claim}

\begin{proof}
    This is an immediate consequence of the following calculation:
    \begin{align*}
            (\bigotimes_{ \partial_{\leftarrow}{\smash{R}} } J
            \otimes
            \bigotimes_{ \partial_{\rightarrow}{{R}} } \mathrm{Id})
        &(\Omega_{(1)}\otimes \Omega_{(2)} \otimes \cdots \otimes \Omega_{(|\partial\smash{R}|)})
        \\
        &=
            (\bigotimes_{ \partial_{\leftarrow}{{R}} } J
            \otimes
            \bigotimes_{ \partial_{\rightarrow}{{R}} } \mathrm{Id})
         (J(\Omega)_{(1)}\otimes J(\Omega)_{(2)} \otimes \cdots \otimes J(\Omega)_{(|\partial\smash{R}|)})
        \\
        \intertext{since $J(\Omega) = \Omega$, as proven in \cref{Item:Omega-And-PT} of \cref{Thm:PT}}
        &=
            (\bigotimes_{ \partial_{\leftarrow}{{R}} } J^2
            \otimes
            \bigotimes_{ \partial_{\rightarrow}{{R}} } J)
         (\Omega_{(1)}\otimes \Omega_{(|\partial\smash{R}|)} \otimes \cdots \otimes \Omega_{(2)})
        \\
        \intertext{by virtue of the antimultiplicativity of $J$, see \cref{Item:Janticomult} of \cref{Prop:J}, and}
        &=
            (\bigotimes_{ \smash{\partial_{\leftarrow}{ R}} } \mathrm{Id}
            \otimes
            \bigotimes_{ \smash{\partial_{\rightarrow}{R}} }  J)
         (\Omega_{(1)}\otimes \Omega_{\smash{(|\partial R|)}} \otimes \cdots \otimes \Omega_{(2)}),
    \end{align*}
    due to the fact that $J$ is involutive, see \cref{Item:Jinvolutive} of \cref{Prop:J}.
\end{proof}
For $R= \Sigma$, this establishes the equivalence between \cref{Eq:LTQOaa} and the expression
\begin{equation}
    \sum_{a=1}^r
\ 
\vcenter{\hbox{\includegraphics[page=113]{haagduality_diagrams.pdf}}}
\ .
\end{equation}
Note that the orientation of the arrows in the virtual inner thick circles have been reversed, and the colors of the tensors involved have also been swapped. Then, by taking into account the multiplicities as described in \cref{Eq:MultiplPsibx}, one can rewrite the previous expression as follows:
\begin{equation}
\sum_{a=1}^r\ 
\vcenter{\hbox{\includegraphics[page=114]{haagduality_diagrams.pdf}}}
\end{equation}
Note that only the outer matrix product operator is associated to the irreducible $*$-re\-pre\-sen\-ta\-tion $\Psi_a$ in the picture above.
Then, by virtue of \cref{Rem:ExamplBic}, and in particular \cref{Eq:OmegaEigBiconn}, the inner circle evaluates to $\mathrm{x}_a\Omega = d_a\Omega$, obtaining
\begin{equation}
    D^2 \sum_{a=1}^r
        \frac{d_a}{D^2 }
\ 
\vcenter{\hbox{\includegraphics[page=115]{haagduality_diagrams.pdf}}}
\ ,
\end{equation}
which is in turn equivalent to the following expression:
\begin{equation}
    D^2 \ 
\vcenter{\hbox{\includegraphics[page=116]{haagduality_diagrams.pdf}}}
\ .
\end{equation}
The linear functional $c_\Sigma$ consists on conjugating the observable on $\Sigma$ with the corresponding tensor network state maps, followed by a trace against the corresponding weights and multiplication by the constant $D^2$.
\end{proof}

This implies the following result.

\begin{corollary}\label{Cor:StateThLimit}
Let $\Lambda$ be a finite rectangular subregion of $\mathbb{Z}^2$ and let $X\in\mathscr{H}_{\partial \Lambda}$ be any boundary condition such that the map $\omega_{\Lambda,X}:\operatorname{End}\mathscr{H}_{\Lambda}\to \mathbb{C}$, defined by
\[
    \omega_{\Lambda,X}(O) \coloneq
    \mybra{\Gamma_{\Lambda}(X) } O \myket{ \Gamma_{\Lambda}(X) },
\]
is a state, i.e.~a positive unital linear functional.
Then,  there exists a unique frustration-free pure state $\omega_0: \caA\to\mathbb{C}$ in the thermodynamic limit, given by
\[ \omega_0(O) = \lim_{n \to \infty} \omega_{\Lambda_n,X_n}(O), \]
for any increasing and absorbing sequence $n\mapsto (\Lambda_n,X_n)$ of rectangular subregions $\Lambda_n\subset \mathbb{Z}^2$ and  boundary conditions  $X_n\in \mathscr{H}_{\partial\Lambda_n}$ such that $\omega_{\Lambda_n,X_n}$ are states.
\end{corollary}

\section{Proof of the sufficient condition via tensor networks}\label{Section:HaagForPEPS}

\subsection{The sufficient condition in tensor networks}

In this section, we prove that the family of two-dimensional tensor network states introduced in the previous section satisfies the conditions in \cref{seta} and \cref{finite} for cone-like regions as defined below in \cref{Def:comblikelkmk}. Together with \cref{Lemma:HaagReduction}, this will prove, as a corollary, our two main results: approximate Haag duality (\cref{Cor:MR1}) and exact Haag duality for a coarse-grained lattice (\cref{Cor:MR2}).
\\[5pt]
First, the interactions given in \cref{Thm:ParentHamiltonian} and the state $\omega_0$ in \cref{Cor:StateThLimit} satisfy conditions in \cref{seta} by \cref{Thm:ParentHamiltonian} and \cref{Cor:StateThLimit}. For each $N\in\bbN$, choose any unit vector $\Omega_N\in \mathcal{G}_{[-N,N]^2}$. From \cref{Thm:ParentHamiltonian}, Item 1 of \cref{finite} holds.
Item 2 of \cref{finite} follows from \cref{Cor:StateThLimit}.
\\[5pt]
It only remains to prove that Items 3, 4 and 5 in \cref{finite} hold for cone-like regions, as we define now:

\begin{definition}\label{Def:comblikelkmk}
Let $\boldsymbol{a}\in\mathbb{R}^2$ be a point in the dual lattice of $\mathbb{Z}^2\subset\mathbb{R}^2$ and consider two sequences $\{(l_i,m_i)\}_{i=1}^\infty,\{(l_i',m_i')\}_{i=1}^\infty\subset 2\mathbb{Z}^2\setminus\{(0,0)\}$ such that for all $i,j\in\mathbb{N}$ and non-zero $l_i,l_j,\ldots, m_i',m_j'$, the following holds:
\begin{equation}\label{Eq:signcond}
\begin{split}
\operatorname{sign}(l_i) = \operatorname{sign}(l_j),
&\quad
\operatorname{sign}(m_i) = \operatorname{sign}(m_j), 
\\
\operatorname{sign}(l_i') = \operatorname{sign}(l_j'), 
&\quad
\operatorname{sign}(m_i') = \operatorname{sign}(m_j').
\end{split}
\end{equation}
Consider a dual lattice path $L_1$ starting from $\boldsymbol{a}$ taking steps of the form $(l_1,0)$, $(0,m_1)$, $(l_2,0)$, $(0,m_2)$, \ldots, a dual lattice path $L_2$ starting from $\boldsymbol{a}$ taking steps of the form $(0,m_1')$, $(l_1',0)$, $(0,m_2')$, $(l_2',0)$, \ldots, such that $L_1\cap L_2 = \emptyset$.
We define the region
\[
\Gamma:=\Gamma_{\boldsymbol{a},l_1,m_1,l_2,m_2,\ldots}^{l_1', m_1',l_2',m_2',\ldots}\subset\mathbb{R}^2
\]
as the open region between $L_1$ and $L_2$, taken in counterclockwise direction from $L_1$ to $L_2$. We say that $\Gamma$ is cone-like if there exists cones $C_\kappa,C_{\kappa'}\subset\mathbb{R}^2$ with opening angles $\kappa$ and $\kappa'$, respectively, such that
\begin{equation}
C_{\kappa'} \subseteq \Gamma_{\boldsymbol{a},l_1,m_1,l_2,m_2}^{l_1', m_1',l_2',m_2',\ldots}\subseteq C_{\kappa'}.
\end{equation}
\end{definition}

\begin{remark}
   For every cone-like region $\Gamma$ as in the previous definition, the intersections $\Gamma \cap [-\ell,\ell]^2 \cap\mathbb{Z}^2$ are comb-like regions.
\end{remark}

The following picture exemplifies this definition. Note that the condition in \cref{Eq:signcond} gives a well-defined direction to each lattice path; in the picture, the first lattice path has direction right-up, while the second lattice path has direction left-up.
\[
\vcenter{\hbox{\includegraphics[page=117]{haagduality_diagrams.pdf}}}
\]

\begin{theorem}\label{thm:haagTN}
    Let $\Gamma\subset\mathbb{Z}^2$ be a cone-like region as in \cref{Def:comblikelkmk}, and set
    \begin{equation*}
        \mathcal{L}:=[-6\ell, 6\ell]^2\cap\mathbb{Z}^2
        ,\quad
        \Lambda := \mathcal{L}\cap \Gamma
        ,\quad
        \Sigma := [-\ell,\ell]^2\cap\Gamma,
    \end{equation*}
    for any $\ell\in\mathbb{N}$.
    Then, for sufficiently large $\ell\in\mathbb{N}$ there exist subregions
    \[ \tilde\Sigma\subset \mathcal{L}\setminus\Lambda,\quad \mathcal{R}\subseteq\Lambda \setminus\Sigma\quad\text{ and }\quad \tilde{\mathcal{R}}\subseteq (\mathcal{L}\setminus\Lambda)\setminus \tilde\Sigma
    \]
    satisfying the following properties:
    \begin{enumerate}
    \setlength\itemsep{3pt}
        \item[(i)] $\Sigma$, $\tilde\Sigma$, $\Lambda$, $\mathcal{L}\setminus\Lambda$, $\mathcal{R}$ and $\tilde{\mathcal{R}}$ are comb-like subregions of $\mathbb{Z}^2$;
        \item[(ii)] $\Sigma\cup\mathcal{R}$ and $\tilde\Sigma\cup \tilde{\mathcal{R}}$ are comb-like regions which are unions of plaquettes,
        \item[(iii)] $\partial\Lambda\cap \partial\Sigma \supseteq \partial \Lambda\cap \partial\tilde\Sigma$;
        \item[(iv)] the boundary of $\tilde\Sigma$ is large enough so that $|\partial\tilde\Sigma \setminus \partial\Lambda | \geq |\partial\Sigma\setminus\partial\Lambda|$,
    \end{enumerate}
    such that, for all $X_\Sigma \in \operatorname{End}\mathscr{H}_\Sigma$ there exists $Y_{\smash{\tilde\Sigma}}\in\operatorname{End}\mathscr{H}_{\smash{\tilde\Sigma}}$ such that
    \begin{equation*}
        \Stab_\Lambda\circ X_\Sigma \circ \Gamma_{\mathcal{L}}
        =
        \Stab_{\mathcal{L}\setminus \Lambda} \circ Y_{\smash{\tilde\Sigma}} \circ \Gamma_{\mathcal{L}}.
    \end{equation*}
    As in previous sections, $\Gamma_{\mathcal{L}}$ denotes the tensor network state map on $L$ as defined in \cref{Subsection:PEPS}, $\Stab_{\Lambda}$ and $\Stab_{\mathcal{L}\setminus\Lambda}$ stand for the corresponding ground-state projections as introduced in \cref{Subsection:PH}. Moreover, the map $X_\Sigma\mapsto Y_{\smash{\tilde\Sigma}}$ is $\bbC$-linear and $*$-preserving.
\end{theorem}
In the following figure we present an example in which we distinguish the regions $\Lambda$, $\Sigma$, $\mathcal{R}$ and $\tilde\Sigma$ on the two-dimensional tensor network states in \cref{Subsection:PEPS}:
\begin{equation}\label{eq:HaagProof-2d-1}
\vcenter{\hbox{\includegraphics[page=118]{haagduality_diagrams.pdf}}}
\ .
\end{equation}

Let us now show how \cref{thm:haagTN} implies Items 3, 4 and 5 in \cref{finite}.  In first place, we consider the bound $6\ell$ instead of $2\ell$, as discussed in \cref{arem}; we postpone this discussion to \cref{Rem:6ell} for the sake of clarity. The set $\Lambda$ in this section corresponds to the subregion denoted $\Gamma_{6\ell}^{(R)}$ in \cref{finite}, i.e.~the augmented truncation of the cone-like region.  The subregion $\Sigma$ corresponds then to $\Gamma_{\ell}^{(R)}$ and $\tilde\Sigma$ is constructed as a subregion of $\Gamma_{6\ell}^{(L)}$. The following picture exemplifies this situation:
\[
\vcenter{\hbox{\includegraphics[page=119]{haagduality_diagrams.pdf}}}
\ .
\]
Moreover, let
\[
S_\ell^{(1)}:(\operatorname{End}\mathscr{H}_{\smash{ \Gamma_{\ell}^{(R)} }})_{\mathrm{h}}
\to (\operatorname{End}\mathscr{H}_{\smash{ \Gamma_{6\ell}^{(R)} }})_{\mathrm{h}},
\quad 
S_\ell^{(2)}:(\operatorname{End}\mathscr{H}_{\smash{ \Gamma_{\ell}^{(R)} }})_{\mathrm{h}}
    \to(\operatorname{End}\mathscr{H}_{\smash{\Gamma_{6\ell}^{(L)}}})_{\mathrm{h}},
\]
be defined by the expressions
\begin{equation}
    S^{(1)}_\ell \equiv 0 \quad\text{ and }\quad S^{(2)}_{\ell}(X_\Sigma) := Y_{\smash{\tilde\Sigma}},
\end{equation}
with $Y_{\smash{\tilde\Sigma}}$ as in \cref{thm:haagTN}.
In addition, for the definition of the map
\begin{equation}\label{tell}
\Theta_{\ell}:\operatorname{End}\mathscr{H}_{\smash{\Lambda_\ell}}\to\operatorname{End}\mathscr{H}_{\smash{\Gamma_{6\ell}^{(R)}}},
\end{equation}
consider the analogous counterpart of $S_\ell^{(2)}$ for the complementary cone,
\begin{equation}\label{Eq:TildeSell}
    \tilde S_{\ell}^{(2)}
    :
    (\operatorname{End}\mathscr{H}_{\smash{\Gamma_{\ell}^{(L)}}})_{\mathrm{h}}
    \to
    (\operatorname{End}\mathscr{H}_{\smash{\Gamma_{6\ell}^{(R)}}})_{\mathrm{h}},
\end{equation}
and define the linear extension of the map that assigns to each tensor product operator $ X\otimes Y\in \operatorname{End}\mathscr{H}_{\smash{\Gamma_{\ell}^{(L)}}}\otimes \operatorname{End}\mathscr{H}_{\smash{\Gamma_{\ell}^{(R)}}}$ the expression
\begin{equation}
    \Theta_\ell(X\otimes Y) := Y  Q_{\smash{\Gamma_{6\ell}^{(R)}}} \tilde{S}_{\ell}^{(2)}(X) \in \operatorname{End}\mathscr{H}_{\smash{\Gamma_{6\ell }^{(R)}}}.
\end{equation}
Set $f:\bbN^2\to[0,\infty)$ so that
\begin{equation}
    f(N,\ell)=
    \begin{cases}
        0& \text{if } N\ge 6\ell,\\
        2+2\lV S_\ell^{(2)}\rV +\lV\Theta_\ell\rV& \text{if } N<6\ell.
    \end{cases}
\end{equation}
These are the choices corresponding to Items 3 and 4 in \cref{finite}. By construction of 
${S}_\ell^{(2)}$ and $\tilde{S}_\ell^{(2)}$,
\begin{align}
    &Q_{\Gamma_N^{(R)}}Z\Omega_N= Q_{\Gamma_N^{(L)}}S_\ell^{(2)}(Z)\Omega_N,\quad Z\in \operatorname{End}\mathscr{H}_{\smash{\Gamma_\ell^{(R)}}},\; N\ge 6\ell,\\
    &Q_{\Gamma_{6\ell}^{(L)}}X\Omega_N=Q_{\Gamma_{6\ell}^{(R)}}\tilde{S}_{\ell}^{(2)}(X)\Omega_N,\quad X\in \operatorname{End}\mathscr{H}_{\smash{\Gamma_{\ell}^{(L)}}}.
\end{align}
Therefore
\begin{align}
    Q_{\smash{\Gamma_{N}^{(L)}}} (X\otimes Y) \Omega_N &=
    Q_{\smash{\Gamma_{6 \ell}^{(L)}}} (X\otimes Y) \Omega_N =
    Y Q_{\smash{\Gamma_{6 \ell}^{(L)}}} X \Omega_N
    \\
    \notag
    & = Y Q_{\smash{\Gamma_{6\ell }^{(R)}}}\tilde S_\ell^{(2)} (X)\Omega_N =
    \Theta_\ell(X\otimes Y) \Omega_N,
\end{align}
for all $ X\otimes Y\in \operatorname{End}\mathscr{H}_{\smash{\Gamma_{\ell}^{(L)}}}\otimes \operatorname{End}\mathscr{H}_{\smash{\Gamma_{\ell}^{(R)}}}$
and $N\ge 6  \ell$
proving \cref{Eq:QQQ,Eq:PropThetaL} for $f=0$.
\cref{QQ} with $f=0$ is given by \cref{Lem:int}. 
\\[5pt]
This concludes that all items in \cref{finite} hold and hence, by \cref{Lemma:HaagReduction}, this proves the following result.

\begin{theorem}\label{Thm:Haaglkmk}
    For every cone-like region $\Gamma$ as in \cref{Def:comblikelkmk}, it holds that \[\pi(\mathcal{A}_{\Gamma})'' = \pi(\mathcal{A}_{\Gamma^c})'.\]
\end{theorem}

As corollaries, we obtain the two main results of this work, which we now restate and prove.

\begin{corollary}\label{Cor:MR1}
    Approximate Haag duality holds for two-dimensional biconnected $C^*$-weak Hopf algebras tensor network states.
\end{corollary}

\begin{proof}
See \cref{Appendix:Deformations}. 
\end{proof}

In the case of the coarse-grained lattice, we consider the scenario in which the new tensor network is constructed as a blocked version of the original one, obtained by grouping disjoint plaquettes of four sites into single sites, as illustrated in the following figure.
\begin{equation}\label{Eq:BlockedLatt}
\vcenter{\hbox{\includegraphics[page=120]{haagduality_diagrams.pdf}}}
\end{equation}

The following is the second main result of this work.

\begin{corollary}\label{Cor:MR2}
    For the coarse-grained lattice model, Haag duality holds.
\end{corollary}

\begin{proof}
Note that cones in the coarse-grained lattice correspond, in the original lattice, to cone-like regions as introduced in \cref{Def:comblikelkmk}. Indeed, the rays defining the cone on the coarse-grained lattice provide well-defined directions satisfying \cref{Eq:signcond}. In the original lattice, this corresponds to a region with the same shape for which the lengths of the steps are doubled, since each site in the coarse-grained lattice corresponds to a plaquette of fours sites in the original lattice. Therefore, \cref{Thm:Haaglkmk} applies.
\end{proof}

\subsection{Sketch of the proof of \texorpdfstring{\cref{thm:haagTN}}{Theorem \ref{thm:haagTN}}}\label{Subsection:Haag-example}

Let us postpone the proof of \cref{thm:haagTN} to the following subsection, and provide now a highly simplified and informal version of the argument to convey the core ideas in a more accessible setting.
We consider a one-dimensional system instead of the full two-dimensional setting, and disregard the weights presented in \cref{Section:Construction}. Note that the latter condition holds in essence if the $C^*$-weak Hopf algebra arises from a finite group.
\\[5pt]
More concretely, let us consider the lattice $\mathcal{L}$ to be a finite interval in $\mathbb{Z}$, the cone-like region $\Lambda$ to consist of two adjacent sites, and the inner region $\Sigma$ to be a single site.
From the formulation of the theorem, our goal is to show that for every observable $X_\Sigma$ supported in $\Sigma$, the following condition is satisfied:
\begin{equation}\label{eq:HaagCondSimplf}
    \Stab_\Lambda \circ X_\Sigma \circ \Gamma_{\mathcal{L}} =
    \Stab_{\smash{\mathcal{L}\setminus \Lambda}} \circ Y_{\smash{\tilde\Sigma}} \circ \Gamma_{\mathcal{L}}
\end{equation}
for certain observable $\tilde\Sigma$, which we will construct explicitly.
The key idea underlying the proof lies in the interplay between the structure of the tensor network state, the form of the parent Hamiltonian, and the bulk-boundary correspondence.

Note that the left-hand side of \cref{eq:HaagCondSimplf} can be interpreted as follows:
\begin{equation}\label{Eq:Sketch1}
\vcenter{\hbox{\includegraphics[page=121]{haagduality_diagrams.pdf}}}
\end{equation}
Here, $\Stab_\Lambda$ is a simplified version of the projector onto the ground state space, introduced in \cref{Subsection:PH}. Since $\Omega^2 = \Omega^* = \Omega$, we can further simplify \cref{Eq:Sketch1}, obtaining:
\begin{equation}
\vcenter{\hbox{\includegraphics[page=122]{haagduality_diagrams.pdf}}}
\end{equation}
Now, one can apply the pulling-through equations in order to ``move'' the matrix product operator lying on $\mathcal{R}:=\Lambda\setminus\Sigma$ against the matrix product operator on top of the observable, obtaining:
\begin{equation}
\vcenter{\hbox{\includegraphics[page=123]{haagduality_diagrams.pdf}}}
\end{equation}
Due to the fact that the boundary operator vanishes when acting on the tensor network map, as in \cref{Lemma:GammaInvarBd}, the previous expression becomes:
\begin{equation}
\vcenter{\hbox{\includegraphics[page=124]{haagduality_diagrams.pdf}}}
\end{equation}
As mentioned above, we reinterpret this as the original tensor network state with a defect on the virtual edges of $\partial\Sigma\cap\partial\tilde{\Sigma}$,
\begin{equation}
\vcenter{\hbox{\includegraphics[page=125]{haagduality_diagrams.pdf}}}
\end{equation}
Now, by \cref{Claim:BvsW} for $R=\tilde\Sigma$, we can reverse simultaneously the colors and arrow orientations of the defect, obtaining 
\begin{equation}
\vcenter{\hbox{\includegraphics[page=126]{haagduality_diagrams.pdf}}}
\end{equation}
By following the same procedure one can reinterpret the defect $\mathcal{V}_{\Sigma,\Lambda}(X_\Sigma)$ as the analogous defect arising from the right-hand side of \cref{eq:HaagCondSimplf}, of the form $\mathcal{V}_{\smash{\tilde\Sigma,\mathcal{L}\setminus\tilde\Sigma}}(Y_{\smash{\tilde\Sigma}})$, from the action of another observable $Y_{\smash{\tilde\Sigma}}$ supported in $\tilde\Sigma$, which can be easily constructed.

\subsection{Proof of \texorpdfstring{\cref{thm:haagTN}}{Theorem \ref{thm:haagTN}}}
The goal is to prove an intermediate result: The action of an operator $X_\Sigma$ supported on $\Sigma$ followed by the projection onto the ground state space will result in the original tensor network state except for a \emph{virtual} matrix product operator supported on the edges $\partial\Sigma\cap\partial\Lambda$, which can be then reconstructed inside $\Lambda\setminus\Sigma$.
\\[5pt]
Let us first describe the construction of the regions $\tilde\Sigma$, $\mathcal{R}$ and $\tilde{\mathcal{R}}$. For that purpose, we first construct  regions $\Sigma\subseteq \Sigma_{+} \subset \Lambda$ and $\tilde\Sigma\subset\mathcal{L}\setminus\Lambda$ for which $\partial\Lambda\cap\partial\Sigma_+ = \partial\Lambda\cap \partial\tilde\Sigma$, i.e., there is a one-to-one correspondence, along the boundary of $\Lambda$, between the subsites of $\Sigma_+$ and those of $\tilde{\Sigma}$. We distinguish two cases.
\\[5pt]
\textbf{Case 1.} Assume in first place that the region $\Sigma$ does terminate (regarded as a cone-like subregion with endpoints at the dual lattice) at an odd length from the origin $\boldsymbol{a}$ of the cone-like region $\Gamma$, measured in the perpendicular direction to that direction in which it ends. Recall that $\Gamma$ is defined by employing steps of even length. In this case, we set $\Sigma_{+} := \Sigma$ and the region $\tilde{\Sigma}$ is defined as the minimal comb-like region in $\mathcal{L}\setminus\Lambda$ containing all sites that are adjacent to $\Sigma$, and possibly other sites (not adjacent to $\Lambda$) in order to guarantee condition (iv) in \cref{thm:haagTN}. This case is exemplified in the figure below.
\begin{equation*}
    \vcenter{\hbox{\includegraphics[page=127,scale=0.95]{haagduality_diagrams.pdf}}}
\end{equation*}
In this particular example, the region $\Sigma$ terminates at a site for which the (vertical) distance from the dual lattice from $\boldsymbol{a}$ is three, an odd integer.
\\[5pt]
In order to guarantee that $\partial \Sigma\cap\partial\Lambda = \partial\tilde\Sigma\cap\partial \Lambda$, note that this is trivially fulfilled in the interior of $[-\ell,\ell]^2$, so we only need to discuss what happens at $\partial\Lambda\cap\partial[-\ell,\ell]^2$. Since the steps in $\Gamma$ were chosen to be even integers, and due to the odd length assumption, $\partial\Lambda\cap\partial[-\ell,\ell]^2$ is exactly of the form illustrated in the zoomed region, up to changes of the relative positions of the sets represented with different colors in the figure. It is clear then, that also at  $\partial\Lambda\cap\partial[-\ell,\ell]^2$ there is a one to one corresponence between the boundaries of $\Sigma$ and $\tilde{\Sigma}$ along the boundary of $\Lambda$. This concludes that  $\partial \Sigma\cap\partial\Lambda = \partial\tilde\Sigma\cap\partial \Lambda$.
\\[5pt]
\textbf{Case 2.} Now assume that $\Sigma$ terminates (regarded as a cone-like subregion with endpoints at the dual lattice) at an even length from the origin $\boldsymbol{a}$ of the cone-like region $\Gamma$, measured in the perpendicular direction to that direction in which it ends. The following picture illustrates this situation.
\begin{equation*}
    \vcenter{\hbox{\includegraphics[page=165,scale=0.95]{haagduality_diagrams.pdf}}}
\end{equation*}
In this particular case, the (vertical) distance between the endpoint of the dual lattice path and $\boldsymbol{a}$ is two, an even integer. In order to guarantee that  $\partial \Sigma\cap\partial\Lambda \subseteq \partial\tilde\Sigma\cap\partial \Lambda$ we need to include in $\tilde{\Sigma}$ the sites in blue in the figure. But then $\partial\tilde\Sigma\cap\partial \Lambda\not  \subseteq  \partial \Sigma\cap\partial\Lambda$ as one can see by looking at the vertex $\boldsymbol{v}$ in the zoomed region. 
\\[5pt]
To fix this potential problem in this case, we define $\Sigma_+$ as the minimal extension of $\Sigma$ inside $\Lambda$ such that $\Sigma_+$ terminates at an odd length from the origin $\boldsymbol{a}$ of the cone-like region $\Gamma$. Since lengths of the lattices paths are chosen to be even integers, one can always fulfill this condition by incorporating additional column and/or row segments. Furthermore, we extend the observable $X_\Sigma$ trivially to the new region as $X_{\Sigma_+}:= X_\Sigma \otimes \mathrm{Id}_{\Sigma_+}$. Similarly, the region $\tilde{\Sigma}$ is defined as the minimal comb-like region in $\mathcal{L}\setminus\Lambda$ containing all sites that are adjacent to the new region $\Sigma_+$, as in the figure below.
\begin{equation*}
    \vcenter{\hbox{\includegraphics[page=166,scale=0.95]{haagduality_diagrams.pdf}}}
\end{equation*}
 Finally, in both Cases 1 and 2, we define the regions $\mathcal{R}\subset\Lambda\setminus\Sigma_+$ and $\tilde{\mathcal{R}}\subset (\mathcal{L}\setminus \Lambda)\setminus \tilde\Sigma$ as any comb-like regions such that $\Sigma_+ \cup \mathcal{R}$ and $\tilde{\Sigma} \cup \tilde{\mathcal{R}}$ are comb-like regions which can be expressed as unions of plaquettes. See \cref{eq:HaagProof-2d-1} for the corresponding example in the previous cases.
\\[5pt]
From now on, we assume without loss of generality that we are in Case 1 after possibly extending $\Sigma$, and therefore $\Sigma = \Sigma_+$.
\\[5pt]
We can now introduce the following quantities, which will be used later in the proof:
\begin{equation}
    p:= |\partial\Sigma\setminus\partial\Lambda|=|\partial\Sigma\cap\Interior{\Lambda}|,\quad q :=|\partial\Sigma\cap \partial \Lambda|,\quad \eta := |\partial\tilde\Sigma\setminus \partial\Lambda|-p \geq 0.
\end{equation}
In other words, $p$ stands for the number of edges of $\partial\Sigma$ which lie on the interior of the cone-like region $\Lambda$, $q$ stands for the number of edges of $\partial\Sigma$ which connect with the outside of $\Lambda$, and $\eta$ is exactly $|\partial\tilde\Sigma|- |\partial\Sigma|$ (note that, by the assumption that $\Sigma = \Sigma_+$,  $\partial \Sigma\cap\partial\Lambda = \partial\tilde\Sigma\cap\partial \Lambda$), which is non-negative by condition (iv) in the statement of \cref{thm:haagTN}.
\\[5pt]
The first step consists on figuring out how to understand the effect of the observable and the projection onto the ground state space only on $\partial\Sigma$, in order to not depend on the structure of the interior of $\Sigma$. For that purpose, we first \emph{rewrite} the expression in terms of $\partial\Sigma$ and $\Sigma^\circ$. As a consequence of \cref{Eq:DefGSStab}, we obtain that
\begin{equation}
\label{Eq:ExplicExprQXG}
 \Stab_{\Lambda}\circ X_\Sigma \circ \Gamma_{\mathcal{L}}
  = \Gamma_{\Lambda}\circ \tilde\kappa_{\partial\Lambda}^{-2} \circ (\Gamma_{\Lambda})^\dagger \circ X_\Sigma \circ \Gamma_{\mathcal{L}},
\end{equation}
and let us focus on simplifying the term $(\Gamma_{\Lambda})^\dagger\circ X_\Sigma\circ \Gamma_{\Lambda}$ on the right-hand side, for which we introduce the following definitions for the first step of the proof.
First, consider the linear map $\tilde\Gamma_{\Sigma}:\mathscr{H}_{\partial\Sigma}\to\mathscr{H}_\Sigma$ defined by
\[
    \tilde\Gamma_{\Sigma} := (\Phi_{\partial\Sigma}(\Omega) \circ \tilde\kappa_{\partial\Sigma})\circ (\mathrm{Id}_{\partial\Sigma}\otimes \myket{u_{\Interior{\Sigma}}}),
\]
where $\myket{u_{\Interior{\Sigma}}}\in \bigotimes_{e\in \Interior{\Sigma}} \mathscr{H}\otimes \mathscr{H}^*$ is any choice of unit vector and $\tilde\kappa_{\partial\Sigma}$ is any map as in \cref{Prop:FactKappaBd}. Second, let us define the linear map
\begin{equation}
    \label{eq:defGammaTilde}
    \tilde\Gamma_{\mathcal{L},\Sigma} 
    \coloneq
    (\tilde\Gamma_\Sigma\otimes \bigotimes_{\boldsymbol{x} \in \mathcal{L}\setminus\Sigma}^{\vphantom{a}} \Gamma_{\boldsymbol{x}})
    \circ (\mathrm{Id}_{\partial\mathcal{L}}\otimes 
    \bigotimes_{e\in \Interior{\mathcal{L}}\setminus\Interior{\Sigma} } \myket{\mathrm{Id}_e}).
\end{equation}
Pictorially, $\tilde\Gamma_{\mathcal{L},\Sigma}$ takes the form:
\begin{equation}\label{eq:HaagProof-3d-1}
\vcenter{\hbox{\includegraphics[page=128]{haagduality_diagrams.pdf}}}
\end{equation}
 As we have just commented, the following claim justifies the fact that the expressions in the statement can be evaluated in terms of this new tensor network.
\begin{stepHD}
    \label{Lemma:HD_Step1}
    There exists a unitary endomorphism $U_\Sigma\in\operatorname{End}\mathscr{H}_{\Sigma}$ such that
    \begin{equation}\label{Eq:GammaSigmaUSigma}
        \Gamma_{\Sigma} = U_\Sigma \circ \tilde\Gamma_{\Sigma},
    \end{equation}
    and hence $\Gamma_{\Lambda} = U_\Sigma \circ \tilde \Gamma_{\Lambda,\Sigma}$, and $\Gamma_{\mathcal{L}} = U_\Sigma \circ \tilde \Gamma_{\mathcal{L},\Sigma}$.
\end{stepHD}

\begin{proof}
    By virtue of \cref{Eq:DefRhoLambda,Thm:BulkBoundary},
    \begin{align*}
        (\Gamma_{\Sigma})^\dagger \circ \Gamma_{ \Sigma }
        \eqqcolon \rho_{\partial \Sigma}
        &=
        \tilde\kappa_{\partial\Sigma}\circ \Phi_{\partial\Sigma}(\Omega) \circ \tilde\kappa_{\partial\Sigma}
        \\
        &=
        ( \Phi_{\partial\Sigma}(\Omega) \circ \tilde\kappa_{\partial\Sigma} \otimes \myket{u_{\Interior{\Sigma}}})^\dagger \circ (\Phi_{\partial\Sigma}(\Omega)\circ \tilde\kappa_{\partial\Sigma} \otimes \myket{u_{\Interior{\Sigma}}})
        \\&=
        (\tilde\Gamma_{\Sigma})^\dagger\circ \tilde\Gamma_{\Sigma}.
    \end{align*}
   Then, \cref{Eq:GammaSigmaUSigma} follows as a consequence of the polar decomposition theorem.
   The remaining part of the statement follows directly by concatenating single-site tensor maps in the complementary region.
\end{proof}

\begin{stepHD}
    \label{Lemma:HD_Step2}
    Let $\Sigma \subset \Lambda \subset \mathcal{L}$ be as above. Then, there exists a linear map
    \[
        \mathcal{V}_{\Sigma,\Lambda}:
            \operatorname{End}\mathscr{H}_{\Sigma}
            \to
            \operatorname{End}\mathscr{H}_{{\partial{\Sigma}} \cap \partial{\Lambda}}
                =
            \bigotimes_{e\in \partial\Sigma\cap\partial\Lambda} \operatorname{End}\mathscr{H},
    \]
    assigning to each observable $X_\Sigma \in\operatorname{End}\mathscr{H}_\Sigma$ a matrix product operator supported on the vir\-tual spaces satisfying the following property
    \begin{equation}\label{eq:partialHaag}
        \Stab_{ \Lambda }\circ X_{ \Sigma } \circ \Gamma_{\mathcal{L}}
        = 
        (\bigotimes_{\boldsymbol{x}\in\mathcal{L}}\Gamma_{\boldsymbol{x}} \circ \mathcal{V}_{\Sigma,\Lambda}(X_\Sigma))
            \bigotimes_{e \in \Interior{\mathcal{L}}} \myket{ \mathrm{Id}_e }
    \end{equation}
    for all observables $X_\Sigma$ supported on $\Sigma$.
\end{stepHD}

\begin{proof}
We outline the proof of this claim using a simple example as a guiding aid throughout the process, which can be readily generalized to any comb-like region.
Note that, due to the definition of $\mathcal{R}$, \cref{Thm:ParentHamiltonian,Lem:int},
\[
\Stab_{\Lambda}\circ X_\Sigma\circ \Gamma_{\mathcal{L}} = \Stab_{\Sigma\cup \mathcal{R}}\circ X_\Sigma\circ \Gamma_{\mathcal{L}}.
\]
The following picture illustrates the right-hand side of this equation:
\begin{equation*}
\vcenter{\hbox{\includegraphics[page=129]{haagduality_diagrams.pdf}}}
\end{equation*}
More specifically, here one can distinguish the gray region on the plane corresponding to $\Lambda$, and the yellow subregion, associated to $\Sigma$. 
For simplicity, that $\mathcal{R}$ consists of the adjacent row as described above.
 The observables have been depicted by the orange nodes, and have been assumed to have a tensor-product expression, since all expressions are linear on the observables. Three possible families of observables can be distinguished:
\begin{itemize}
\setlength{\itemsep}{3pt}
\item[(i)] observables $X_k$ supported on a subsite $k\in \partial\Sigma\cap \Interior{\Lambda}$,
\item[(ii)] observables $X_\ell$ supported on a subsite $\ell\in \partial\Sigma\cap \partial\Lambda$,
\item[(iii)] observables $X_m$ supported on a subsite $m\in \Interior{\Sigma}$.
\end{itemize}
As a consequence of \cref{Lemma:HD_Step1,eq:defGammaTilde}, 
\begin{align*}
    Q_{\Sigma\cup \mathcal{R}}\circ X_\Sigma {} & {} \circ  \Gamma_{\mathcal{L}}
    =
    \Gamma_{\Sigma\cup \mathcal{R}}\circ \tilde\kappa_{\partial(\Sigma\cup \mathcal{R})}^{-2}
    \circ
    (\Gamma_{\Sigma\cup \mathcal{R}})^{\dagger} \circ X_\Sigma \circ \Gamma_{\mathcal{L}} 
    \\ &=
    \Gamma_{\Sigma\cup \mathcal{R}}\circ \tilde\kappa_{\partial(\Sigma\cup \mathcal{R})}^{-2}
    \circ (\tilde\Gamma_{\Sigma\cup \mathcal{R},\Sigma})^{\dagger}
    \circ \operatorname{Ad}U_\Sigma(X_\Sigma)
    \circ \tilde\Gamma_{\mathcal{L},\Sigma}
    \\ &=
    (\Gamma_{\Sigma\cup \mathcal{R}}\circ \tilde\kappa_{\partial(\Sigma\cup \mathcal{R})}^{-1})
    \circ ( \tilde\Gamma_{\Sigma\cup \mathcal{R},\Sigma}
    \circ \tilde\kappa_{\partial(\Sigma\cup \mathcal{R})}^{-1})^\dagger
    \circ \operatorname{Ad}U_\Sigma (X_\Sigma)
    \circ \tilde\Gamma_{\mathcal{L},\Sigma}.
\end{align*}
Here, the term $\Gamma_{\Sigma\cup \mathcal{R}} \circ \tilde\kappa^{-1}_{\partial(\Sigma\cup \mathcal{R})}$ corresponds to the tensor network map on the region $\Sigma\cup \mathcal{R}$ without weights along $\partial(\Sigma\cup \mathcal{R})$, nor the intrinsic weights $\Psi(\hat{\xi}_R^{-1})$ corresponding to the plaquettes on $L \setminus (\Sigma\cup \mathcal{R})$. In turn, the term $\tilde\Gamma_{\Sigma\cup \mathcal{R},\Sigma}\circ \tilde\kappa_{\partial(\Sigma\cup \mathcal{R})}^{-1}$ corresponds to the same tensor network map as before but after ``renormalizing'' the region as described in \cref{Lemma:HD_Step1}. The term $\operatorname{Ad} U_\Sigma(X_\Sigma)$ is the new observable, which we denote again $X_k\otimes X_\ell\otimes \cdots$ and without loss of generality assume that it is a tensor product. Finally, $\tilde\Gamma_{\Sigma\cup \mathcal{R},\Sigma}$ corresponds to the original tensor network state,  after ``renormalizing'' the region $\Sigma$.
In summary, the previous picture takes the following form:
\begin{equation}\label{eq:HaagProof3d-2}
\vcenter{\hbox{\includegraphics[page=130]{haagduality_diagrams.pdf}}}
\!\!\!\!\!\!
\end{equation}
Above the plane, the tensor network expression corresponds to the adjoint of the renormalized tensor network in $\Sigma$ as defined in \cref{eq:defGammaTilde}, followed by the full projected entangled pair state map in $\Sigma\cup \mathcal{R}$. We note that, for the sake of clarity, the expectation values of the observables found in $\Interior{\Sigma}$ have not been depicted and recall that $\mathcal{R}$ is a row for simplicity.
Now, for the lower tensor network state map $\tilde\Gamma_{\mathcal{L},\Sigma}$ consider the distribution of the weights presented in \cref{Rem:DistribW} along $\partial(\Sigma\cup \mathcal{R})$, and use \cref{Eqs:xiExchGraphBlack,Eqs:xiExchGraphBlack2,Eqs:xiExchGraphWhite,Eqs:xiExchGraphWhite2} along $\partial\Sigma\cap\partial \mathcal{R}$ in such a way that we obtain the following equality:
\begin{align*}
\vcenter{\hbox{\includegraphics[page=131]{haagduality_diagrams.pdf}}}
 \ = \ \qquad\qquad
    \\
    \qquad\qquad \ = \  
\vcenter{\hbox{\includegraphics[page=132]{haagduality_diagrams.pdf}}}
\end{align*}
By virtue of the previous observations, \cref{eq:HaagProof3d-2} now becomes:
\begin{equation}\label{eq:HaagProof3d-3}
\vcenter{\hbox{\includegraphics[page=133]{haagduality_diagrams.pdf}}}
\end{equation}
Let us simplify this expression on $\mathcal{R}$ as a consequence of the bulk-boundary correspondence. First, by virtue of \cref{Item:Omega-And-PT} of \cref{Thm:PT}, one can simplify the products of on-site operators $\Phi_{\partial{\boldsymbol{x}}}(\Omega)$ for all $\boldsymbol{x}\in \mathcal{R}$ in \cref{eq:HaagProof3d-3}, obtaining the following equivalent expression:
\begin{equation}\label{eq:HaagProof3d-4}
\vcenter{\hbox{\includegraphics[page=134]{haagduality_diagrams.pdf}}}
\end{equation}
and, since $\mathcal{R}$ is constructed as a comb-like region, by \cref{Eq:IdentityBlackVsWhiteFull}, this equals to:
\begin{equation}\label{eq:HaagProof3d-5}
\vcenter{\hbox{\includegraphics[page=135]{haagduality_diagrams.pdf}}}
\end{equation}
Now, the operator $\Phi_{\partial\mathcal{R}}(\Omega)$ can be pulled through the operator $\Phi_{\partial\Sigma}(\Omega)^\dagger$:
\begin{equation}\label{eq:HaagProof3d-6}
\vcenter{\hbox{\includegraphics[page=136]{haagduality_diagrams.pdf}}}
\end{equation}
This is a consequence of an extension of the pulling-through identity presented in \cref{Eq:PT_4sites_vs2} (see e.g.~the proof of \cref{Lemma:GammaInvarBd}).
The operator $\Phi_{\partial(\Sigma\cup\mathcal{R})}(\Omega)$ eventually vanishes as on top we can find the tensor network state, as justified by \cref{Lemma:GammaInvarBd}. Thus, we obtain the following equivalent expression:
\begin{equation}\label{eq:HaagProof3d-7}
\vcenter{\hbox{\includegraphics[page=137]{haagduality_diagrams.pdf}}}
\end{equation}
This expression corresponds to \cref{eq:partialHaag}.
Here, let us note that all operators $X_k$ supported on the subsites $k\in \partial\Sigma\cap\Interior{\Lambda}$ are enclosed in a trace (with weights $b(\omega)$, which we have omitted for clarity), while all operators $X_\ell$ supported on the subsites $\ell\in \partial\Sigma\cap \partial\Lambda$ are connected to the boundary, as we continue analyzing below.
\\[5pt]
We have obtained that $\Stab_\Lambda \circ X_\Sigma\circ \Gamma_{\mathcal{L}}$ is equal to
\begin{equation}\label{eq:HaagProof2d-8}
\vcenter{\hbox{\includegraphics[page=138]{haagduality_diagrams.pdf}}}
\end{equation}
where the matrix product operator is denoted $\mathcal{V}_{\Sigma,\Lambda}(X_\Sigma)$, and is supported on the virtual indices $\partial{\Sigma} \cap \partial{\Lambda}$. Note that the operators in the upper part $\partial\Sigma \setminus \partial\Lambda$ are not linked to the virtual indices of the original tensor network state.
\\[5pt]
More concretely, the matrix product operator $\mathcal{V}_{\Sigma,\Lambda}(X_\Sigma)$ takes the form
\begin{equation}\label{eq:HaagProofVirtOp-1}
\vcenter{\hbox{\includegraphics[page=139]{haagduality_diagrams.pdf}}}
\end{equation}
where for convenience we will read the composition in the virtual indexes from the upper right corner and counterclockwise.
Let us elaborate more on the structure of this matrix product operator.
On the one hand, if $X_k$ stands for a local observable supported on a subsite $k\in \partial\Sigma \cap \Interior{\Lambda}$, then $\mathcal{V}_{\Sigma,\Lambda}(X_\Sigma)$ is locally composed, as described above, by rank-two tensors of the form
\begin{equation}
\vcenter{\hbox{\includegraphics[page=140]{haagduality_diagrams.pdf}}}
    \ \coloneq \
\vcenter{\hbox{\includegraphics[page=141]{haagduality_diagrams.pdf}}}
    \quad \text{ or }  \quad 
\vcenter{\hbox{\includegraphics[page=142]{haagduality_diagrams.pdf}}}
    \ \coloneq \
\vcenter{\hbox{\includegraphics[page=143]{haagduality_diagrams.pdf}}}
    \ ,
\end{equation}
depending on whether $k$ corresponds to the top/right subsite of some site or a bottom/left subsite of some site, respectively. In the cases in which the matrix product operator involves $\Phi(\xi_{L/R})$, it is defined as
\begin{equation}
\vcenter{\hbox{\includegraphics[page=144]{haagduality_diagrams.pdf}}}
    \ \coloneq \
\vcenter{\hbox{\includegraphics[page=145]{haagduality_diagrams.pdf}}}
    \quad \text{ or }  \quad 
\vcenter{\hbox{\includegraphics[page=146]{haagduality_diagrams.pdf}}}
    \ \coloneq \
\vcenter{\hbox{\includegraphics[page=147]{haagduality_diagrams.pdf}}}
\ .
\end{equation}
On the other hand, if $X_\ell$ stands for the local observable supported on $\ell\in \partial\Sigma \cap \partial\Lambda $, then $\mathcal{V}_{\Sigma,\Lambda}(X_\Sigma)$ is locally composed by rank-four tensors of the form
\begin{equation}
\vcenter{\hbox{\includegraphics[page=148]{haagduality_diagrams.pdf}}}
    \ \coloneq \ 
\vcenter{\hbox{\includegraphics[page=149]{haagduality_diagrams.pdf}}}
    \quad \text{ or } \quad  
\vcenter{\hbox{\includegraphics[page=150]{haagduality_diagrams.pdf}}}
    \ \coloneq \ 
\vcenter{\hbox{\includegraphics[page=151]{haagduality_diagrams.pdf}}}
    \ ,
\end{equation}
depending on whether $\ell$ corresponds to a top/right subsite of some site or a bottom/left subsite of some site, respectively. This concludes the proof of this claim.
\end{proof}
The next step focus on constructing an operator $Y_{\smash{\tilde\Sigma}}$ on the region $\tilde\Sigma$.

\begin{stepHD}
    \label{Lemma:HD_Step4}
    Let $\Sigma \subset \Lambda \subset \mathcal{L}$ be as above and let $X_\Sigma$ be an observable supported on $\Sigma$. Then, there exists an observable $Y_{\smash{\tilde\Sigma}}$ supported on $\tilde\Sigma$ such that
    \begin{equation}
        \mathcal{V}_{\Sigma,\Lambda}(X_{\Sigma})=
        \mathcal{V}_{\smash{\tilde\Sigma},\mathcal{L}\setminus\Lambda}(Y_{\smash{\tilde\Sigma}}),
    \end{equation}
    where $\mathcal{V}_{\Sigma,\Lambda}$ and $\mathcal{V}_{\smash{\tilde\Sigma},\mathcal{L}\setminus\Lambda}$ are the maps introduced in \cref{Lemma:HD_Step2}.
\end{stepHD}

A first challenge arises by noticing that the observable in $\Sigma$ enters the virtual plane in the previous step with a structure determined by whether the tensors at the boundary are white or black. On the opposite side, these tensors have complementary colors. However, in \cref{Claim:BvsW} with $R=\tilde\Sigma$, we checked that this issue can be circumvented, as the tensors are simultaneously interchangeable.
\\[5pt]
More specifically, the following is a diagrammatic representation of the renormalized tensor network map $\tilde \Gamma_{ \mathcal{L}, \smash{\tilde{\Sigma}} }$:
\begin{equation}\label{Eq:RenTN-b}
\vcenter{\hbox{\includegraphics[page=152]{haagduality_diagrams.pdf}}}
\end{equation}
Note that the black and white tensors have been exchanged along $\partial\tilde\Sigma$ and the order of composition along the virtual indices has been reversed, as commented above.

\begin{proof}
Let us consider the observable $X_\Sigma$ after renormalization from \cref{Lemma:HD_Step1}, i.e.~we simply denote $X_\Sigma$ instead of $U_\Sigma \circ X_\Sigma \circ (U_\Sigma)^\dagger$ for simplicity, is of the form
\begin{equation}
X_\Sigma = X_{\partial\Sigma\setminus\partial\smash{\tilde\Sigma}} \otimes X_{\smash{\partial\tilde\Sigma}\cap\partial\Sigma} \otimes X_{\Interior{\Sigma}},
\end{equation}
where all the terms are supported on the corresponding subindices, and moreover
\[
X_{\partial\Sigma\setminus\partial\smash{\tilde\Sigma}} = X_{1}\otimes\cdots\otimes X_{p}
,\quad\text{ and }
\quad
X_{\smash{\partial\Sigma}\cap\partial\tilde\Sigma} = X_{p+1}\otimes \cdots \otimes X_{p+q},
\]
with $p=|\partial\Sigma\setminus\partial\tilde\Sigma|$ and $q = |\partial\tilde\Sigma\cap\partial\Sigma|$. This can be assumed without loss of generality as any operator can be expanded as a sum of tensor-product operators, and the identity above is clearly linear on the observable. Here we enumerated the subsites of $\partial\Sigma$ in counterclockwise order. In this step, we construct an observable $Y_{\smash{\tilde\Sigma}}$ of the form
\begin{equation}
    Y_{\smash{\tilde\Sigma}}
        =
        Y_{\partial\Sigma\cap\partial\smash{\tilde\Sigma}}
        \otimes Y_{ \partial\smash{\tilde\Sigma}\setminus\partial\Sigma }
        \otimes Y_{\smash{\Interior{\tilde\Sigma}}}.
\end{equation}
for which the first part can be expanded as
\[
    Y_{\partial\Sigma\cap\partial\smash{\tilde\Sigma}} = Y_{\eta+p+1}\otimes \cdots \otimes  Y_{\eta+p+q},
\]
while the second part can be written as
\[
    Y_{\partial\smash{\tilde\Sigma} \setminus \partial\Sigma}
        =
        Y_{(1,\ldots,\eta+1)}\otimes Y_{\eta+2}  \otimes \cdots \otimes Y_{\eta+p},
\]
where we recall that $\eta := |\partial\tilde\Sigma\setminus \partial\Lambda|-p$ and note that the term $Y_{(1,\ldots,\eta+1)}$ is not necessarily a tensor product operator.
Following the same notation, we let $\hat{Y}_{(1,\ldots,\eta+1)}$ stand for the matrix product operator arising from the same procedure from $Y_{(1,\ldots,\eta+1)}$.
Here, we enumerated the subsites of $\partial\Sigma$ in clockwise order.
Specifically, the tensor network and the virtual matrix product operator $\mathcal{V}_{\smash{\tilde\Sigma},\mathcal{L}\setminus\Lambda}(Y_{\smash{\tilde\Sigma}})$ are as follows:
\begin{equation}
\vcenter{\hbox{\includegraphics[page=153]{haagduality_diagrams.pdf}}}
\end{equation}
Let us elaborate more on the choices and the structure of the operators.
Recall that our goal to find $Y_j$'s such that
\begin{align*}
\hat{X}_1 = \hat{Y}_{(1,\ldots, \eta+1)}, \quad   \hat{X}_{k} = \hat{Y}_{\eta+k}, \quad \text{for all } k=2,\ldots,p+q.
\end{align*}
Based on the decomposition outlined above, we consider four distinct cases.

\begin{caseHD}
    Let us first construct $Y_{\smash{(1,\ldots,\eta+1)}}$.
\end{caseHD}

\noindent{}For the specification of $\tilde\Sigma$, we required the part $\partial\tilde\Sigma\setminus\partial\Sigma$ of its boundary to contain at least as many edges as $\partial\Sigma\setminus\partial\tilde\Sigma$.
Therefore, in order to achieve a reconstruction of the original observable the idea here is to recreate, for example, the first observable $X_1$ by comultiplying it and adapting each cofactor to the boundary of the new region.
We therefore define $Y_{(1,\ldots,\eta+1)}$ in such a way that the corresponding matrix product operator $\hat{Y}_{(1,\ldots,\eta+1)}$ that arises coincides with $\hat{X}_1$.
Now, in order to construct the observable adapted to the (corresponding colors of the) new boundary, we first have to prove that one can assume without loss of generality that
\[ X_1 = \Phi(x) \text{ for some } x\in A. \]
For that purpose, let us note that by Artin-Wedderburn's theorem,
\[ A\cong \bigoplus_{\alpha=1}^{s} \mathrm{M}_{n_\alpha}(\mathbb{C}),\quad \mathscr{H}\cong \bigoplus_{\alpha=1}^{s} \mathbb{C}^{n_\alpha m_\alpha},\quad  \operatorname{Im}\Phi \cong \bigoplus_{\alpha=1}^{s} \mathrm{M}_{n_\alpha}(\mathbb{C}) \otimes \mathrm{Id}_{m_\alpha}.
\]
Moreover, all unitaries of the form 
\[
    \bigoplus_{\alpha=1}^{s} \mathrm{Id}_{n_\alpha} \otimes U_\alpha \in\bigoplus_{\alpha=1}^{s}  \mathrm{Id}_{n_\alpha}  \otimes \mathrm{M}_{m_\alpha}(\mathbb{C}),
\]
commute with all elements of $\operatorname{Im}\Phi$,
and recall that all tensors involved satisfy
\[
b(\omega),\Phi(\xi_{L}),\Phi(\xi_{R})\in \operatorname{Im}\Phi,\quad
T_{\Phi,\Psi} \in \operatorname{Im}\Phi\otimes\operatorname{Im}\Psi.
\]
This immediately implies that 
\begin{equation}
\!\!\!\!\!\!\!\!\!\!
    \hphantom{\bigoplus_\alpha(Id_{n_\alpha}\otimes U_\alpha)}
\vcenter{\hbox{\includegraphics[page=154]{haagduality_diagrams.pdf}}}
    = \
\vcenter{\hbox{\includegraphics[page=155]{haagduality_diagrams.pdf}}}
\end{equation}
for any choice of unitaries $U_\alpha$, $\alpha\in \operatorname{Irr}A$, and therefore one can replace
\[ X_1=\bigoplus_{\alpha=1}^{s} X_{1,\alpha}  \quad \text{ with } \quad \bigoplus_{\alpha=1}^{s} \int \operatorname{Ad}(\mathrm{Id}\otimes U_\alpha) (X_{1,\alpha}) \, dU_\alpha.  \]
Note that the latter term belongs to $\operatorname{Im}\Phi$, by virtue of the well-known identity
\[
\int \operatorname{Ad} U (X) \, dU = \frac{\mathrm{Tr}(X)}{\operatorname{dim}\mathscr{H}} \,\mathrm{Id}.
\]
In other words, we have projected out the part of $X_1$ in $(\operatorname{Im}\Phi)'$. More concretely, let $X_1 = \Phi(x)$ for an element $x\in A$. 
Now, note that the following identities
\begin{equation}
\vcenter{\hbox{\includegraphics[page=156]{haagduality_diagrams.pdf}}}
    \ = \
\vcenter{\hbox{\includegraphics[page=157]{haagduality_diagrams.pdf}}}
    \text{ and }
\vcenter{\hbox{\includegraphics[page=158]{haagduality_diagrams.pdf}}}
    = 
\vcenter{\hbox{\includegraphics[page=159]{haagduality_diagrams.pdf}}}
\end{equation}
hold for all $x\in A$, by since $\xi_R = J(\xi_L)$ as introduced in \cref{Item:XiFactor} of \cref{Thm:PT}, and $J$ is anti-multiplicative by \cref{Prop:J}.
Then, for all $j=1,\ldots, \eta+1$, let
\begin{equation*}
    Y_j \coloneq \begin{cases}
        \Phi(x_{(j)}) & \text{if $X_1$ and $Y_{j}$ correspond to subsites of the same color,}\\
        (\Phi\circ J)(x_{(j)}) & \text{if ${X}_1$ and ${Y}_{j}$ correspond to subsites of different colors.}\\
    \end{cases}
\end{equation*}
Indeed, as a consequence of the following calculation:
\begin{align*}
    \eval{ \omega }{  x_{(1)} }\cdots \eval{ \omega }{ x_{(\eta+1)} }
= \eval{ \omega^{\eta+1} }{ x } = \eval{\omega}{x},
\end{align*}
where the first equality is due to \cref{Rem:DualCWHA}, and the second equality follows since $\omega$ is an idempotent of $A^*$, see \cref{Thm:PT}, this implies
\begin{equation}
\vcenter{\hbox{\includegraphics[page=160]{haagduality_diagrams.pdf}}}
    =\!\!\!
\vcenter{\hbox{\includegraphics[page=161]{haagduality_diagrams.pdf}}}
    = \hat{Y}_{(1,\ldots,\eta+1)},
\end{equation}
since $\xi_L\in A^L$ and hence it satisfies $\Delta^{(\eta)}(\xi_L) = (\xi_L\otimes 1\otimes \cdots \otimes 1)\Delta^{(\eta)}(1)$, see \cref{Rem:CharALAR,Thm:PT}. Therefore, the following holds by construction for our choices:
\begin{equation}
    \hat{Y}_{(1,\ldots,\eta+1)} = \hat{X}_{1}.
\end{equation}
Let us finally note that $Y_{\smash{(1,\ldots,\eta+1)}}$ is self-adjoint whenever  $X_1=\Phi(x)$ is self-adjoint, since $\Phi$, $J$ and $\Delta$ are $*$-preserving linear maps.

\begin{caseHD}
    Let us now construct the observables $Y_{\eta+2},\ldots,Y_{\eta+p}$ on $\partial\tilde\Sigma\setminus\partial\Sigma$.
\end{caseHD}

For this purpose, we can also assume that $X_{j} = \Phi(x)$ for a certain $x\in A$, as previously discussed, and then simply define, for all $j = 2,\ldots, p$:
\begin{equation}\label{Eq:HaagRecYetaj}
    Y_{\eta+j} \coloneq \begin{cases}
        \Phi(x)  & \text{if ${X}_{j}$ and ${Y}_{\eta+j}$ correspond to different colors,}\\
        (\Phi\circ J)(x) &\text{if ${X}_{j}$ and ${Y}_{\eta+j}$ correspond to the same colors,}
    \end{cases}\!\!\!\!\!\!
\end{equation}
when the colors (i.e.~black or white) are taken into consideration in the context of the tensor networks renormalized on $\Sigma$ and $\tilde\Sigma$ in \cref{eq:HaagProof-3d-1,Eq:RenTN-b}, i.e.~note that by \cref{Claim:BvsW} we have reversed the color of the white and black indices.
\\[5pt]
Let us remark that, if $X_j$ is a self-adjoint operator then $Y_{\eta+j}$ it is also self-adjoint. Indeed, in the first case of \cref{Eq:HaagRecYetaj} it is exactly the same operator, but in the second case one simply makes use of the fact that $J$ is $*$-preserving and linear by \cref{Prop:J} and $\Phi$ is a $*$-homomorphism.

\begin{caseHD}
Let us now construct the observables $Y_{\eta+p+1},\ldots, Y_{\eta+p+q}$ corresponding to the subsites in $ \partial\tilde\Sigma\cap\partial\Sigma$.
\end{caseHD}

Since these are in one-to-one correspondence and the colors of the tensors in both pictures coincide by virtue of \cref{Claim:BvsW}, we can simply choose, for all $j=1,\ldots,q$:
\begin{equation}
Y_{\eta+p+j}\coloneq X_{p+j} \text{, \quad and therefore \quad }\hat{Y}_{\eta+p+j} = \hat{X}_{p+j}.
\end{equation}

\begin{caseHD}
Finally, for the observables in the bulk of the region, simply let
\begin{equation}
    Y_{\smash{\Interior{\tilde\Sigma}}} \coloneq \mybra{ u_{\Interior{\Sigma}} } X_{\Interior{\Sigma}} \myket{ u_{\Interior{\Sigma}} } \, \mathrm{Id},
\end{equation}
and recall that $\myket{u_{\Interior{\Sigma}}}$ was chosen to be a unit vector.
\end{caseHD}

In summary, we have constructed the same virtual operator in both pictures:
\begin{equation}
    \mathcal{V}_{\smash{\tilde\Sigma,\mathcal{L}\setminus\Lambda}}(Y_{ \smash{\tilde\Sigma} })
    =
    \mathcal{V}_{\Sigma,\Lambda}(\smash{X_{\smash{\tilde\Sigma}}}),
\end{equation}
as we wanted to conclude. Note that each $X_j$ is mapped to a self-adjoint element above if $X_j$ is itself self-adjoint.
\end{proof}
This concludes the proof of \cref{thm:haagTN}.

\begin{remark}\label{Rem:6ell}
 The following picture illustrates the construction of $\tilde\Sigma$ in the case in which $\Lambda$ corresponds to a non-convex cone, as required in the construction of $\tilde S_\ell^{(2)}$ in \cref{Eq:TildeSell}.
\[
\vcenter{\hbox{\includegraphics[page=162]{haagduality_diagrams.pdf}}}
\]
To justify the bound \(6\ell\), we proceed with the following reasoning. In the worst-case scenario, in which the cone is non-convex with the maximal opening angle allowed by the assumptions, the correlations that must be recovered lie along the intersection \(\partial \Sigma \cap \Interior{\Lambda}\), which can span up to \(8\ell\) sites, corresponding to the perimeter of the square \([-\ell, \ell]^2\). This estimate is subject to a finite-size correction, independent of $\ell$, as e.g.\ corner sites contribute to two boundary segments while regions intersected by the complement \(\Gamma^c\) are excluded. Consider the following exemplification:
\[
\vcenter{\hbox{\includegraphics[page=163]{haagduality_diagrams.pdf}}}
\]
The observables along $\partial\Sigma\cap\Interior{\Lambda}$ can then be reconstructed in a narrow strip outside the cone, with length approximately \(4\ell\), up to corrections independent of \(\ell\). To account for these corrections, the support needed is at most \(\ell+4\ell=5\ell\) sites from the center of the square, and hence it suffices to consider  \(\mathcal{L}=[ -6\ell, 6\ell ]^2\).
\end{remark}

\begin{acknowledgements}
    Y.O. acknowledges financial support from JSPS KAKENHI (Grant Numbers 19K03534 and 22H01127) and JST CREST (Grant Number JPMJCR19T2).
    D.P-G.~and A.R-d-A.~acknowledge funding from the Spanish Ministry of Science and Innovation MCIN/AEI/10.13039/501100011033 (CEX2023-001347-S, PID2020-113523GB-I00, PID2023-146758NB-I00), Universidad Complutense de Madrid (FEI-EU-22-06), Comunidad Aut\'onoma de Madrid (TEC-2024/COM-84-QUITEMAD-CM), and the Ministry for Digital Transformation and of Civil Service of the Spanish Government through the QUANTUM ENIA project call -- Quantum Spain project, and by the European Union through the Recovery, Transformation and Resilience Plan -- NextGenerationEU within the framework of the Digital Spain 2026 Agenda.
    D.P-G.~also acknowledges support from Perimeter Institute for Theoretical Physics. Research at Perimeter Institute is supported by the Government of Canada through the Department of Innovation, Science, and Economic Development, and by the Province of Ontario through the Ministry of Colleges and Universities.
    A.R-d-A.~also acknowledges financial support from the Deutsche Forschungsgemeinschaft (DFG, German Research Foundation) -- TRR~352 with Project-ID 470903074.
\end{acknowledgements}

\begin{conflictofinterest}
On behalf of all authors, the corresponding author states that there is no conflict of interest.
\end{conflictofinterest}

\begin{dataavailability}
Data sharing is not applicable to this article as no new data were created or analyzed in this study.
\end{dataavailability}

\bibliographystyle{abbrv}
\bibliography{haagduality.bib}

\newcommand{\gR}{g_R}
\newcommand{\gL}{g_L}
\newcommand{\dualgR}{\hat{g}_R}
\newcommand{\dualgL}{\hat{g}_L}

\appendix

\section{Cone deformations and Haag duality}
\label{Appendix:Deformations}

\begin{lemma}\label{ea}
Let $\pi$ be a representation of $\caA$ on a Hilbert space $\caH$.
Let $\Lambda$ be a cone and $S$ a finite subset of $\Lambda$.
Suppose that 
$
\pi(\caA_{\Lambda\setminus S})''=\pi\lmk\caA_{(\Lambda\setminus S)^c}\rmk'
$.
Then we have $\pi(\caA_{\Lambda})''=\pi\lmk\caA_{\Lambda^c}\rmk'$.
\end{lemma}

\begin{proof}
Let $|S|$ be the number of elements in $S$ and set  $n:=d^{|S|}$.
Because $\caA_S$ is a size $n$-matrix algebra,
there is a Hilbert space $\mathscr{K}$ and a representation $\pi_1$ of $\caA_{S^c}$
on $\mathscr{K}$
and a unitary $U: \caH\to\bbC^n\otimes \mathscr{K}$
such that
\begin{align}
U\pi(A\otimes B)U^*=A\otimes \pi_1(B),\quad
\text{ for all }
A\in \caA_S,\quad B\in \caA_{S^c}.
\end{align}
See Proposition~1.8  IV of \cite{takesaki_theory_2001}.
With this, we have
\begin{align*}
&U\pi(\caA_{\Lambda})U^*=\caA_S\otimes \pi_1(\caA_{\Lambda\setminus S}),\\
&U\pi(\caA_{\Lambda^c}) U^*=\bbC \mathds{1}\otimes \pi_1\lmk \caA_{\Lambda^c}\rmk,\\
&U\pi\lmk \caA_{\Lambda\setminus S}\rmk U^*
=\bbC\mathds{1}\otimes \pi_1\lmk\caA_{\Lambda\setminus S}\rmk,\\
&U\pi\lmk \caA_{\lmk\Lambda\setminus S\rmk^c}\rmk U^*
=\caA_S\otimes\pi_1\lmk \caA_{\Lambda^c}\rmk.
\end{align*}
From the condition $\pi(\caA_{\Lambda\setminus S})''=\pi\lmk\caA_{(\Lambda\setminus S)^c}\rmk'$,
we get
\begin{align*}
&\bbC\mathds{1}\otimes \pi_1(\caA_{\Lambda\setminus S})''
=U\pi\lmk \caA_{\Lambda\setminus S}\rmk'' U^*\\
&=U \pi\lmk\caA_{(\Lambda\setminus S)^c}\rmk' U^*
=\lmk U \pi\lmk\caA_{(\Lambda\setminus S)^c}\rmk U^*\rmk'\\
&=\lmk \caA_S\otimes\pi_1\lmk \caA_{\Lambda^c}\rmk\rmk'
=\bbC\mathds{1}\otimes \pi_1\lmk\caA_{\Lambda^c}\rmk',
\end{align*}
which gives $\pi_1\lmk\caA_{\Lambda\setminus S}\rmk''= \pi_1\lmk\caA_{\Lambda^c}\rmk'$.
Here we used Theorem~5.9 of IV \cite{takesaki_theory_2001}. 
Substituting this, we have
\begin{align*}
U\pi(\caA_{\Lambda})''U^*
&=\caA_S\otimes \pi_1(\caA_{\Lambda\setminus S})''
=\caA_S\otimes \pi_1\lmk\caA_{\Lambda^c}\rmk'\\
&=\lmk \bbC\mathds{1}\otimes \pi_1(\caA_{\Lambda^c})\rmk'
=\lmk U\pi(\caA_{\Lambda^c}) U^*\rmk '
=U \pi\lmk \caA_{\Lambda^c}\rmk' U^*.
\end{align*}
Hence we have $\pi(\caA_{\Lambda})''=\pi\lmk \caA_{\Lambda^c}\rmk'$.
\end{proof}

\begin{lemma}
    \label{one2}
For every cone $\Lambda\subset \bbR^2$ and every $\varepsilon>0$ there exists a cone-like region $\Gamma_{\varepsilon,\Lambda}\subset\mathbb{R}^2$ as in \cref{Def:comblikelkmk}, such that
$\Lambda\cap \Gamma_{\varepsilon,\ld}^c\cap\bbZ^2$ and $\Gamma_{\varepsilon,\ld}\cap \Lambda_{\varepsilon}^c\cap\bbZ^2$ are finite.
\end{lemma}

\begin{proof}
Let $\ld:=\Lambda_{\bm a, \theta,\varphi}$ and $\varepsilon>0$.
Choose $0<\varepsilon_1,\varepsilon_2<\varepsilon$
such that 
\begin{align}
\tan(\theta-\varphi-\varepsilon_1), \; \tan(\theta+\varphi+\varepsilon_2)\in\bbQ.
\end{align}
Furthermore, choose $l,m,l',m'\in 2\bbZ$, satisfying
\begin{equation}
\frac{m}{l}=\tan(\theta-\varphi-\varepsilon_1),\quad
\frac{m'}{l'}=\tan(\theta+\varphi+\varepsilon_2).
\end{equation}
Since $\theta-\varphi-\varepsilon_1 \in (\theta-\varphi-\varepsilon,\theta-\varphi)$ and $\theta+\varphi+\varepsilon_2\in(\theta+\varphi,\theta+\varphi+\varepsilon)$,  one can find points $\bm{b},\bm{c}\in\mathbb{R}^2$ in the dual lattice of $\mathbb{Z}^2$, such that:
\begin{itemize}
\item the dual lattice path from $\bm{c}$ with steps $l,m,l,m,\ldots$ lies in $\Lambda_{\bm{a},\theta,\varphi+\varepsilon}\setminus \Lambda_{\bm{a},\theta,\varphi}$,
\item the dual lattice path from $\bm{b}$ with steps $l',m',l',m',\ldots$ lies in $\Lambda_{\bm{a},\theta,\varphi+\varepsilon}\setminus \Lambda_{\bm{a},\theta,\varphi}$, 
\item there exists an additional dual lattice path joining $\bm{b}$ and $\bm{c}$ providing a cone-like region $\Gamma_{\varepsilon,\ld}$, as introduced in \cref{Def:comblikelkmk}.
\end{itemize}
This is illustrated in the subsequent figure:
\begin{equation}
\vcenter{\hbox{\includegraphics[page=164]{haagduality_diagrams.pdf}}}
\end{equation}
By construction, it is clear that
$\Gamma_{\varepsilon,\ld}^c\cap \ld\cap\bbZ^2$ and $\Gamma_{\varepsilon,\ld}\cap \ld_\varepsilon^c\cap\bbZ^2$ are finite.
\end{proof}

\begin{proof}[Proof of Corollary~\ref{Cor:MR1}]
We show that approximate Haag duality, as introduced in \cref{Section:Intro}, holds with $R_{\varphi,\varepsilon}=0$, $f_{\varphi,\varepsilon,\delta}=0$
for any $\varphi\in (0,2\pi)$ and $\varepsilon>0$ with
$\varphi+4\varepsilon<2\pi$, and $\delta>0$.
For each cone $\ld$ with $|\arg\ld|=\varphi$, choose  $\Gamma_{\varepsilon,\ld}$
as in Lemma \ref{one2}.
Let
\begin{align}
\tilde\Gamma:=
\lmk \Gamma_{\varepsilon,\ld}\cup
\lmk \Lambda\cap \Gamma_{\varepsilon,\ld}^c\rmk\rmk\cap \ld_\varepsilon.
\end{align}
By definition, we have $\ld\subset \tilde\Gamma\subset\ld_\varepsilon$.
Because $ \Lambda\cap \Gamma_{\varepsilon,\ld}^c\cap\bbZ^2$
and $\Gamma_{\varepsilon,\ld}\cap \Lambda_{\varepsilon}^c\cap\bbZ^2$ are finite,
the difference between
$\tilde\Gamma\cap\bbZ^2$ and $\Gamma_{\varepsilon,\ld}\cap\bbZ^2$
is finite. Therefore, by virtue of \cref{ea}, $\pi( \caA_{\Gamma_{\varepsilon,\ld}^c})'=\pi(\caA_{\Gamma_{\varepsilon,\ld}})''$ implies
$\pi(\caA_{\tilde\Gamma^c})'=\pi\lmk \caA_{\tilde \Gamma}\rmk''$.
From the inclusion $\ld\subset \tilde\Gamma\subset\ld_\varepsilon$, we have
\begin{align}
\pi\lmk\caA_{\Lambda^c}\rmk'\subset 
\pi\lmk\caA_{\tilde\Gamma^c}\rmk'
=\pi\lmk \caA_{\tilde \Gamma}\rmk''
\subset 
\pi\lmk \caA_{ \Lambda_\varepsilon}\rmk'',
\end{align}
Namely, condition (i) in the definition of approximate Haag duality holds with $U_{\Lambda,\varepsilon}:=\unit_{\caH}$
and $R_{\varphi,\varepsilon}=0$, and condition (ii)
holds with $\tilde U_{\Lambda,\varepsilon,\delta,t}=\unit$
and $f_{\varphi,\varepsilon,\delta}=0$.
\end{proof}

\section{Proof of \texorpdfstring{\cref{Thm:PT}}{Theorem \ref{Thm:PT}}}

\label{Appendix:Omega}

Let $A$ denote a $C^*$-weak Hopf algebra for the remainder of this appendix. We begin by reviewing two fundamental results in the theory of $C^*$-weak Hopf algebras: the existence of a Haar integral, and the existence and factorization of the canonical grouplike element.

\subsection{Integrals and grouplike elements}
\label{Subsection:Integrals_and_g}

\begin{definition}
    \label{Def:Integrals}
    \begin{enumerate}
        \item An element $\ell\in A$ is a \emph{left integral} if 
    for all $x\in A$,
    \begin{equation*}
        x\ell = \varepsilon^L(x) \ell;
    \end{equation*}
    \item an element $r\in A$ is a \emph{right integral} if 
    for all $x\in A$
    \begin{equation*}
        r x  = r\varepsilon^R(x);
    \end{equation*}
    \item a \emph{two-sided integral} is both a left and a right integral.
    \end{enumerate}
\end{definition}

\begin{proposition}
\label{Prop:LeftRightInt}
\begin{enumerate}
    \item $\ell\in A$ is a left integral if and only if for all $x\in A$,
\begin{equation*}
    \label{eq:LeftIntegral2}  
    S(x) \ell_{(1)} \otimes \ell_{(2)} =
    \ell_{(1)} \otimes x \ell_{(2)};
\end{equation*}
\item $r\in A$ is a right integral if and only if for all $x\in A$,
\begin{equation*}
    \label{eq:RightIntegral2}  
     r_{(1)} S^{-1}(x) \otimes r_{(2)} =
     r_{(1)} \otimes r_{(2)} x;
\end{equation*}
\item $r\in A$ is a right integral if and only if $S(r)$ is a left integral.
\end{enumerate}
\end{proposition}

\begin{theorem}
    \label{Thm:ExistenceHaar}
    There exists a unique non-degenerate two-sided integral $h\in A$ satisfying $\varepsilon^L(h)=\varepsilon^R(h) = 1$, called the Haar integral of $A$. In particular, $h^2 = h = h^* = S(h)$.
\end{theorem}

See Theorem 4.5 in \cite{bohm_weak_1999} for a proof. The Haar integral of $A^*$ is denoted $\hat{h}\in A^*$.

%


The Haar integral of $A^*$ induces a non-degenerate conditional expectation
\begin{equation}
    \label{Eq:DefEL}
    E^L:A\to A^L,\quad E^L (x) \coloneq \eval{\hat{h}}{x_{(2)}} x_{(1)},
\end{equation}
for the unital inclusion $A\supset A^L$ of $C^*$-algebras,
known as the \emph{left Haar conditional expectation} on $A$. Analogously, it induces a 
non-degenerate conditional expectation,
\begin{equation}
    \label{Eq:DefER}
    E^R:A\to A^R,\quad E^R (x) \coloneq \eval{\hat{h}}{x_{(1)}} x_{(2)},
\end{equation}
for the unital inclusion $A\supset A^R$ of $C^*$-algebras,
known as the \emph{right Haar conditional expectation} on $A$.

\begin{remark}
The images of both conditional expectations are $A^L$ and $A^R$ respectively due to the characterization of the left and right integrals; see Lemma~3.2 (a) and (c) in \cite{bohm_weak_1999}. In addition, the fact that they are non-degenerate is due to the fact that the Haar integral is a non-degenerate element; see e.g.~Lemma~3.2 in \cite{bohm_weak_1999}.
\end{remark}

\begin{remark}
\label{Rem:RelationELER}
Both Haar conditional expectations are related by means of
\begin{equation*}
    E^R = S^{\mp 1} \circ E^L \circ S^{\pm 1},
\end{equation*}
by virtue of the invariance of the Haar integral of $A^*$ with respect to the antipode.
\end{remark}

In previous sections we noted that group $C^*$-algebras are specific instances of $C^*$-weak Hopf algebras, distinguished by their trivial coalgebra structure. The following definition generalizes these types of structures within this broader framework.

\begin{definition}
\label{Def:Grouplike}
An invertible element $x\in A$ is said to be \emph{grouplike} if it satisfies
\[ 
    x_{(1)}\otimes x_{(2)} =
    x 1_{(1)}\otimes x1_{(2)}=
    1_{(1)}x\otimes 1_{(2)}x.
\]
\end{definition}

Specifically, we focus on the following notable element.

\begin{theorem}
\label{Thm:g}
There exists a unique invertible positive grouplike element $g\in A$ that implements $S^2$ as an inner automorphism, i.e.
\begin{equation*}
    S^2(x) = g x g^{-1}
\end{equation*}
for all $x\in A$, and is compatible with the Haar integral $h\in A$ in the sense that
\begin{equation*}
    h_{(1)} \otimes g h_{(2)} g = h_{(2)} \otimes h_{(1)}.
\end{equation*}
\end{theorem}

This unique element is known as the \emph{canonical grouplike element}.

\begin{proof}
    See Proposition 4.9 in \cite{bohm_weak_1999}.
\end{proof}

In \cref{Prop:J} we stated that there exists a linear map $J:A\to A$ which allows to define a dual $*$-representation; more concretely,
\begin{equation}
    \label{Eq:ExplicitExpressionJ}
    J(x) \coloneq g^{-\frac{1}{2}} S(x) g^{\frac{1}{2}}
\end{equation}
defines the map, for all $x\in A$.
\begin{proof}[Proof of \cref{Prop:J}]
First, note that $J$ is antimultiplicative and unital since $S$ is antimultiplicative and unital.
Moreover, $J$ is involutive since $S$ is an antihomomorphism with $S(g) = g^{-1}$ and $S^2(x) = gxg^{-1}$ for all $x\in A$.
This implies that the notion of dual $*$-representation is reflexive.
In addition, since $g^{\pm 1/2}$ are grouplike elements, it is also simple to prove that $J$ is anticomultiplicative.
\end{proof}

We refer the reader to Subsection~3.1 and Proposition~3.5 from \cite{bohm_weak_2000}.

\begin{proposition}
\label{Prop:gFactorization}
The canonical grouplike element can be factorized in the form
\begin{equation*}
g = \gL \gR^{-1} \quad\text{ for }\quad \gL \coloneq E^L(h)^{\frac{1}{2}} \;\text{ and }\; \gR \coloneq E^R(h)^{\frac{1}{2}}.
\end{equation*}
In addition, note that $g_L$ and $g_R$ are positive and they are related by $g_R = S^{\pm 1}(g_L)$.
\end{proposition}

\begin{proof}
    See Lemma~4.12 in \cite{bohm_weak_1999}.
\end{proof}

Let $\hat{g}$, $\dualgL$ and $\dualgR\in A^*$ stand for their dual analogues.

\begin{proposition}
\label{Rem:gRgL_Exchange}
The following equalities hold for all $x\in A$:
        \begin{align*}
            \gR x
                &= \eval{\dualgL}{x_{(2)}} x_{(1)}
            , \quad 
            \gL x
                = \eval{\dualgL}{x_{(1)}}  x_{(2)}
            , \\ 
            x \gL
                &=  \eval{\dualgR}{x_{(1)}} x_{(2)}
            , \quad 
            x \gR
                =   \eval{\dualgR}{x_{(2)}} x_{(1)}
            ,
        \end{align*}
The dual analogues, which hold for all $\phi\in A^*$ ar given as follows:
        \begin{align*}
            \dualgR \phi
                &= \eval{\phi_{(2)}}{\gL} \phi_{(1)} = \eval{\phi}{(\,\cdot\,)\gL}
            , \quad 
            \dualgL \phi
                = \eval{\phi_{(1)}}{\gL} \phi_{(2)}= \eval{\phi}{\gL(\,\cdot\,)}
            , \\ 
            \phi \dualgL
                &=  \eval{\phi_{(1)}}{\gR} \phi_{(2)}= \eval{\phi}{\gR(\,\cdot\,)}
            , \quad 
            \phi \dualgR
                =   \eval{\phi_{(2)}}{\gR} \phi_{(1)}= \eval{\phi}{(\,\cdot\,)\gR}
            .
        \end{align*}
\end{proposition}

\begin{proof}
    See Lemma~4.13 and Scholium~2.7 in \cite{bohm_weak_1999}.
\end{proof}


\subsection{Index of the Haar conditional expectations}

\label{Subsection:Watatani}

Having introduced the previous elements, we are now ready to present an element of technical interest, as it provides the appropriate normalization weights required for the construction of the pulling-through element in \cref{Thm:PT}.

This element is identified as the Watatani index of both the left and right Haar conditional expectations. In his seminal paper~\cite{watatani_index_1990}, Watatani introduced the concept of a quasi-basis for conditional expectations $E:\mathcal{M}\to \mathcal{N}$  arising from uni\-tal in\-clu\-sions of finite-dimensional $C^*$-algebras $\mathcal{M}\supset \mathcal{N}$, and an element currently known as the \emph{Watatani index}.
More concretely, a \emph{quasi-basis} is any element of the form $Q = p_1\otimes q_1+\cdots +p_k\otimes q_k\in \mathcal{M}\otimes \mathcal{M}$ satisfying
\begin{equation*}
    (E\otimes \mathrm{Id})(Q(1\otimes x)) = (E\otimes \mathrm{Id})((x\otimes 1)Q)
\end{equation*}
for all $x\in \mathcal{M}$. Then, the index of $E$ is defined as the element
\begin{equation*}
    \operatorname{Index}E \coloneq p_1 q_1 + \cdots + p_k q_k\in \mathcal{A}.
\end{equation*}

\begin{remark}
\label{Rem:WatataniProps}
The element $\operatorname{Index}E$ is well-defined, i.e.~it does not depend on the choice quasi-basis. Moreover, it is invertible, positive, and central.
\end{remark}

\begin{proof}
See Proposition~1.2.8, Corollary~2.1.7 and Lemma~2.3.1 in \cite{watatani_index_1990}.
\end{proof}

\begin{definition}\label{Def:Hypercenter}
Let $A$ be a C*-weak Hopf algebra. The \emph{hypercenter} of $A$ is de\-fi\-ned as the $*$-subalgebra
\[
    \operatorname{Hyp}{A}\coloneq A^L\cap A^R \cap\operatorname{Center}{A},
\]
and their elements are said to be \emph{hypercentral}.
\end{definition}

\begin{lemma}\label{Lemma:Watatani}
Let $A$ be a $C^*$-weak Hopf algebra. The element
\begin{equation*}
    \imath \coloneq \varepsilon^L( h \gL^{-2} ) = \varepsilon^R(\gR^{-2}h)
\end{equation*}
is invertible, positive, invariant under the antipode, and hypercentral.
\end{lemma}

The following proof is based on the discussion in Proposition~4.3 of \cite{bohm_weak_2000}.

\begin{proof}
First, let us note that the Haar integral $\hat{h}\in A^*$ induces a left integral $\ell\in A$ and a right integral $r\in A$ such that
\begin{align}
\label{Eq:DualLeftInt}
    \eval{\hat{h}}{\ell_{(2)}}\ell_{(1)} = 1, 
    \quad
    \eval{\hat{h}_{(2)}}{\ell}\hat{h}_{(1)} = \eval{\hat{h}}{(\,\cdot\,)\ell} = \varepsilon,
    \\
\label{Eq:DualRightInt}
    \eval{\hat{h}}{r_{(1)}}r_{(2)} = 1, 
    \quad
    \eval{\hat{h}_{(1)}}{r}\hat{h}_{(2)} =\eval{\hat{h}}{r (\,\cdot\,)} = \varepsilon;
\end{align}
known as the dual left and dual right integrals of $\hat{h}$, respectively; see Theorem~3.18 in \cite{bohm_weak_1999}. Moreover, it is possible to check that they are given by the expressions
\begin{equation}
\label{Eq:DualIntExpr}
    \ell = h g_L^{-2}\quad\text{ and }\quad r = g_R^{-2}h = S^{\pm 1}(\ell);
\end{equation}
see Proposition~4.14 in \cite{bohm_weak_1999}. Let us prove now that $E^L$ has quasi-basis 
\[
    Q \coloneq S(h_{(1)}) \otimes \gR^{-2} h_{(2)}
\]
On the one hand, note that by~\cref{Prop:LeftRightInt}, for any $x\in A$ it holds that
\begin{align*}
    S(h_{(1)}) E^L(\gR^{-2}h_{(2)} x)
    &=
    S(h_{(1)} S^{-1}(x)) {E}^{L}(\gR^{-2}h_{(2)})
    \\
\intertext{now, due to the fact that $S$ is antimultiplicative, see~\cref{Def:CWHA}, this equals}
    & =
    x S(h_{(1)}) {E}^{L}(\gR^{-2}h_{(2)})
    \\
\intertext{and, by virtue of~\cref{Eq:DualIntExpr}, where the explicit form of $r$ is provided,}
    &=
    x S(r_{(1)}) {E}^{L}(r_{(2)})
    \\
\intertext{by simply substituting the expression of $E^L$ from~\cref{Eq:DefEL},}
    &=
    x S(r_{(1)}) r_{(2)} \eval{\hat{h}}{r_{(3)}}
    \\
\intertext{as a consequence of the definition of $\varepsilon^R$ in \cref{Eq:DefEpsilonLR},}
    &=
    x \varepsilon^R(r_{(1)})  \eval{\hat{h}}{r_{(2)}}
    \\
\intertext{due to the identity in Equation~2.8a in \cite{bohm_weak_1999},}
    &=
    x 1_{(1)} \eval{\hat{h}}{ r 1_{(2)} }
    \\
\intertext{by the expression of $r$ in terms of the dual left integral $\ell$ of $\hat{h}$, see \cref{Eq:DualIntExpr},}
    &=
    x 1_{(1)} \eval{\hat{h}}{ S^{-1}(\ell) 1_{(2)} }
    \\
\intertext{since $\hat{h}$ is invariant under the antipode, see \cref{Thm:ExistenceHaar},}
    &=
    x 1_{(1)} \eval{\hat{h}}{  S(1_{(2)}) \ell }
    \\
\intertext{due to the fact that $\ell$ is a left integral of $\hat{h}$, see \cref{Eq:DualLeftInt},}
    &=
    x 1_{(1)} \eval{ \varepsilon }{ S( 1_{(2)} ) } 
    \\
\intertext{by the defining property of the counit $\varepsilon\in A^*$, and clearly:}
    &=
    x 1_{(1)} \eval{\varepsilon}{  1_{(2)}}
    = 
    x\\
\intertext{On the other hand, for any $x\in A$, by simply using that $S$ is antimultiplicative,}
    {E}^L(x S(h_{(1)})) \gR^{-2} h_{(2)}
    &=
    {E}^L( S(h_{(1)} S^{-1}(x)) ) \gR^{-2} h_{(2)}
    \\
\intertext{since $h\in A$ is in particular a right integral, see \cref{Prop:LeftRightInt},}
    &=
    {E}^L( S(h_{(1)}) ) \gR^{-2} h_{(2)} x
    \\ 
\intertext{by substituting the explicit expression of $r$ in \cref{Eq:DualIntExpr},}
    &=
    {E}^L( S(r_{(1)}) ) r_{(2)} x
    \\
\intertext{as $r$ can be rewritten in terms of $\ell$ as in \cref{Eq:DualIntExpr},}
    &=
    {E}^L( S( S^{-1}(\ell)_{(1)} ) ) S^{-1}(\ell)_{(2)} x
    \\
\intertext{as a consequence of the antimultiplicativity of $S$,}
    &=
    {E}^L (\ell_{(2)} ) S^{-1}(\ell_{(1)}) x
    \\
\intertext{by simply substituting the expression of $E^L$ in \cref{Eq:DefEL},}
    &= 
    \ell_{(2)} \eval{\hat{h}}{\ell_{(3)}} S^{-1}(\ell_{(1)}) x
    \\
\intertext{by virtue of the fact that $\ell$ is a dual left integral of $\hat{h}$, see \cref{Eq:DualLeftInt},}
    &=
    1_{(2)}  S^{-1}( 1_{(1)} )x
    && 
    \\
\intertext{by simply rearranging terms}
    &=
    S^{-1}( 1_{(1)} S(1_{(2)}) )x
    \\
\intertext{as $\varepsilon^L$ and $S^{-1}$ preserve the unit:}
    &=
    (S^{-1} \circ \varepsilon^L)(1)x = x.
\end{align*}
Together with \cref{Lemma:Watatani}, this proves that
\begin{equation*}
    \operatorname{Index}E^L = S(r_{(1)}) r_{(2)} = \varepsilon^R(g_R^{-2}h)
\end{equation*}
and hence it is a positive central element.
,It is straightforward to check by the antimultiplicativity of $S$ and the relation between the Haar conditional expectations presented in \cref{Rem:RelationELER} that, given an expansion $Q \coloneq p_1\otimes q_1+\cdots + p_k \otimes q_k$, the element $\tilde{Q}\coloneq S(q_1)\otimes S(p_1) +\cdots + S(q_k)\otimes S(p_k)$ is in turn a quasi-basis for $E^R$. Thus, in particular,
\begin{equation*}
    \operatorname{Index}E^R = \varepsilon^L(h g_L^{-2}) = S^{\pm 1}(\operatorname{Index}E^L).
\end{equation*}
Now, let us prove that $\operatorname{Index}E^L = \operatorname{Index}E^R$, which concludes the proof.
First, let us note that by Artin-Wedderburn's theorem,
\[ A = A e_{1}+\cdots + A e_{r}, \quad A e_a\cong \mathrm{M}_{n_a\times n_a}(\mathbb{C}),\] where $e_1,\ldots, e_r\in A$ are the minimal central idempotents of $A$. Let $\myket{ i , a}\mybra{ j , a }\in A$ denote the matrix units within $\mathrm{M}_{n_a\times n_a}(\mathbb{C})$, where $i,j=1,\ldots, n_a$ and $a=1,\ldots,r$.
Then, the minimal central idempotents of $A$ and the characters $\chi_a:A\to \mathbb{C}$ of the irreducible representation are given by
\[
e_a = \sum_{ i = 1 }^{ n_a } \myket{ i , a }\mybra{ i , a }
\quad
\chi_a(x)=
    \sum_{ j = 1 }^{ n_a } \mybra{ j, a } x \myket{ j, a },
\]
respectively.
Then, Equations~4.13 and 4.14 in \cite{bohm_weak_1999} state that
    \begin{align*}
        S(h_{(1)}) \otimes h_{(2)} &=
            \sum_{a=1}^r \frac{1}{\chi_a(g^{-1})}
            \sum_{i,j=1}^{n_a} \myket{ i, a } \mybra{ j , a } g^{-\frac{1}{2}}
            \otimes g^{-\frac{1}{2}} \myket{ j , a} \mybra{ i , a } ,\\
        h_{(1)} \otimes S(h_{(2)}) &=
            \sum_{a=1}^r \frac{1}{\chi_a(g^{- 1})}
            \sum_{i,j=1}^{n_a} \myket{ i , a } \mybra{ j , a } g^{\frac{1}{2}}
            \otimes g^{\frac{1}{2}} \myket{ j , a } \mybra{ i , a },
    \end{align*}
where $g$ is the canonical grouplike element of $A$, and therefore, on the one hand,
\begin{align*}
    \operatorname{Index}E^L &=
        \sum_{a=1}^r \frac{1}{\chi_a(g^{- 1})}
        \sum_{i=1}^{n_a} \myket{i , a } \big( \sum_{j=1}^{n_a} \mybra{ j , a }
        g^{-\frac{1}{2}} \gR^{-2} g^{-\frac{1}{2}}   \myket{ j , a } \big) \mybra{ i , a }
    \\ &=
        \sum_{a=1}^r \frac{1}{\chi_a(g^{- 1})} \sum_{i=1}^{n_a} \myket{ i , a }
        \chi_a(\gL^{-1} \gR^{-1})   \mybra{ i ,  a }
    =
    \sum_{a=1}^r \frac{ \chi_a(\gL^{-1} \gR^{-1})}{ \chi_a(g^{- 1})} e_a,
\end{align*}
and, on the other hand,
    \begin{align*}
        \operatorname{Index}E^R &= \sum_{a=1}^r \frac{1}{\chi_a(g^{- 1})}
            \sum_{i=1}^{n_a} \myket{ i , a } \big( \sum_{j = 1}^{ n_a } \mybra{ j , a }
            g^{\frac{1}{2}} \gL^{-2} g^{\frac{1}{2}} \myket{ j , a } \big) \mybra{ i , a }
        \\ &=
        \sum_{a=1}^r \frac{1}{\chi_a(g^{- 1})} \sum_{i=1}^{n_a} \myket{ i , a }
            \chi_a(\gL^{-1} \gR^{-1})   \mybra{ i ,  a }
        =
        \sum_{a=1}^r \frac{ \chi_a(\gL^{-1} \gR^{-1})}{ \chi_a(g^{- 1})} e_a.
    \end{align*}
The two expressions coincide, as we wanted to prove.
\end{proof}

\begin{remark}\label{Rem:TrivialHyp}
    (i) If $A$ has trivial hypercenter, i.e.~$\operatorname{Hyp}A=\mathbb{C}1$, then
    \begin{equation*}
        \imath = \eval{ \varepsilon }{ \gR^{-2} } \eval{ \varepsilon }{ 1 }^{-1} 1.
    \end{equation*}
    For instance, this happens in the particular case in which $A$ is biconnected.
    (ii) Furthermore, if $A$ has involutive antipode, then
    $
        \eval{\varepsilon}{\gR^{-2}} = \dim A
    $.
    (iii) In particular, for a group $C^*$-algebra, $\imath = |G|1$.
\end{remark}

\begin{proof}
    The first statement follows from the last parts of the previous proof. The last statement follows as a consequence of Lemma~5.1 in \cite{bohm_weak_2000}.
\end{proof}

\begin{lemma}
\label{Lemma:Existence_j}
The element
$ \hat\jmath
        \coloneq \eval{\varepsilon_{(2)}}{\imath} \varepsilon_{(1)}
         \in \operatorname{Hyp}(A^*) $
is such that
\begin{equation*}
    \hat\jmath
        = \eval{\varepsilon_{(2)}}{\imath} \varepsilon_{(1)}
        = \eval{\varepsilon_{(1)}}{\imath} \varepsilon_{(2)},
    \quad
    \imath
        = \eval{\hat\jmath}{1_{(1)}} 1_{(2)}
        = \eval{\hat\jmath}{1_{(2)}} 1_{(1)};
\end{equation*}
moreover, $\hat{\jmath}$ is positive, invariant under the antipode,  invertible, and
\begin{equation*}
    \hat{\jmath}^{-1} = \eval{\varepsilon_{(2)}}{\imath^{-1}} \varepsilon_{(1)}
            = \eval{\varepsilon_{(1)}}{\imath^{-1}} \varepsilon_{(2)}.
\end{equation*}
\end{lemma}

\begin{proof}
The first statement is a particular case of Lemma~2.14 from \cite{bohm_weak_1999}.
Also,
\begin{align*}
    \hat{S}(\hat{\jmath})
    &=
    \eval{\varepsilon_{(1)}}{\imath} \hat{S}(\varepsilon_{(2)})
    =
    \eval{\varepsilon_{(1)}}{S(\imath)} \hat{S}(\varepsilon_{(2)})
    \\
    &=
    \eval{\hat{S}(\varepsilon_{(1)})}{\imath} \hat{S}(\varepsilon_{(2)})
    =
    \eval{\hat{S}(\varepsilon)_{(2)}}{\imath} \hat{S}(\varepsilon)_{(1)}
    =
    \eval{\varepsilon_{(2)}}{\imath} \varepsilon_{(1)}
    =
    \hat{\jmath}
\end{align*}
where the first equality simply uses the definition of $\hat\jmath$, the second follows by the definition of the dual antipode, see \cref{Rem:DualCWHA}, the third step holds since $S(\imath)= \imath$, as proven in \cref{Lemma:Watatani}, the fourth equality is due to the anticomultiplicativity of $S$, the fifth is due to the fact that $\hat{S}$ is unit-preserving, i.e.~$\hat{S}(\varepsilon) = \varepsilon$, 
and the last one holds by virtue of the expression of $\jmath$ in \cref{Lemma:Existence_j}.
It is also simple to prove that $\hat\jmath$ is positive, as $\imath$ is positive and hypercentral.
The second formula follows as a consequence of the following calculation
\begin{align*}
    \eval{\varepsilon_{(1)}}{\imath}
    \eval{\varepsilon_{(2')}}{\imath^{-1}}\varepsilon_{(2)} \varepsilon_{(1')} 
    &=
    \eval{\varepsilon_{(1)}}{\imath}
    \eval{\varepsilon_{(3)}}{\imath^{-1}} \varepsilon_{(2)} 
    \\&=
    \eval{\varepsilon_{(1)}}{\imath}
    \eval{\varepsilon_{(2)}}{\imath^{-1}} \varepsilon_{(3)}
    =
    \eval{\varepsilon_{(1)}}{\imath \imath^{-1}}  \varepsilon_{(2)} = \varepsilon,
\end{align*}
where the first equality holds by using the dual analogue of \cref{Item:UnitWeaklyComult} in \cref{Def:CWHA},  the second is due to the fact that $\imath^{-1}$ is hypercentral, and the last steps are a consequence of \cref{Rem:DualCWHA} and the dual analogue of the unit axiom.
\end{proof}

Let $\hat\imath$ and $\jmath$ denote the corresponding dual analogues of $\imath$ and $\hat\jmath$. The following result provides an explicit expression of the latter one in elementary terms.

\begin{lemma}
\label{Lemma:Expression_j}
$
    \jmath = \eval{\hat{h}}{(\gL^{-2})_{(1)}} (\gL^{-2})_{(2)}
           = \eval{\hat{h}}{(\gR^{-2})_{(2)}} (\gR^{-2})_{(1)}.
$
\end{lemma}

\begin{proof}
Let $\hat\varepsilon^L,\hat\varepsilon^R:A^*\to A^*$ denote the corresponding dual analogues of the source and target counital maps, introduced in \cref{Eq:DefEpsilonLR}.
Then, on the one hand,
\begin{align*}
    \eval{\hat{h}}{(\gR^{-2})_{(2)}} (\gR^{-2})_{(1)}
    &
    =
    \eval{ \hat{h}\dualgL^{-2} }{ 1_{(2)} } 1_{(1)} 
    =
    \eval{ \hat{h}\dualgL^{-2} }{\varepsilon^L(1_{(2)}) }1_{(1)}
    \\&
    =
    \eval{ \hat\varepsilon^L(\hat{h} \dualgL^{-2}) }{ 1_{(2)} }1_{(1)}
    =
    \eval{\hat\imath }{ 1_{(2)} } 1_{(1)} 
    =
    \jmath,
\end{align*}
where in the first step we have used the exchange properties for $g_R$ in \cref{Rem:gRgL_Exchange}, the second equality holds since $1_{(1)}\otimes 1_{(2)}\in A^R\otimes A^L$ and $\varepsilon^L:A\to A^L$ is a projection, for the third it is easy to check that $ \eval{\hat\varepsilon^{L}(\phi)}{y} = \eval{\phi}{\varepsilon^{L}(y)}$ for all $y\in A$ by \cref{Rem:DualCWHA}, and the last follows from the expression of the index in \cref{Lemma:Watatani} in conjunction with \cref{Lemma:Existence_j}.
The second equality follows similarly.
\end{proof}


\subsection{Proof of \texorpdfstring{\cref{Thm:PT}}{Theorem \ref{Thm:PT}}}
The following rephrases the aforementioned result in elementary terms; note that the order of presentation is slightly different.

\begin{theorem*}
    Let $A$ be a $C^*$-weak Hopf algebra.
    Then,
    \begin{enumerate}
    \item\label{Item-PT:Omega}
    the element
    \begin{equation*}
        \Omega \coloneq \imath^{-1} \gL^{-1} h \gL^{-1} \in A
    \end{equation*}
    is cocentral, non-degenerate, and satisfies
    \[
        \Omega^2 = \Omega^* = S(\Omega) = J(\Omega) = \Omega;
    \]
    \item\label{Item-PT:omegaReexpr}
    let $\omega \in A^*$ be the dual analogue of $\Omega\in A$, then 
    for all $x\in A$,
    \begin{equation*}
        \eval{\hat{h}}{\jmath^{-1} \gL^{-1} \gR^{-1} x}
        =
        \eval{\omega}{x}
        =
        \eval{\hat{h}}{ x \jmath^{-1} \gL^{-1} \gR^{-1}}
    \end{equation*}
    \item\label{Item-PT:JLJRpt}
    the linear maps
    $J^L,J^R:A\to A$, defined for all $x\in A$ by the expressions
    \begin{equation*}
        J^L(x) \coloneq \gL^{-1} S(x)  \gL
        ,\quad
        J^R(x) \coloneq \gL  S^{-1}(x) \gL^{-1},
    \end{equation*}
    fulfill the following identities, for all $x,y\in A$:
    \begin{align*}
        J^L(x) \Omega_{(1)}\otimes\Omega_{(2)}
        &=
        \Omega_{(1)} \otimes x \Omega_{(2)}
        ,\\
        \Omega_{(1)} J^R(y) \otimes\Omega_{(2)}
        &=
        \Omega_{(1)} \otimes  \Omega_{(2)}y;
    \end{align*}
    \item\label{Item-PT:JLJRprops}
    in particular, $J^L$ and $J^R$ are involutive algebra antihomomorphisms with
    \begin{equation*}
        \omega \circ J^L = \omega \circ J^R = \omega \circ J = \omega \circ S = \omega;
    \end{equation*}
    \item\label{Item-PT:xi}
    the elements
    \begin{equation*}
        \xi_{L} \coloneq \imath^{\frac{1}{2}}\jmath^{\frac{1}{2}} \gL \in A^L, \quad \xi_{R} \coloneq \imath^{\frac{1}{2}}\jmath^{\frac{1}{2}} \gR \in A^R, \quad \xi \coloneq \xi_L\xi_R,
    \end{equation*}
    are invertible and positive, and it is simple to check that, for all $x\in A$,
    \begin{equation*}
        J^L(x) = \xi^{-\frac{1}{2}} J(x) \xi^{\frac{1}{2}}
        , \quad
        J^R(x) = \xi^{\frac{1}{2}} J(x) \xi^{-\frac{1}{2}}
        , \quad 
        g = \xi_L\xi_R^{-1},
    \end{equation*}
    \item\label{Item-PT:xiExch}
    let $\hat\xi$, $\hat\xi_{L}$ and $\hat\xi_R$ stand for the dual analogues, then, 
    for all $x\in A$,
    \begin{align*}
        \xi_R x &= \eval{\hat{\xi}_L}{x_{(2)}} x_{(1)}
        , \quad 
        \xi_L x = \eval{\hat{\xi}_L}{x_{(1)}}  x_{(2)}
        , \\ 
        x \xi_L &=  \eval{\hat{\xi}_R}{x_{(1)}} x_{(2)}
        , \quad 
        x \xi_R  =  \eval{\hat{\xi}_R}{x_{(2)}} x_{(1)}
        ;
    \end{align*}
    \item\label{Item-PT:omegaXi}
    the following identity holds for all $x\in A$:
    \begin{equation*}
        \eval{\omega}{\xi^{\frac{1}{2}} J(\Omega_{(1)}) \xi^{\frac{1}{2}} x} \Omega_{(2)} = x;
    \end{equation*}
    or, equivalently,
    \begin{equation*}
        \eval{\omega}{\Omega_{(1)}} \Omega_{(2)} = \xi^{-1}.
    \end{equation*}
    \item\label{Item-PT:XiClosing}
    the following identities hold
    for all $x\in A$
    \begin{align*}
        \eval{\omega}{g^{\frac{1}{2}} x_{(1)} g^{\frac{1}{2}} J(x_{(2)})} x_{(3)}  &= x,
        \\
        \eval{\omega}{g^{-\frac{1}{2}} J(x_{(1)}) g^{-\frac{1}{2}} x_{(2)}} x_{(3)}  &= x.
    \end{align*}
    \end{enumerate}
\end{theorem*}

\begin{proof}
First, note that
\begin{equation}
\label{Eq:gLhgRh}
    \gL^\theta h = \gR^\theta h
    ,\quad
    h \gL^\theta = h \gR^{\theta} 
\end{equation}
for all $\theta\in\mathbb{R}$. Indeed, this is a consequence of the following calculation:
\[
    \gR h = \varepsilon^L(\gR)h = S(\gR) h = \gL h ,
\]
since $h$ is in particular a left integral, see \cref{Thm:ExistenceHaar,Def:Integrals}, and $\varepsilon^L\upharpoonright A^R = S\upharpoonright A^R$.
A similar argument can be employed to prove the other equation, and both are extended to all powers of $\theta\in \mathbb{R}$, since $S$ is antimultiplicative. Then, the following are equivalent definitions of $\Omega$:
\begin{equation}
    \label{eq:EquivDefsOmega}
    \Omega = \imath^{-1} \gR^{-1} h \gR^{-1} = (\imath\gL\gR)^{-\frac{1}{2}} h (\imath\gL\gR)^{-\frac{1}{2}}.
\end{equation}

\begin{stepPTproof}
In first place, note that $\Omega$ is positive since $h$, $\gL$ and $\imath$ are positive elements and $\imath$ is central. Moreover, it is an idempotent since
\begin{equation*}
    \Omega^2
    =
    \imath^{-2} \gL^{-1} h \gL^{-2} h \gL^{-1}
    =
    \imath^{-2} \gL^{-1} \varepsilon^L(h \gL^{-2})h \gL^{-1}
    =
    \Omega,
\end{equation*}
where in the first equality we have employed that $\imath$ is central by \cref{Lemma:Watatani}, the second equality follows from the fact $h$ is a left integral, see \cref{Thm:ExistenceHaar,Def:Integrals}, and the last equality simply employs the explicit expression of $\imath$ in \cref{Lemma:Watatani}.
Furthermore, $\Omega$ is cocentral since
\begin{equation}\label{eq:exprOmegaCoprod}
    \Omega_{(1)} \otimes \Omega_{(2)}
    =
    h_{(1)} \otimes \imath^{-1} \gR^{-1} h_{(2)} \gR^{-1}
    =
    h_{(2)} \otimes \imath^{-1} \gL^{-1} h_{(1)} \gL^{-1}
    =
    \Omega_{(2)} \otimes \Omega_{(1)},
\end{equation}
where in the first step we have used \cref{eq:EquivDefsOmega}, and that $\imath,g_R\in A^R$ and thus satisfy \cref{Rem:CharALAR}, the second equality follows from the compatibility formula in \cref{Thm:g}, and for the third we have applied \cref{Item-PT:Omega,Rem:CharALAR}; this proves that $\Omega$ is cocentral.
That $\Omega$ is non-degenerate follows from the fact that $h$ is non-degenerate and the elements $\imath$ and $\gR$ are both invertible. Indeed, for all elements $x\in A$, let $f_x\in A^*$ be such that $\eval{f_x}{h_{(1)}}h_{(2)} = x$, hence
\begin{equation}
\eval{\phi_x}{\Omega_{(1)}} \Omega_{(2)} = x
\quad\text{ for }\quad
\phi_x \coloneq f_{\imath \gR x \gR}.
\end{equation}
by virtue of the first equality in \cref{eq:exprOmegaCoprod}. Finally, $S(\Omega) = \Omega$ since $S$ is antimultiplicative, $S(\imath) = \imath$, $S(h) = h$, and due to \cref{Eq:gLhgRh} since $\gR = S(\gL)$. It is then straightforward to check that $J(\Omega) =\Omega$ as a consequence of \cref{Eq:gLhgRh}.
\end{stepPTproof}


\begin{stepPTproof}
It is an straightforward consequence of the fact that $\omega$ is trace-like, as we have just proven in \cref{Item-PT:Omega}, the identities in \cref{Rem:gRgL_Exchange}, and \cref{Lemma:Watatani,Lemma:Existence_j}, that
\begin{equation}
    \label{eq:iminusExch}
    \hat\imath^{-1} \phi = \phi \hat\imath^{-1}  = \eval{\phi}{\jmath^{-1}(\,\cdot\,)}  =  \eval{\phi}{(\,\cdot\,)\jmath^{-1}},
\end{equation}
for all $\phi\in A^*$. Together with the dual analogues presented in \cref{Rem:gRgL_Exchange},
\begin{align*}
    \eval{\omega}{x}
    &= \eval{\hat\imath^{-1} \hat{g}_L^{-1} \hat{h}\hat{g}_L^{-1}}{x}
    = \eval{\hat\imath^{-1}\hat{g}_L^{-1}\hat{h}}{g_R^{-1} x}
    \\
    &
    = \eval{\hat{g}_L^{-1}\hat{h}}{\jmath^{-1} g_R^{-1} x }
    = \eval{\hat{h}}{ g_L^{-1} \jmath^{-1}  g_R^{-1} x },
\end{align*}
as we wanted to prove, since $\jmath$ is central. The other equation follows analogously.
\end{stepPTproof}


\begin{stepPTproof}
Let $x\in A$ be arbitrary, then
\begin{align*}
    J^L(x) \Omega_{(1)} \otimes \Omega_{(2)}
    &=
    J^L(x) \gL^{-1} h_{(1)} \otimes h_{(2)} \imath^{-1}\gR^{-1}
    \\ &
    =
    \gL^{-1} S(x)h_{(1)} \otimes h_{(2)} \imath^{-1}\gR^{-1}
    \\ &
    =
    \gL^{-1} h_{(1)} \otimes x h_{(2)} \imath^{-1}\gR^{-1}
    =
    \Omega_{(1)} \otimes x \Omega_{(2)}\vphantom{\gR^{-1}}
\end{align*}
where the first step follows by the definition of $\Omega$ in \cref{Item-PT:Omega}, \cref{Eq:gLhgRh}, and the properties of $\imath$, $g_L$ and $g_R$ presented in \cref{Rem:CharALAR}, the second equality is immediate by simply substituting and simplifying the expression of $J^L$ in \cref{Item-PT:JLJRpt},  the third is a consequence of the fact that $h\in A$ is a left integral, see \cref{Prop:LeftRightInt,Thm:ExistenceHaar}.
The second identity in the statement follows similarly.
\end{stepPTproof}


\begin{stepPTproof}
Let us prove the last equality. Note that by rewriting $\omega$ as in \cref{Item-PT:omegaReexpr} of this proof:
\begin{align*}
    \eval{\omega}{S(x)} 
    &=
    \eval{\hat{h}}{S(x)\jmath^{-1}g_L^{-1}g_R^{-1}}
    \\
\intertext{now, due to \cref{Rem:DualCWHA} and the fact that $\hat{h}\circ S^{-1}=\hat{h}$ by \cref{Thm:ExistenceHaar},}
    &=
    \eval{\hat{h}}{S^{-1}(S(x)\jmath^{-1}g_L^{-1}g_R^{-1})}
    \\
\intertext{and since $S(\jmath) = \jmath$, $g_R = S^{\pm 1}(g_L)$ and $S$ is antimultiplicative,}
    &=
    \eval{\hat{h}}{\jmath^{-1}g_L^{-1}g_R^{-1} x }
    \\
\intertext{again by virtue of \cref{Item-PT:omegaReexpr} of this proof:}
    &=
    \eval{\omega}{x}.
\end{align*}
The other equalities follow from the fact $\omega$ is tracelike and $J$, $J^L$ and $J^R$ are defined as adjoints of certain elements with respect to $S$.
\end{stepPTproof}


\begin{stepPTproof}
The statements about the invertibility and positivity of $\xi_L$, $\xi_R$ and $\xi$ follow trivially from the fact that $\imath$, $\jmath$ are positive and central, and $g_L$ and $g_R$ are positive. The other statements follow by simply substituting \cref{Eq:ExplicitExpressionJ} and \cref{Item-PT:JLJRpt} of this proof, and due to the fact that $\imath$ and $\jmath$ are central.
\end{stepPTproof}


\begin{stepPTproof}
Let us recall that $\imath,\jmath\in \operatorname{Hyp}A = A^L\cap A^R \cap\operatorname{Center}A$, and hence
\begin{align*}
    \imath^{\frac{1}{2}} x &= x \imath^{\frac{1}{2}}  = \eval{\hat\jmath^{\frac{1}{2}}}{x_{(1)}} x_{(2)} =  \eval{\hat\jmath^{\frac{1}{2}}}{x_{(2)}} x_{(1)},
    \\
    \jmath^{\frac{1}{2}} x &= x \jmath^{\frac{1}{2}}  = \eval{\hat\imath^{\frac{1}{2}}}{x_{(1)}} x_{(2)} =  \eval{\hat\imath^{\frac{1}{2}}}{x_{(2)}} x_{(1)},
\end{align*}
for all $x\in A$; see \cref{Rem:CharALAR}. By \cref{Rem:gRgL_Exchange},
\begin{align*}
    \xi_R x
    &=
    \imath^{\frac{1}{2}} \jmath^{\frac{1}{2}} \gR x
    = 
    \eval{\dualgL}{x_{(1)}}
        \imath^{\frac{1}{2}} \jmath^{\frac{1}{2}}x_{(2)}
    \\
\intertext{and using that $\jmath^{1/2}$ is hypercentral, and in particular in $A^L$,}
    &=
    \eval{\dualgL}{x_{(1)}}
    \eval{\hat\imath^{\frac{1}{2}}}{x_{(2)}}
    \imath^{\frac{1}{2}} x_{(3)} 
    \\
\intertext{using that $\imath^{1/2}$ is hypercentral, and in particular in $A^L$,}
    &=
    \eval{\dualgL}{x_{(1)}}
    \eval{\hat\imath^{\frac{1}{2}}}{x_{(2)}}
    \eval{\hat\jmath^{\frac{1}{2}}}{x_{(3)}} x_{(4)} 
    \\
\intertext{by simply rewriting it in terms of the multiplication in the dual, as in \cref{Rem:DualCWHA},}
    &=
    \eval{ \dualgL\hat\imath^{\frac{1}{2}}\hat\jmath^{\frac{1}{2}}}{x_{(1)}} x_{(2)}
    =
    \eval{\hat\xi_L}{x_{(1)}} x_{(2)}
\end{align*}
as we wanted to prove. The rest of the identities follow similarly.
\end{stepPTproof}


\begin{stepPTproof}
The equivalence between the first and the second identities follows by virtue of the following calculation, valid for any $x\in A$:
\begin{align}
    \eval{ \omega }{ \xi^{\frac{1}{2}} J(\Omega_{(1)}) \xi^{\frac{1}{2}} x  } \Omega_{(2)} 
    &=
    \eval{ \omega }{ \xi J^L(\Omega_{(1)}) x } \Omega_{(2)} 
    \\ &=
    \eval{ \omega }{  J^L(\Omega_{(1)}) (x\xi) } \Omega_{(2)} 
    \\ &=
    \eval{ \omega }{ J^L(\Omega_{(1)}) } (x  \xi) \Omega_{(2)} 
    \\ &=
    \eval{ \omega }{ \Omega_{(1)} }  x \xi \Omega_{(2)} 
    =
    x,
\end{align}
where the first step follows by simply substituting the expressions relating $J$ and $J^L$ in \cref{Item-PT:xi} of this proof, the second equality is due to the fact that $\omega$ is tracelike (i.e.~cocentral in $A^*$), as we proved in \cref{Item-PT:omegaReexpr} of this proof, the third is a consequence of the pulling-through identities in \cref{Item-PT:JLJRpt} of this proof, the fourth step holds since $\omega = \omega \circ J^L$, as we proved in \cref{Item-PT:JLJRprops} of this proof, and the last equality follows if and only if the second identity in the statement holds. Let us check now that this is the case:
\begin{align}
    \eval{\omega}{\Omega_{(1)}}\Omega_{(2)}
    &=
    \eval{\hat{h}}{\jmath^{-1} g_L^{-1}g_R^{-1} h_{(1)}} \imath^{-1} g_R^{-1} h_{(2)} g_R^{-1}
    \\
    &=
    \eval{\hat{h}}{ h_{(1)}} \imath^{-1} g_R^{-1} (\jmath^{-1} g_L^{-1}g_R^{-1})  h_{(2)} g_R^{-1}
    \\
    &= \imath^{-1} g_R^{-1} (\jmath^{-1} g_L^{-1}g_R^{-1}) g_R^2 g_R^{-1}
    = (\imath\jmath g_L g_R)^{-1} = \xi^{-1}
\end{align}
where in the first step we have made use of the characterizations of $\Omega = \imath^{-1} g_{R}^{-1} h g_{R}^{-1}$ in \cref{eq:EquivDefsOmega} and of $\omega = \eval{\hat{h}}{\jmath^{-1}g_L^{-1}g_R^{-1}(\,\cdot\,)}$ in \cref{Item-PT:omegaReexpr} of this proof, in the second step we have employed that $h$ is a left-integral and $S(\jmath) = \jmath$, $S(g_L) = g_R$ and $S(g_R) = g_L$, as proven previously, for the third equality we have applied the definition of $g_R = \eval{\hat{h}}{h_{(1)}}^{1/2}h_{(2)} = E^R(h)^{1/2}$ in \cref{Prop:gFactorization}, and the last steps holds because all the terms $g_L$, $g_R$, $\imath$ and $\jmath$ commute and $\xi$ was defined in this way in \cref{Item-PT:xi} of this proof.
\end{stepPTproof}

\begin{stepPTproof}
Let us prove only the first identity in the statement. By the explicit expression of $J$ in \cref{Eq:ExplicitExpressionJ} and the fact that $\omega$ is tracelike,
\begin{align}
     \eval{ \omega }{ g^{\frac{1}{2}} x_{(2)} g^{\frac{1}{2}} J(x_{(3)})  } x_{(1)}
    &=
     \eval{ \omega }{ g x_{(2)} S(x_{(3)}) } x_{(1)}\vphantom{g_R^{-2}}
    \\
\intertext{now by simply considering the definition of $\varepsilon^L$ in \cref{Eq:DefEpsilonLR},}
    &=
     \eval{ \omega }{ g\varepsilon^L(x_{(2)}) } x_{(1)}\vphantom{g_R^{-2}}
    \\
\intertext{since $x_{(1)}\otimes \varepsilon^L(x_{(2)}) = 1_{(1)} x \otimes 1_{(2)}$ by virtue of Lemma~2.3 in \cite{bohm_weak_1999},}
    &=
    \eval{ \omega }{ g 1_{(2)} } 1_{(1)} x\vphantom{g_R^{-2}}
    \\
\intertext{as a consequence of rewriting the expression in \cref{Item-PT:omegaReexpr} in this proof,}
    &=
    \eval{ \hat{h} }{ \jmath^{-1} \gR^{-2} 1_{(2)} } 1_{(1)} x
    \\
\intertext{due to the fact that $\jmath^{-1}$ is hypercentral,}
    &=
     \eval{ \hat{h} }{  \gR^{-2} 1_{(2)} } \jmath^{-1} 1_{(1)} x
    \\
\intertext{by considering the characterization of $A^R$ in \cref{Rem:CharALAR}, and finally}
    &=
    \eval{ \hat{h} }{  (\gR^{-2})_{(2)} } \jmath^{-1} (\gR^{-2})_{(1)} x 
    \\
\intertext{and as a consequence of the explicit expression of $\jmath$ in \cref{Lemma:Expression_j},}
    &=
    \jmath^{-1} \jmath \, x\vphantom{g_R^{-2}}.
\end{align}
The second identity in the statement follows using analogous calculations.
\end{stepPTproof}
This step concludes the proof of the theorem.
\end{proof}

\end{document}